\documentclass[pdftex,a4paper]{article}
\usepackage[top=2.5cm, bottom=2.5cm, left=2.0cm, right=2.0cm]{geometry}

\usepackage{authoraftertitle} 	
\usepackage{authblk} 			

\usepackage[square,sort,comma,numbers]{natbib}
\usepackage[breaklinks,
	colorlinks,
	urlcolor=black,	
	citecolor=red,
	linkcolor=blue
]{hyperref}

\usepackage{bm}											
\usepackage{amssymb}									
\usepackage{amsmath}									
\usepackage{geometry,mathtools}							
\usepackage{graphicx}									
\usepackage{float}										
\usepackage{subcaption}									
\usepackage{import}

\usepackage{booktabs}
\usepackage{multirow}

\usepackage{pgfplots}
\usepackage{pgfplotstable}
\usepackage{tikz}
\usepackage{tikzscale}
\usepackage{siunitx}
\pgfplotsset{compat=1.14}

\usepackage{xcolor}
\newcommand{\blue}[1]{\textcolor{black}{#1}}
\newcommand{\red}[1]{\textcolor{black}{#1}}

\title{\vspace{-2em}Finite Cell Method for functionally graded materials based on V-models and homogenized microstructures}

\author[1]{Benjamin Wassermann\thanks{benjamin.wassermann@tum.de, Corresponding Author}}
\author[1]{Nina Korshunova\thanks{nina.korshunova@tum.de}}
\author[2]{Stefan Kollmannsberger\thanks{stefan.kollmannsberger@tum.de}}
\author[1,3]{Ernst Rank\thanks{ernst.rank@tum.de}}
\author[4]{Gershon Elber\thanks{gershon@cs.technion.ac.il}}

\affil[1]{Chair for Computation in Engineering, Technical University of Munich, Arcisstr. 21, 80333 M\"unchen, Germany}
\affil[2]{Chair of Computational Modeling and Simulation, Technical University of Munich, Arcisstr. 21, 80333 M\"unchen, Germany}
\affil[3]{Institute for Advanced Study, Technical University of Munich, Lichtenbergstr. 2a, 85748 Garching, Germany}
\affil[4]{Center for Graphics and Geometric Computing, Technion Israel Institute of Technology, Haifa 3200003, Israel}

\begin{document}
\maketitle

\hrule
\vspace*{0.2cm}
\noindent \blue{This paper proposes an extension of the finite cell method (FCM) to V-rep models, a novel geometric framework for volumetric representations. This combination of an embedded domain approach (FCM) and a new modeling framework (V-rep) forms the basis for an efficient and accurate simulation of mechanical artifacts, which are not only characterized by complex shapes but also by their non-standard interior structure. These objects gain more and more interest in the context of the new design opportunities opened by additive manufacturing, mainly when graded or micro-structured material is applied. Two different types of functionally graded materials (FGM) are considered: The first one, multi-material FGM, is described using the V-rep models' inherent property to assign different properties throughout the interior of a domain. The second, single-material FGM -- which  is heterogeneously micro-structured -- characterizes the effective material behavior of representative volume elements by homogenization and performs large-scale simulations using the embedded domain approach.}
\vspace*{0.2cm}

\hrule
\vspace*{0.2cm}
\vspace*{0.2cm}
\hrule
\vspace*{0.2cm}
\noindent\textcopyright 2020. This manuscript version is made available under the CC-BY 4.0 license.\\
Preprint submitted to \textit{Advanced Modeling and Simulation in Engineering Sciences} (23. March 2020)

\begingroup
\hypersetup{linkcolor=black}
\tableofcontents
\endgroup
\vspace*{1cm}

\section{Introduction} \label{introduction}

Functionally graded materials (FGM) are advanced materials that offer the possibility to exploit various desired physical properties within one component. FGMs allow manufacturing
'high-performance' and 'multi-functional' artifacts which can resist environmental exposures that could not be withstood by a single material \cite{Suresh1998}. The idea of combining different materials goes back more than 4000 years – the development of the composite bow – and has led to modern carbon fiber reinforced polymers. These composite materials change their material properties step-wise and are consequently prone to delamination. On the other hand, in FGM, material properties vary continuously inside the volume and avoid material interfaces \cite{Bohidar2014}. Specific material properties are achieved by continuous changes in the microstructures, grain sizes, crystal structure, or composition of different materials such as metal, ceramics, polymers, or biological tissues \cite{Noda1999, Zhang2019a}. Prototypes, especially for functionally graded microstructures, can be found in nature, such as in bones, seashells, skin, or wood \cite{Meyers2013} or obtained using topology optimization \cite{Paulino2005,Cheng2017,Liu2018}. Fields of application are, amongst many others, corrosion resistance of chemically exposed components \cite{Chmielewski2016}, bone-like lightweight porous medical implants \cite{Studart2013}, or heat resistance of load-bearing parts such as spacecraft thermal shielding, jet turbine blades, or nuclear reactors \cite{Lee2014, Noda1999}. 

Additive manufacturing (AM) or 3D printing is a generic term for various production techniques in which an object is created by layer-wise material deposition. This material deposition allows the fabrication of objects of almost arbitrary shape. AM is the method of choice for the manufacturing of FGM, as it can (i) resolve tiny structures, (ii) manufacture internal structures which could not be created with any other method, and (iii) the layer-wise material deposition gives control over the composition of the processed material, as well as over the grain size \cite{Zhang2018, Yan2017}. With functionally graded additive manufacturing (FGAM), it is possible to create different single- and multi-material FGM \cite{Loh2018}. \textit{Single-material FGM} specimens consist only of one material that changes its properties due to an adaption of the microstructure, density, or grain size \cite{Aremu2017}. As AM allows the creation of free form structures, a single-material FGM \blue{in the form of a continuously changing microstructure} can be fabricated with any printable material \cite{Ngo2018}. \textit{Multi-material FGMs}, which \blue{blend two or more materials into each other within a volume, }
have recently been under intensive research \cite{Bandyopadhyay2018}. A particular focus was placed on metal-metal combinations, see e.g., ~\cite{Zhang2019a}, where steel and titanium-based combinations are investigated. More complex is the combination of materials of a different kind, such as ceramic-metal compositions \cite{Koopmann2019}. However, these compositions might carry the most potential, as the underlying material properties are very distinct. 

\blue{Material testing is the industry standard to determine the behavior of FGM components. Yet, physical test series are often elaborate and expensive. Therefore, the goal of simulation supported development is to reduce testing to only calibrating data for functionally graded materials and then numerically analyze different shapes and compositions of artifacts. Within this paper's scope, we present two distinct, novel approaches to perform numerical simulations on both single- and multi-material FGMs, respectively.}  
To this end, an analysis-suitable geometrical model needs to be provided, which is naturally created with computer-aided design (CAD) and then transformed into a mesh. This transition process from CAD to an analysis-suitable mesh is error-prone. Depending on the model's quality, manual work must be invested to heal the original geometry before mesh generation can be carried out successfully. Furthermore, the most used CAD representations, i.e., boundary representation (B-rep) or solid-based procedural models, are not well suited for an accurate FGM description. B-rep models represent their volume implicitly by the boundary surfaces, which are modeled either with linear primitives (e.g., triangles and quads) or trimmed spline patches \cite{Cohen2001}. Consequently, B-rep models offer no possibility to represent a heterogeneous material distribution inside the body directly. A workaround is to create vector functions that carry the material properties for each point. These functions can be classified into four different categories: (i) geometrically-independent, e.g., in Cartesian coordinates, (ii) distance-based, (iii) blending composition, and (iv) sweeping composition functions (for a detailed explanation refer to~\cite{Shin2001, Wu2008}). However, except (i), these functions only allow a smooth transition of material properties between the different surfaces, which is not suitable for all material distributions. On the other hand, geometrically-independent functions are cumbersome as they are not related to the object itself. CAD systems using solid-based procedural models follow the constructive solid geometry (CSG) idea \cite{Shah1995}. Here, models are composed of simple primitives: \textit{spheres, cuboids, cylinders, etc.} and more complex primitives: \textit{sweeps, lofts, extrusions, solid of revolution, etc.}  
These primitives are combined with the classical Boolean operations: \textit{union, intersection, difference, negation}, and their derivations: \textit{fillet, chamfer, holes, etc.} Material properties can easily be assigned to the respective primitives. Of course, this requires special treatment in regions with overlapping primitives \cite{Zhang2018}. Furthermore, as primitives are typically provided as implicit functions, they offer, similar to B-rep models, no possibility to further resolve the internal volume. Again, vector functions applied to the primitives are a possible workaround. Another possible geometrical representation is offered by spatial decomposition, such as voxelized models. Here, each voxel can carry its material properties. These voxel models mostly originate from CT scans (e.g., of bones) and provide only a coarse approximation while requiring an extensive amount of storage capacity. Nevertheless, voxel-based models have been used to resolve fine microstructures and quasi-continuous changes of the material properties \cite{Doubrovski2015, Chandru1995}.

Massarwi and Elber \cite{Massarwi2016} recently proposed a novel volumetric representation technique (V-rep) for 3D models, which allow full control over the model's interior. V-reps consist of trimmed, trivariate B-spline patches, which can be combined into V-models using Boolean operations. By extending the control points' dimension, it is possible to assign material parameters directly to the model. \blue{This property can be used to model and simulate multi-material FGM.} Potentially critical overlapping regions of the V-model are resolved by trimming the involved splines and creating new trivariate primitives for the respective overlapping volume. 
Due to the non-singularity of trivariate B-splines, V-models are predestined for a subsequent simulation with the isogeometric analysis (IGA)~\cite{Hughes2005}. However, \blue{a direct application of IGA is often not feasible since, in overlapping regions, the spline patches must be trimmed. Moreover, the respective spline parameterizations -- i.e., the control point meshes, knot vectors, and polynomial degrees -- do not coincide at adjacent faces. Hence,} special techniques are required to glue them together, e.g. Mortar methods \blue{or T-splines} \cite{Brivadis2014,Zuo2015}. \blue{By contrast, embedded domain methods require no special treatment of overlapping regions and pose far fewer requirements on the underlying geometric model.}

Apart from the possibility of controlling the interior of the volume, which can be used to model multi-material FGM, the V-rep framework also offers the option to create single-material FGM, such as continuously changing microstructures. Although easy to fabricate with AM, these multiscale structures are critical from a simulation point of view. Due to the complexity of the underlying CAD models, the meshing becomes difficult. Additionally, attempts to resolve the structure sufficiently accurate may result in over-refined meshes, leading to an additional but unnecessary computational effort. This is where numerical homogenization provides an efficient tool to estimate an overall mechanical behavior of such structures. The basic idea of homogenization is to define a representative volume element (RVE), which is sufficiently large to represent the overall material behavior in the specific region~\cite{Fritzen2013, Gross2017, nemat2013micromechanics}. In the case of periodic microstructures, a unit cell can be extracted for further material characterization. Periodic boundary conditions are then applied at their boundaries, which leads to the best possible estimate of the effective behavior~\cite{Pahr2003, Sanchez-Palencia1987}. The resulting material characterization can then be used to simulate a complete structure under complex loading. The computational cost is reduced considerably by 'smearing out' the detailed complex geometrical features of a microstructure and expressing them in terms of the effective behavior. Still, on the microscopic level of the RVE, the structure needs to be fully resolved in a boundary conforming fashion to account for all geometrical details. Here, embedded domain methods offer an elegant and reliable alternative over classical FEA for non-periodic AM structures~\cite{Korshunova2019}.

Embedded domain methods, such as the finite cell method (FCM)~\cite{Duster2017}, avoid a tedious and error-prone meshing process by embedding the complex geometrical model into a fictitious domain that can be easily meshed into regular simple elements. These methods are known under different names, e.g. fictitious domain~\cite{Burman2010, Heikkola1998, Auricchio2015}, immersed FEM/boundary~\cite{Liu2006,Mittal2005}, or Cartesian grid method~\cite{Nadal2013}. The FCM \cite{Duster2008} uses besides the embedded domain approach also high-order finite elements, deploying hierarchical Legendre, spectral, or B-Spline shape functions \cite{Joulaian2014, Rank2012}. Initially developed for 2D and 3D linear elasticity, it was extended to various fields of applications, such as topology optimization \cite{Groen2017, Cai2014}, local enrichment for material interfaces \cite{Joulaian2013}, elastodynamics and wave propagation \cite{Joulaian2014, Duczek2014}, or additive manufacturing \cite{Ozcan2018}. Further investigations include efficient integration techniques \cite{Fries2016, Joulaian2016}, or homogenization \cite{Korshunova2019}. FCM was successfully applied to various geometrical representations, such as B-rep, CSG \cite{Wassermann2017}, voxel domains \cite{Nguyen2017}, point clouds \cite{Kudela2020}, and defective, mathematically invalid B-rep models \cite{Wassermann2019}.

In this contribution, \blue{three novel methodologies are introduced:
\begin{itemize}
	\item The FCM is extended to V-models as a new CAD representation form.
	\item Based on the trivariate spline description of the V-models, a method for the simulation of multi-material FGM is introduced.
	\item Finally, a distinct approach is proposed that allows numerical analyses on large-scale continuously changing microstructures -- i.e. single-material FGM -- using homogenization.
\end{itemize}} The paper is structured as follows: Sections \ref{sec:fcm} and \ref{sec:vRep} provide a brief overview over the FCM and V-reps, respectively. The methodologies \blue{for the simulation of V-reps, single- and multi-material FGM are} described in Section \ref{sec:method}. Section \ref{sec:Examples} presents and discusses several numerical examples before conclusions are drawn in Section \ref{sec:conclusion}.

\section{Methods} \label{sec:methods}
\subsection{Finite cell method} \label{sec:fcm}

\begin{figure}[H]
	\centering
	\includegraphics[width=\textwidth]{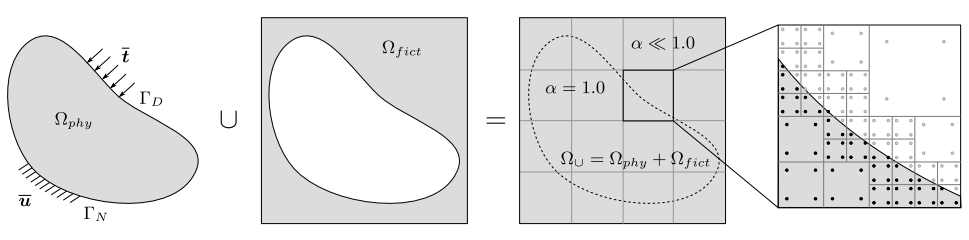}
	\caption{The concept of the finite cell method. \blue{The integration on the cut cell requires special quadrature rules -- here indicated by a composed integration on a quadtree reconstruction.}}
	\label{fig:conceptFCM}
\end{figure}

\blue{In the following, the basic concepts of the finite cell method are briefly summarized for linear elasticity. A detailed description of the FCM can, e.g., be found in~\cite{Duster2008}.} The FCM embeds a physical domain $\Omega_{phy}$ into a fictitious domain $\Omega_{fict}$ forming an extended domain $\Omega_\cup$, as illustrated in Figure~\ref{fig:conceptFCM} for two dimensions. The weak form of the equilibrium equation for the extended domain $\Omega_\cup$ reads as follows 

\begin{equation}
\int_{\Omega_U} [\mathrm{L}\bm{v}]^T \; \alpha \mathrm{C} \;[\mathrm{L}\bm{u}] \; \mathrm{d}\Omega_\cup = \int_{\Omega_U} \bm{v}^T \; \alpha \bm{b} \; \mathrm{d}\Omega_\cup + \int_{\Gamma_N} \bm{v}^T \bm{t} \; \mathrm{d}\Gamma_N \, ,
\label{eq:WeakForm}
\end{equation}
where $\bm{u}$ is the unknown deflection, $\bm{v}$ is a test function, $\mathrm{L}$ is the kinematic differential operator and $\mathrm{C}$ is the constitutive material tensor. The body load and the prescribed tractions on the Neumann boundary $\Gamma_N$ are denoted by $\bm{b}$ and $\bm{t}$, respectively. To resolve the complex domain correctly, an indicator function $\alpha(\bm{x})$ is introduced which weights the material tensor $\mathrm{C}$
\begin{equation}
\alpha(\bm{x}) = \left\{ \begin{array}{ll}
1 & \forall \, \bm{x} \in \Omega_{phy} \\
10^{-q} & \forall \, \bm{x} \in \Omega_{fict} 
\end{array} \right. .
\end{equation}
In the limit $q = \infty$, Equation~\ref{eq:WeakForm} recovers the standard weak form for $\Omega_{phy}$. In a finite element-like discretization, however, it leads to ill-conditioned systems. This can be avoided by choosing a finite $q$ (in practice $q=6...10$) in combination with a suitable preconditioning and/or orthogonalization of the shape functions~\cite{dePrenter2017a}. This choice introduces a modeling error \cite{Dauge2015} but limits the conditioning number of the stiffness matrix. Further improvement on the conditioning can be obtained using preconditioning, orthogonalization of shape functions, and/or the increase of continuity between the cut cells \cite{Prenter2019}.

\noindent The extended domain $\Omega_\cup$ is of simple shape and can be easily meshed into regular cells, e.g., rectangles in 2D and cuboids in 3D, respectively. These cells can be locally refined into sub-cells or with respect to the order of the shape function \cite{Schillinger2011, Zander2016}.

\subsubsection{Geometry treatment} \label{sec:geometricalTreatment}
The FCM resolves the physical domain $\Omega_{phy}$ (i.e., the geometric model) by the discontinuous scalar field $\alpha(\bm{x})$, which is then queried during the integration of the system matrices and load vectors. Consequently, the resolution of the complex geometry is shifted from the discretization (conforming meshing) to the integration level. The only information the FCM requires from the geometry is an unambiguous point inclusion statement, i.e. it must be possible to decide for any point  $\bm{x}$ whether $ \bm{x} \in \Omega_{phy}$ or  $\bm{x} \in \Omega_{fict}$. Due to the discontinuity of $\alpha(\bm{x})$ on cut cells, the integration needs to be carried out using special quadrature rules. Common variants are \blue{composed integration on a space-tree reconstruction (see Figure~\ref{fig:conceptFCM}), smart quadtree/octree, or moment fitting} \cite{Kudela2016, Hubrich2017, Abedian2013}. Another approach uses dimensional reduction, i.e., the integration is not performed over the entire domain, but only along the boundary \cite{Duczek2015a}.

\subsubsection{Boundary conditions} \label{sec:boundaryConditions}
As the boundary of the physical domain $\Omega_{phy}$ typically does not coincide with the edges/faces of the finite cells, essential (Dirichlet) boundary conditions need to be applied in a weak sense. For this, several methods have been investigated – such as the Nitsche method, Lagrange multipliers, and the penalty method \cite{Kollmannsberger2015, Ruess2012, Ruess2013, Guo2015a}. Natural (Neumann) boundary conditions are applied on $\Gamma_N$ following Equation~(\ref{eq:WeakForm}). Homogeneous natural boundary conditions are automatically resolved by $\alpha(\bm{x}) \approx 0$. Inhomogeneous natural and essential boundary conditions require an explicit integrable boundary description, which is either provided by the geometrical model or extracted directly from the volume using, e.g., the marching cubes algorithm, see e.g., ~\cite{Bog2018}.

\subsection{Volumetric representation} \label{sec:vRep}
The V-rep framework \cite{Massarwi2016} provides methods and algorithms for the construction of V-models by combining simple (e.g. cylinder, sphere, etc.) or complex primitives (e.g. ruled primitives or solids of revolution) with the Boolean operations, thus following the idea of constructive solid modeling. Furthermore, it is possible to migrate spline-based B-rep models to V-rep models.
The V-rep framework is embedded in the Irit geometry library~\cite{Elber2020}, developed by Elber et al. 
Irit provides a vast amount of various geometric modeling and analysis functionalities. It can be accessed as a C(++) library, via a scripting language, or graphically with the GuIrit CAD environment \cite{Elber2019}.

\subsubsection{Trivariate B-splines} \label{sec:trivarBspline}
A trivariate B-spline is a parametric function that allows to span a volume over a three-dimensional parameter space. It is typically represented as follows
\begin{equation}
	\bm{V}(\bm{u}) = \sum_{i=1}^l \sum_{j=1}^m \sum_{k=1}^n  B_{i,p}(u)B_{j,q}(v) B_{k,r}(w) \bm{P}_{i,j,k}\, ,
	\label{eq:TrivariateBspline}
\end{equation}
where $\bm{V}(\bm{u})$ is a point inside the volume and $\bm{u} = (u,v,w)^T$ the corresponding three-dimensional parameter position in the parameter space $\bm{u} \in U \times V \times W \subseteq \mathbb{R}^3$. $B_{i,p}$ denotes the $i^{th}$ one-dimensional B-spline basis function of polynomial degree $p$ and $\bm{P}_{i,j,k} \in \mathbb{R}^{k}$ are the $l\times m \times n$ control points. The dimension of the control points is $k=3+s$, where $k=3$ corresponds to the three geometric coordinates $\bm{x}^T = [x,y,z]$. Further information can be represented by additional dimensions $s>0$.

\subsubsection{V-rep primitives}
\begin{figure}[H]
	\centering
	\includegraphics[width=\textwidth]{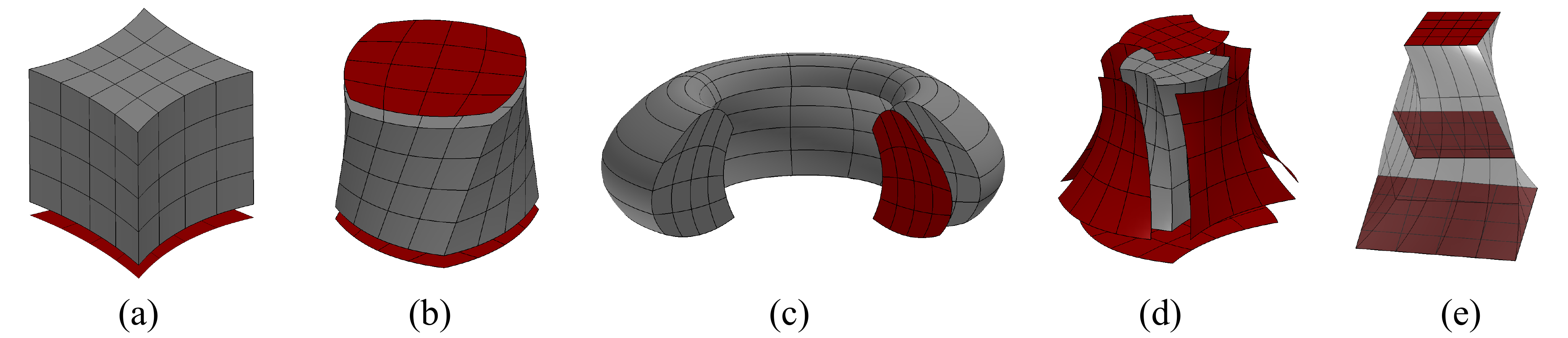}
	\caption{High-level primitives: (a) extrusion, (b) ruled solid, (c) volume of revolution, (d) boolean sum, and  (e) sweep/loft.}\label{fig:complexPrimitives}
\end{figure}

\begin{figure}[H]
	\centering
	\includegraphics[width=\textwidth]{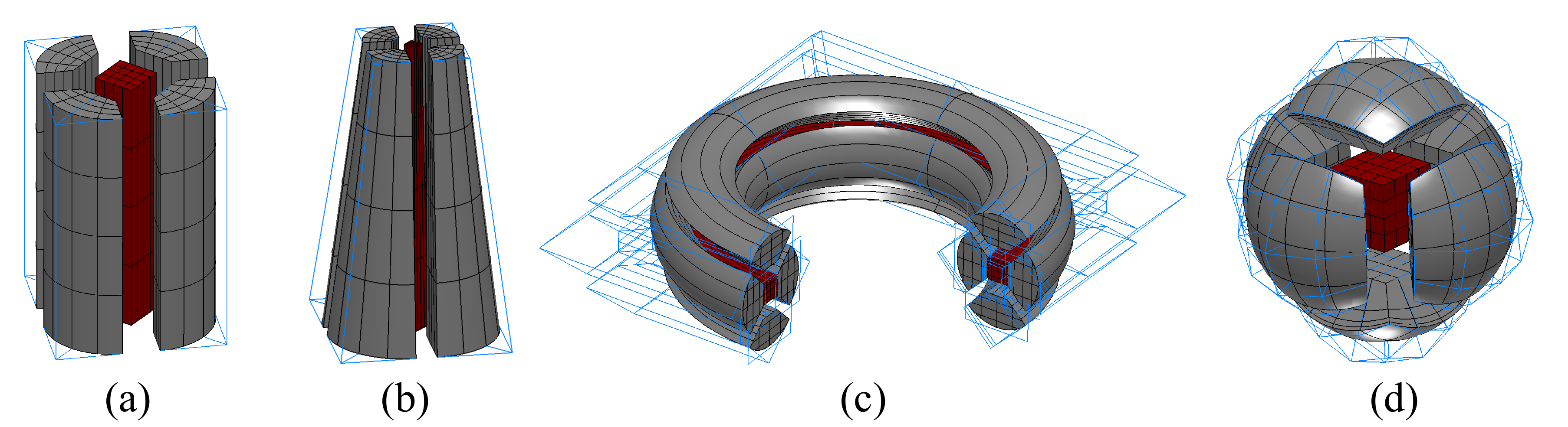}
	\caption{Non-singular primitives composed of trivariate B-splines: (a) A cylinder is composed by five extruded solids, whereas (b) a cone is composed of five ruled solids. (c) A torus is constructed using five solids of revolution, and (d) a sphere is composed of six ruled solids and one cuboid in its center.}\label{fig:nonSingularPrimitives}
\end{figure}

Apart from the trivial case of a cuboid, the V-rep framework offers various high-level and simple primitives. Implemented are several high-level primitive constructors, all of which yield one trivariate patch (see Figure \ref{fig:complexPrimitives}):
\begin{enumerate}
\item Extrusion: A surface is extruded along a vector.
\item Ruled solid: A volume is defined as a linear interpolation between two surfaces.
\item Solid of revolution: A volume is constructed by rotating a surface around an axis.
\item Boolean sum: A volume is created from six boundary surfaces \cite{Elber2012}.
\item Sweep/Loft: A sweep or loft interpolates several surfaces along a sweeping path.
\end{enumerate}
Simple primitives – such as spheres, cylinders, tori, and cones – can not be represented by a single trivariate patch without introducing singularities (e.g., at the mid axis of a sphere, the Jacobi matrix vanishes $\mathrm{det}(J_{\bm{V}}(r=0)) = 0$.) To this end, singular primitives are composed of several non-singular trivariate patches (see Figure \ref{fig:nonSingularPrimitives}). 

\subsubsection{V-model construction}
\begin{figure}[H]
	\centering
	\includegraphics[width=0.8\textwidth]{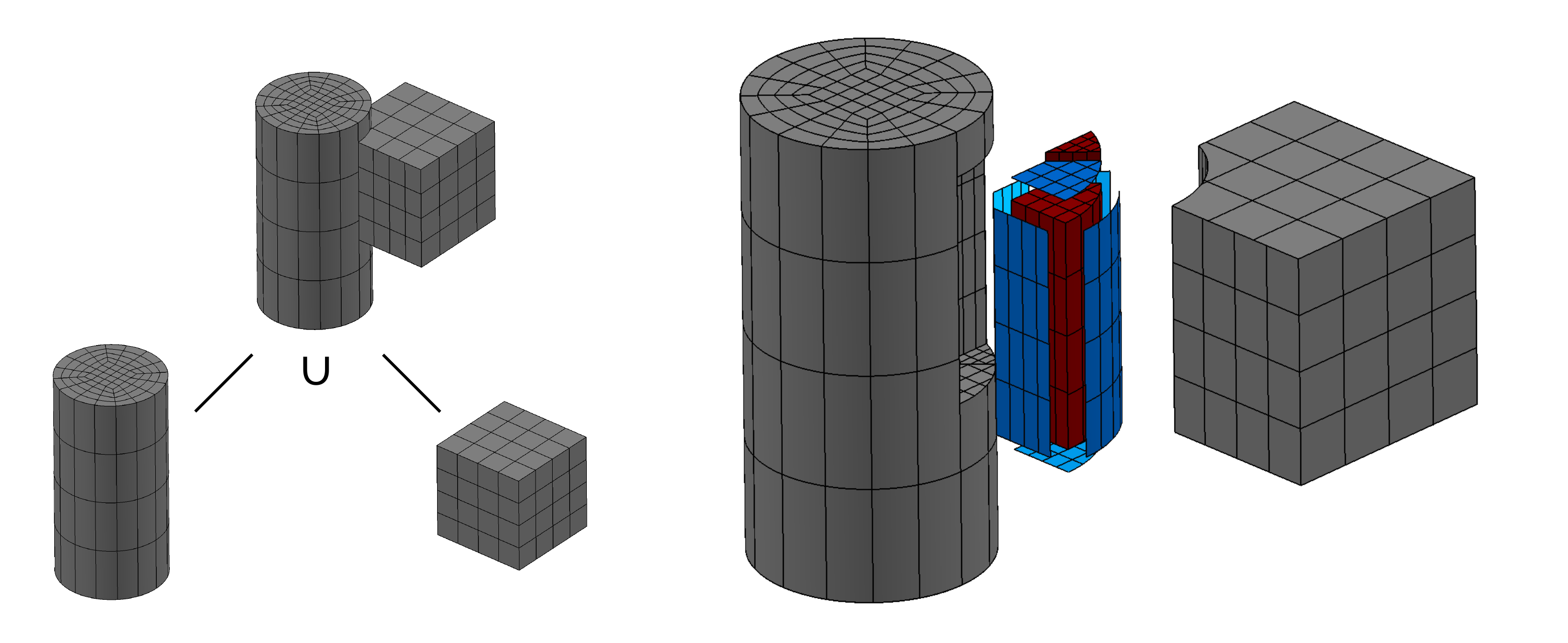}
	\caption{V-Model created as the union of a trivariate cuboid and a trivariate, non-singular cylinder. The intersected volume yields two V-cells (marked in red) which are constructed based on the trimming surfaces (highlighted in blue).}             	
	\label{fig:V-model}
\end{figure}

A trivariate B-spline is limited to a cuboid topology. In order to represent general volumetric shapes, so-called 3-manifold V-cells $\nu_C^i$ are introduced, which correspond to trimmed trivariate B-splines. \blue{A V-model $V_{m}$ is composed of $n$ V-cells: $V_{m} = \bigcup_i \nu_C^i \;,\; i\in\{1,...,n\}$. These V-cells originate firstly from the primitives that constitute the CAD model. New} V-cells occur due to the combination of the Boolean operations in the regions of overlapping primitives. Here, trivariate B-splines are trimmed at intersecting surfaces. Depending on the Boolean operation, the intersection volume is then remodeled from the trimming surfaces using the Boolean constructor (see Figure \ref{fig:V-model}). \blue{Consequently, the V-cells of a V-model} are non-intersecting $\nu_C^i \cap  \nu_C^j = \emptyset\;,\;\forall i\neq j$\blue{$\;,\;\ i,j \in \{1,...,n\}$ and the parametrizations of the new 'intersection' V-cells are different from their parent primitives. This makes the use of IGA more complex.}

\noindent V-cells store additional topological and adjacency information, which allows an efficient model inquiry. Adjacent V-cells share common trimming/boundary surfaces. Analogously to B-Rep, the boundary of the V-model $\partial V_{m}$ forms a closed 2-manifold.

\section{Discussion and Results}\label{sec:discussionAndResults}
\subsection{\blue{Extension of the FCM to V-reps}}\label{sec:pmcFcm}
In the context of the finite cell method, at first, without considering functionally graded materials, the V-model only needs to provide a point inclusion test. To this end, an inverse mapping is carried out on each V-cell. 
\begin{equation}
	f: \bm{x} = \nu_C^i(\bm{u}) \rightarrow \bm{u}
	\label{inverseMapping}
\end{equation}
As splines can generally not be inverted analytically, the corresponding parameter position $\bm{u}$ must be determined iteratively using the Newton-Raphson algorithm. Yet, one should note that – since the splines are regular, i.e. the Jacobian never vanishes – a solution is always unique if one exists. In the case $\bm{x} \cap \nu_C^i\neq \emptyset$ a parameter position can be found in the V-cell $\nu_C^i$ and the respective point $\bm{x}$ is inside the V-model. The \blue{number of required Newton-Raphson iterations for} the inverse mapping can be substantially decreased providing a good guess as an initial value. This is efficiently exploited by the finite cell method as, \blue{due to the Cartesian grid-based data structure,} consecutive integration points are very often geometrically adjacent. Therefore, the last inner point on each V-cell is cached and used as an initial guess for the next query.

\blue{Since, the underlying Irit library \cite{Elber2020} offers already a robust point inclusion test, the extension of the FCM to V-models is straight-forward. It is noteworthy that -- in contrast to IGA -- trimmed splines and non-coinciding spline parameters at adjacent faces, require no special treatment since the adaptive quadrature rules automatically recover the actual shape of the geometry.}

\subsection{\blue{Extension of the FCM to single- and multi-material FGMs}} \label{sec:method}
The V-Rep framework provides two different ways to realize functionally graded materials \blue{which can be produced by additive manufacturing techniques}: (a) the material properties can either be encoded directly into the volume of the V-cells (see Section~\ref{sec:FGMinVcell})\blue{, which is ideally suited to model multi-material FGMs}, or (b) an FGM can be created in a constructive manner \blue{in the form of a continuously changing microstructures, which corresponds to single-material FGMs} (see Section~\ref{sec:constructiveFGM}).

\subsubsection{\blue{Analyzing multi-material FGM with the finite cell method}}\label{sec:FGMinVcell}
\blue{A simulation of multi-material FGM using the FCM requires -- apart from the point-inclusion statement -- also the corresponding material properties. To this end, the spline-based description of the V-cells -- as the smallest, non-intersecting building blocks -- is extended to carry additional material information.} 

\paragraph{V-Rep material representation}
Material properties such as Young's modulus, Poisson's ratio, thermal conductivity, density, etc. can easily be represented on the V-cells by simply extending the dimension of the control points $\mathbb{R}^{3+s}$, with $s>0$ being the additional material parameters (see Equation~(\ref{eq:TrivariateBspline})). Consequently, evaluating the V-cell yields, in addition to the geometric coordinates, also the respective material values
\begin{equation}
	\bm{V}^T = [x,y,z, m^1,...,m^\sigma,...,m^s] \in \mathbb{R}^{3+s} \,.
 	\label{eq:extendedControlPoint}
\end{equation}
As an example, consider a control point that carries additional material properties for the Young's modulus $E$, Poisson's ratio $\nu$, and thermal conductivity $\kappa$ as needed for Example~\ref{sec:Example2}: $\bm{P}^T_{i,j,k} = [x,\,y,\,z,\,E,\,\nu,\,\kappa]_{i,j,k}$.

The material properties of a V-cell, created from the overlap of two or more trivariate B-splines carrying different material information, require additional handling. Either one of the initial trivariate B-spline can be set prevailing. Thus, its properties are inherited to the V-cell. Or some sort of blending scheme interpolates the material properties. For detailed information, refer to \cite{Massarwi2016}. 

\paragraph{Spline based material approximation}
Inside a patch, splines are typical of higher continuity, which renders them perfectly suitable for modeling smooth geometries. However, this restricts the material function to be of the same continuity. A remedy to also represent $C^0$ or discontinuous material distributions is given by knot-insertion, as the continuity depends on the multiplicity of the knots $C^{p-m}$, where $p$ is the polynomial degree and $m$ the number of multiple knots. Naturally, knot-insertion also reduces the potential continuity of the geometry. However, the original higher continuity is preserved in a geometrical sense\footnote{\red{Remark:  This is only the case for the undeformed, initial CAD model. The deformed shape can be of $C^{p-m}-$continuity, for instance, a kink in the case of $C^0$.}}. Hence, the model keeps its geometrical shape, whereas the material is allowed to have material kinks, or even to be discontinuous. Nevertheless, due to the global influence of the knots' position and multiplicity, splines are not the method of choice to represent highly discontinuous material distributions, as e.g., underlying voxel data provided by CT-scans.

\begin{figure}[H]
    \centering
    \includegraphics[width=0.7\textwidth]{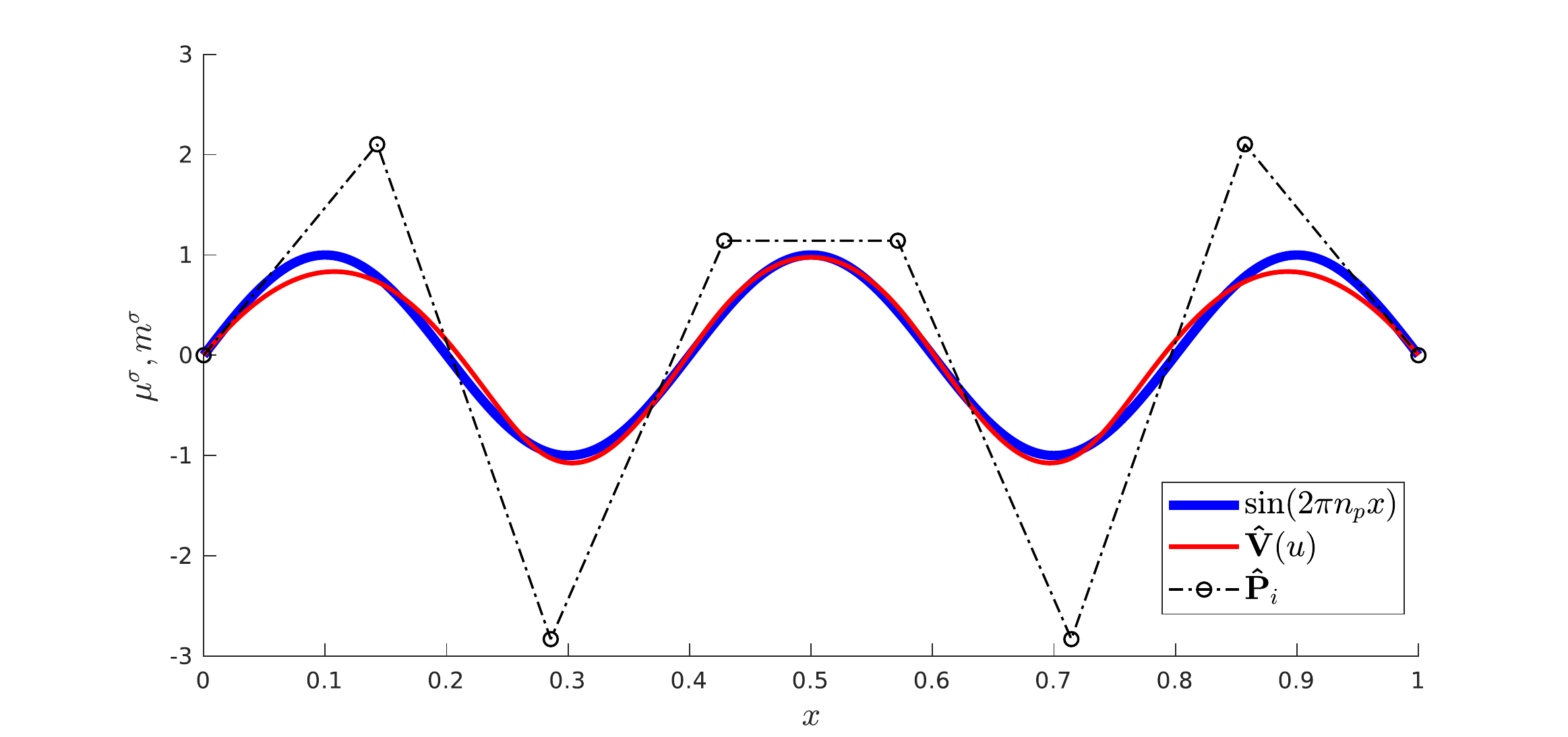}
	\caption{One-dimensional least squares approximation of a hypothetical sinusoidal material function $m^\sigma(x) = sin(2\pi n_p x)$, with $n_p = 2.5$ being the number of periods, yields the material 'coordinates' $\mu_i^\sigma$. Note that the rather large deviation between the curves comes from the fact that the locations (i.e. $x-$coordinates) of the material control points as well as the knot vector are fixed.}
	\label{fig:leastSquares}
\end{figure}

Given a sufficiently smooth material distribution, the material 'coordinates' of the control points can be obtained using least-squares approximation (see Figure~\ref{fig:leastSquares}). For each material property, the least-squares problem reads
\begin{equation}
	\min_{\bm{\mu^\sigma}} \sum_{\lambda=1}^{n_{LS}} r_\lambda^2 = \min_{\bm{\mu^\sigma}} \Vert \bm{V}\left(\bm{u}_\lambda, \bm{\mu}^\sigma\right)-f_m^\sigma\left(\bm{x}_\lambda\right)\Vert ^2_2 = \min_{\bm{\mu^\sigma}} \Vert \mathrm{A}(\bm{u_\lambda})\bm{\mu}^\sigma-f_m^\sigma\left(\bm{x}_\lambda\right)\Vert ^2_2 \,,
	\label{eq:leastSquares}
\end{equation}
where $n_{LS}$ is the number of sample points and $\bm{\mu}^\sigma = \mu^{\sigma}_{i,j,k} \in \mathbb{R}^{l \cdot n \cdot m}$ are the minimization variables (see Equation~(\ref{eq:TrivariateBspline}) for $l,m,n$). The least squares problem is then solved for each material function $f_m^{\sigma}$ and the respective material 'coordinate' $\bm{\mu}^\sigma, \,\sigma \in [1,s]$ of the control mesh $\bm{P}_{i,j,k} = \big[ x,y,z, \mu^1,..., \mu^\sigma ,..., \mu^s\big]^T_{i,j,k}$. Matrix $\mathrm{A} \in \mathbb{R}^{\nu \times \left(l \cdot n \cdot m\right)} $ contains the spline basis functions. The sample points are evaluated in the parameter space $\bm{u}_{\lambda} = [u,v,w]^T_\lambda \in \mathbb{R}^3$. Consequently, the material function needs to be evaluated in the same space (see Equation~(\ref{eq:TrivariateBspline}))
\begin{equation}
	f_m^\sigma\left(\bm{x}\right) = f_m^\sigma(\bm{V}(\bm{u})) \, .
	\label{eq:spaceTransform}
\end{equation}

\subsubsection{\blue{Analyzing single-material FGM with the finite cell method}}\label{sec:constructiveFGM}
\blue{Single-material} FGM structures change their material properties due to adaptions in the microstructure, density, grain size, etc. A prominent example in nature is the trabecular bone. The size and alignment of thin rods and plates of bone tissue create stiffness trajectories that follow the principal stresses for the most common load cases~\cite{Geraldes2013}.

\noindent Today, additive manufacturing (AM) offers the possibility to create similarly complex structures. To this end, AM uses porous infill structures to support the outer hull. However, this infill is typically a repetitive lattice and is either not taken into account for the load transfer, or is assumed to be isotropic~\cite{Jiang2018}. Nonetheless, recent approaches in topology optimization try to exploit the contribution of the infill to the load transfer~\cite{Clausen2016}. Problem-fitted complex 3D anisotropic microstructures can reduce the printing time and material consumption substantially and, at the same time, improve the load-carrying properties and buckling behavior. 

\begin{figure}[H]
    \centering
    \begin{tabular}{c c}
    	\begin{tabular}{c}
    		\begin{subfigure}{0.25\textwidth}
    			\centering
    			\includegraphics[width=0.9\textwidth]{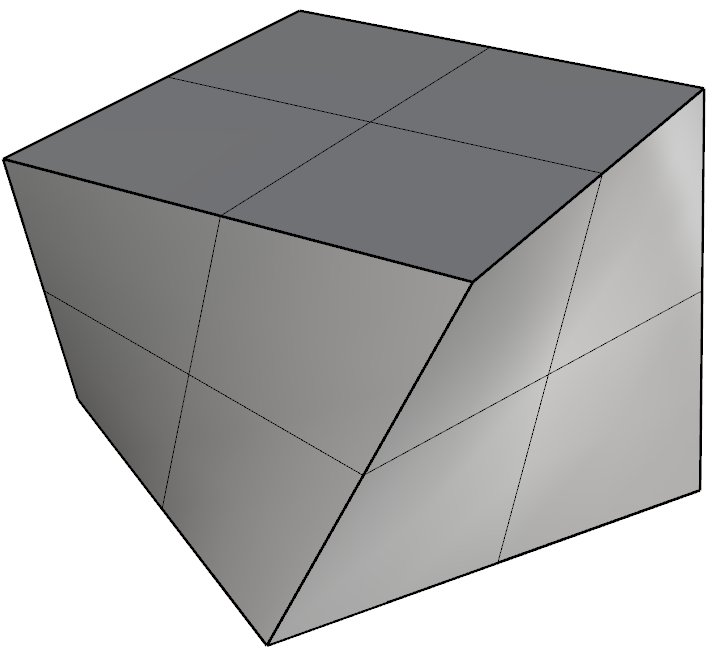}
    			\caption{Ruled base volume}
			\end{subfigure}\\
            \begin{subfigure}{0.35\textwidth}
                \centering
                \includegraphics[width=0.9\textwidth]{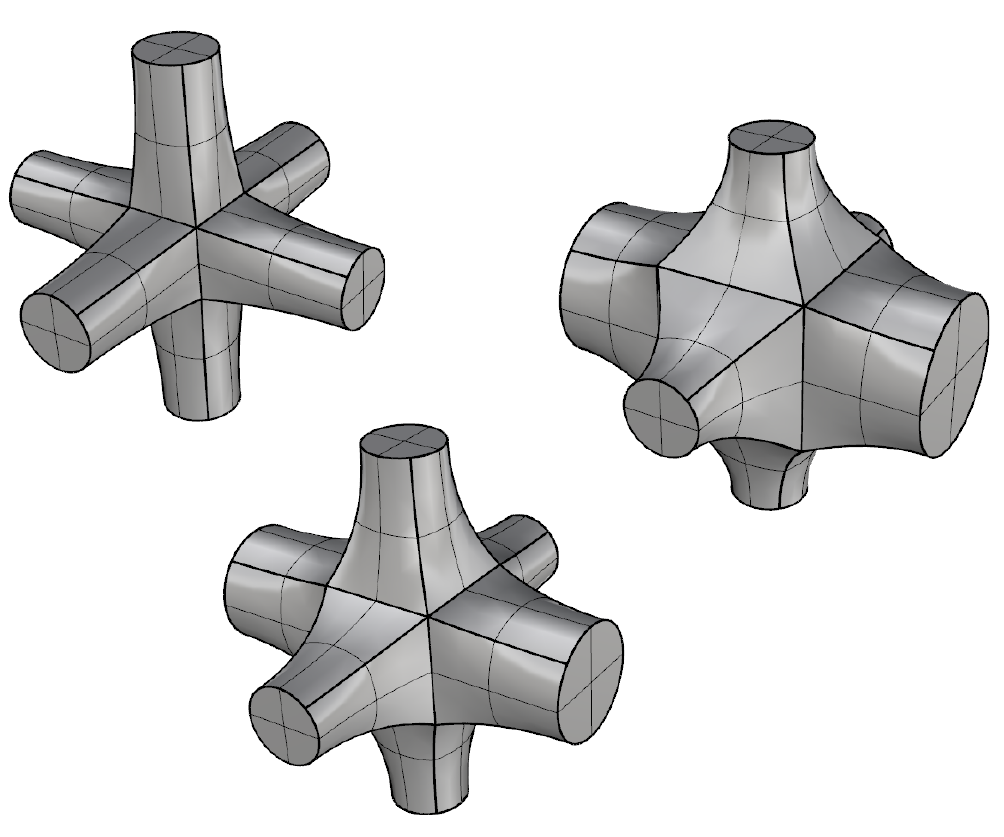}
                \caption{Unit tiles}
                \label{fig:unitTiles}
            \end{subfigure}\\
        \end{tabular}    
	&
	    \begin{subfigure}{0.55\textwidth}
        	\centering
			\includegraphics[width=0.9\textwidth]{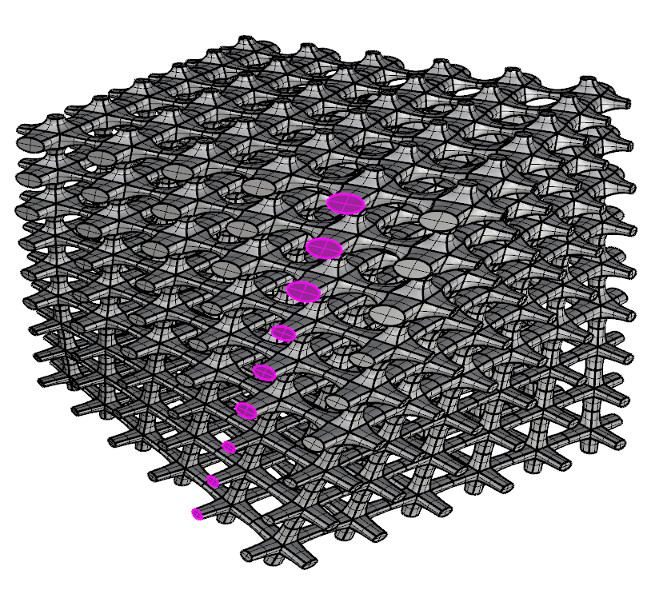}
			\caption{Microstructure}
			 \label{fig:MS}
        \end{subfigure}
	\end{tabular}
    \caption{Functionally graded microstructure: (b) Three different anisotropic tiles, with a changing stiffer direction, are used to tile (a) a rotating ruled volume. (c) The entire resulting microstructure exhibits a continuously changing anisotropic stiffness.}
	\label{fig:microStructure}
\end{figure}

\paragraph{Gradually changing microstructure}
The V-rep framework offers the possibility to create complex anisotro- pic microstructures with its tiling operation. With this, copies of a unit structure are consecutively created inside a base volume. Following the shape of the base volume and by using layers of different unit cells, a complex constructive FGM can be built. As the resulting microstructure is composed of several V-cells, it is again a V-model. Naturally, each V-cell can again represent a heterogeneous material distribution within its volume. \blue{Even for complex tile-based structures, like the example shown in Figure~\ref{fig:MS}, the point inclusion test can be carried out by inverse mapping, as described in Section~\ref{sec:pmcFcm}. Yet in the case of single-material structures, it turns out that a conversion into an auxiliary B-rep and a consecutive ray-tracing based test (see~\cite{Shimrat1962}) is computationally more efficient. In our implementation, the B-rep surface is subdivided into a fine triangular mesh and stored in a $kd-$tree \cite{Bindick2011}. Certainly, in contrast to the inverse mapping on trivariate B-splines, the surface triangulation causes a further approximation error, which can yet be controlled by refining the surface subdivision.}

\paragraph{\blue{Simulation of large-scale single-material FGM with the FCM}} \label{sec:homogenization}
Detailed geometrical features of microstructures require a fine numerical resolution to achieve reliable simulation results. \blue{Hence, large-scale structures necessitate high-performance computers, or might even not be computable at all.}
In order to reduce the computational cost \blue{and, thus allowing the computation of large microstructures}, a numerical homogenization can be used to evaluate a macroscopic mechanical behavior under specified loadings. This method's basic idea is to approximate the solution of a macroscopic boundary value problem by solving less complicated microscopic problems~\cite{nemat2013micromechanics}. This idea relies on the existence of a representative volume element (RVE). This microstructural domain is large enough to represent macroscopic behavior and small enough to ensure the scale separation. The mechanical quantities can then be transferred from the micro- to the macro-scale using the Hill-Mandel condition, also called ‘macro-homogeneity condition’. This mean-field numerical homogenization provides reliable estimates for the effective mechanical behavior if appropriate boundary conditions are chosen. \blue{For the herein considered microstructures that are created with Irit's tiling operation, periodic boundary conditions provide the best effective material properties.}

\begin{figure}[H]
    \centering
    \includegraphics[width=0.8\textwidth]{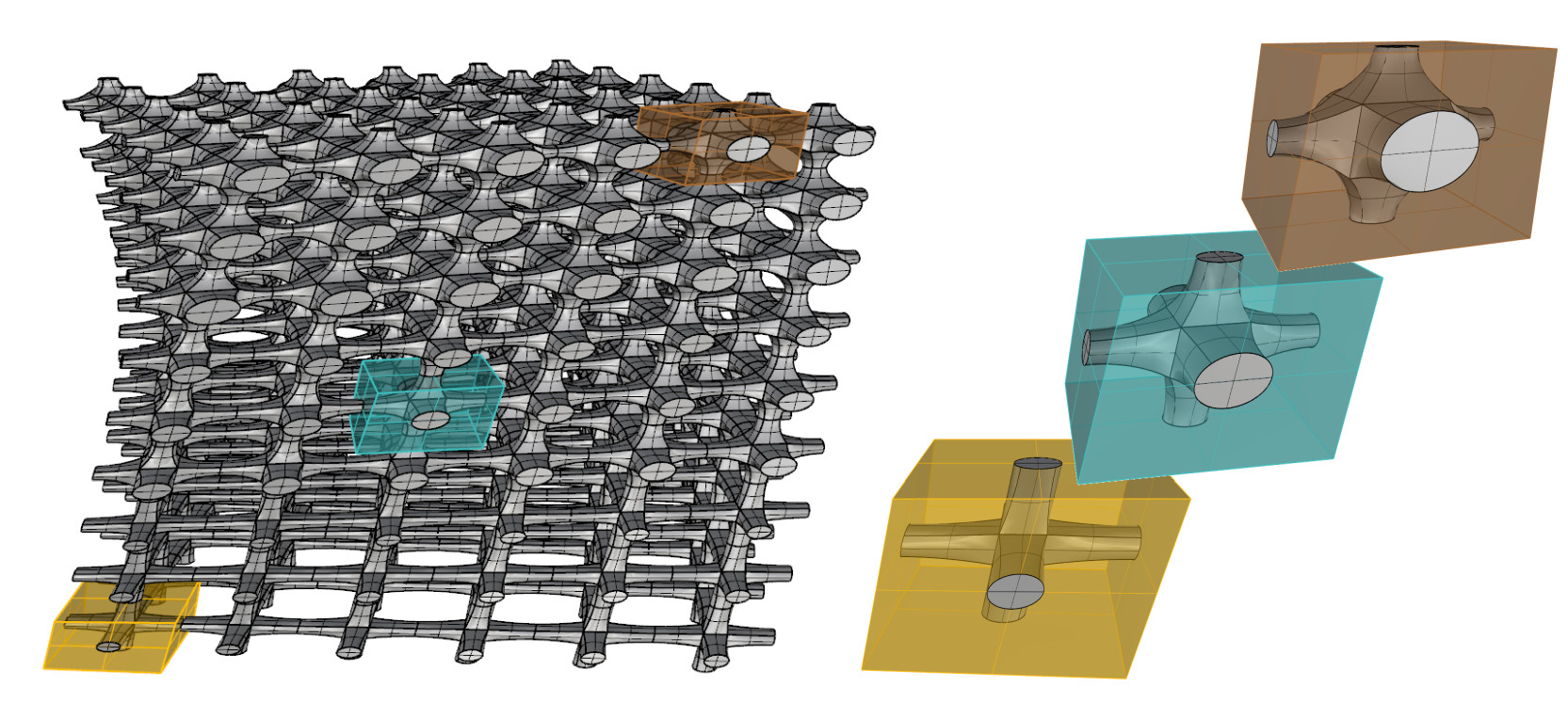}
	\caption{\blue{Continuously changing microstructure with different representative volume elements. The material properties in-between can be interpolated.}}
	\label{fig:RveChanging}
\end{figure}

\blue{Certainly, a functionally graded microstructure cannot be represented by one single RVE. However, since the parameter-based construction plan of the FGM is known apriori, it is sufficient to compute the effective material tensors $\bm{C}_{i}^*$ for several 'representative' RVEs (see Figure~\ref{fig:RveChanging}). At any realization in-between, the material is then be interpolated from corresponding adjacent representative tensors $\bm{C}_{i}^*$. }

\noindent \blue{For the microstructures, considered in this paper, the RVEs correspond to the constituting unit tiles, which can have different properties, for instance, a stiffer direction, a rotation around some axis, or a material composition. All these properties are defined using suitable construction parameters. The following approach then allows the efficient computation of large-scale functionally graded microstructures with the FCM:
\begin{itemize}
 \item Using the parametric description of the microstructure, several representative unit tiles are selected in different configurations.
 \item For the unit tiles, the effective material tensors $\bm{C}_{Ti}^*$ are computed with a combination of numerical homogenization and the finite cell method~\cite{Korshunova2019} and stored in a look-up table (see Example~\ref{sec:Example4}, Tab.~\ref{tab:lookUpTable}).
 \item During the simulation of a large-scale FGM microstructure, the effective material tensor at each integration point is determined by an interpolation of the values from the look-up table (see Example~\ref{sec:Example5}).
\end{itemize}
Based on the model of Example~\ref{sec:Example3} this approach is illustrated in the Examples~\ref{sec:Example4} and~\ref{sec:Example5}.}

\subsection{Numerical examples} \label{sec:Examples}
To demonstrate the variety of simulatable functionally graded materials using a combination of V-reps and the FCM, five examples are presented. The first example serves as a verification of the \blue{extension of the FCM to multi-material FGMs}. To this end, a linear elastic simulation of a simple cuboid with a prescribed material distribution is performed. The second example, a coupled heat, thermo-elastic simulation of a curved thermal protection tile, underlines the applicability to examples of engineering relevance. The third example shows a simulation of a fully resolved single-material FGM -- i.e. a continuously changing microstructure. In the fourth example, the underlying tiles of the third example are evaluated in terms of a homogenization, which are then used in the fifth example to perform a simulation on a \blue{large-scale} homogenized single-material FGM.

\subsubsection{Example 1: Cuboid with sinusoidal material distribution}\label{sec:Example1}
\begin{figure}[H]
	\begin{tabular}{c c}
	 	\begin{subfigure}{0.30\textwidth}
             \centering
             \includegraphics[width=\textwidth]{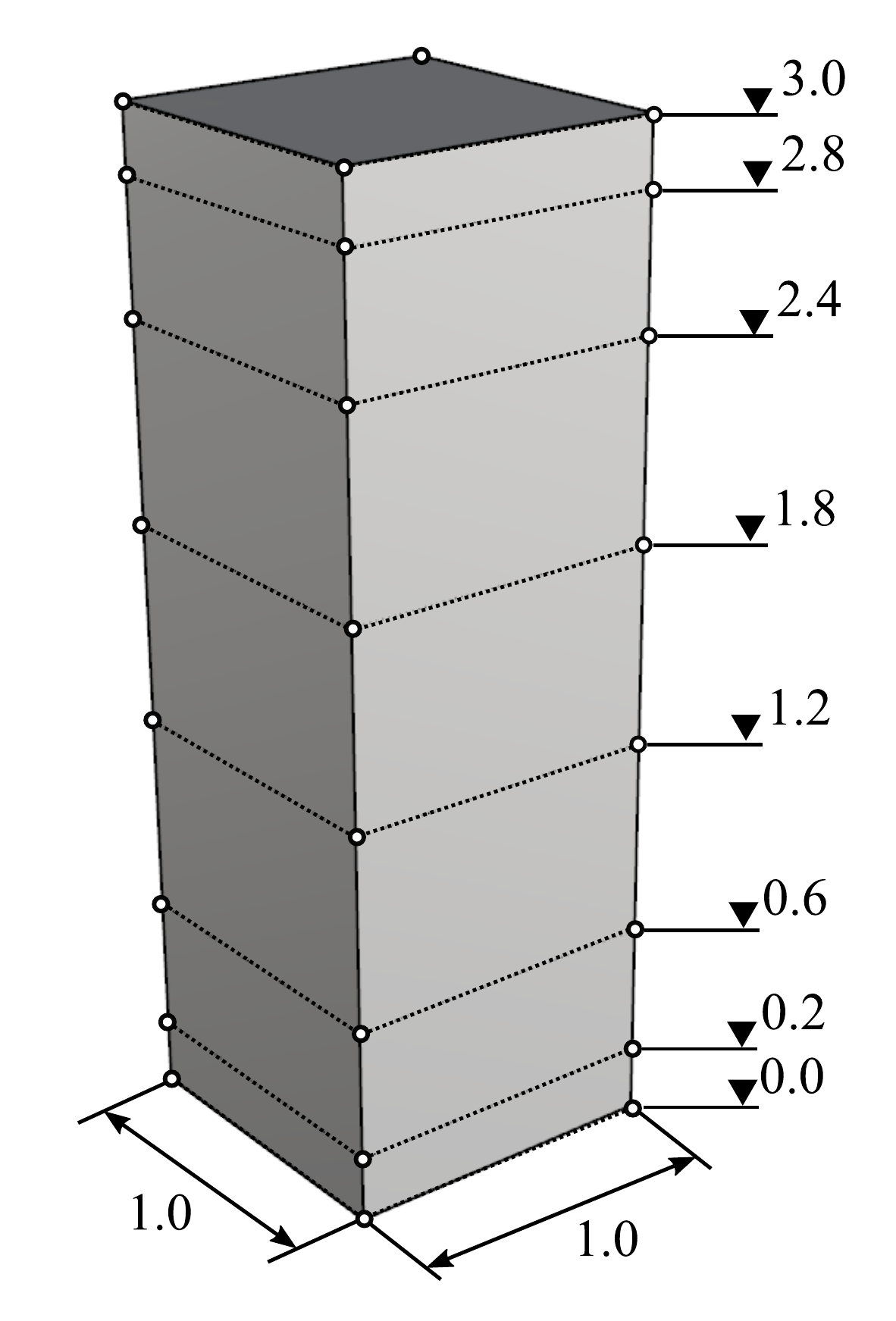}
             \caption{}
             \label{fig:cuboidDimension}
        \end{subfigure}	 
	&
		\begin{subfigure}{0.60\textwidth}
             \centering
             \includegraphics[width=\textwidth]{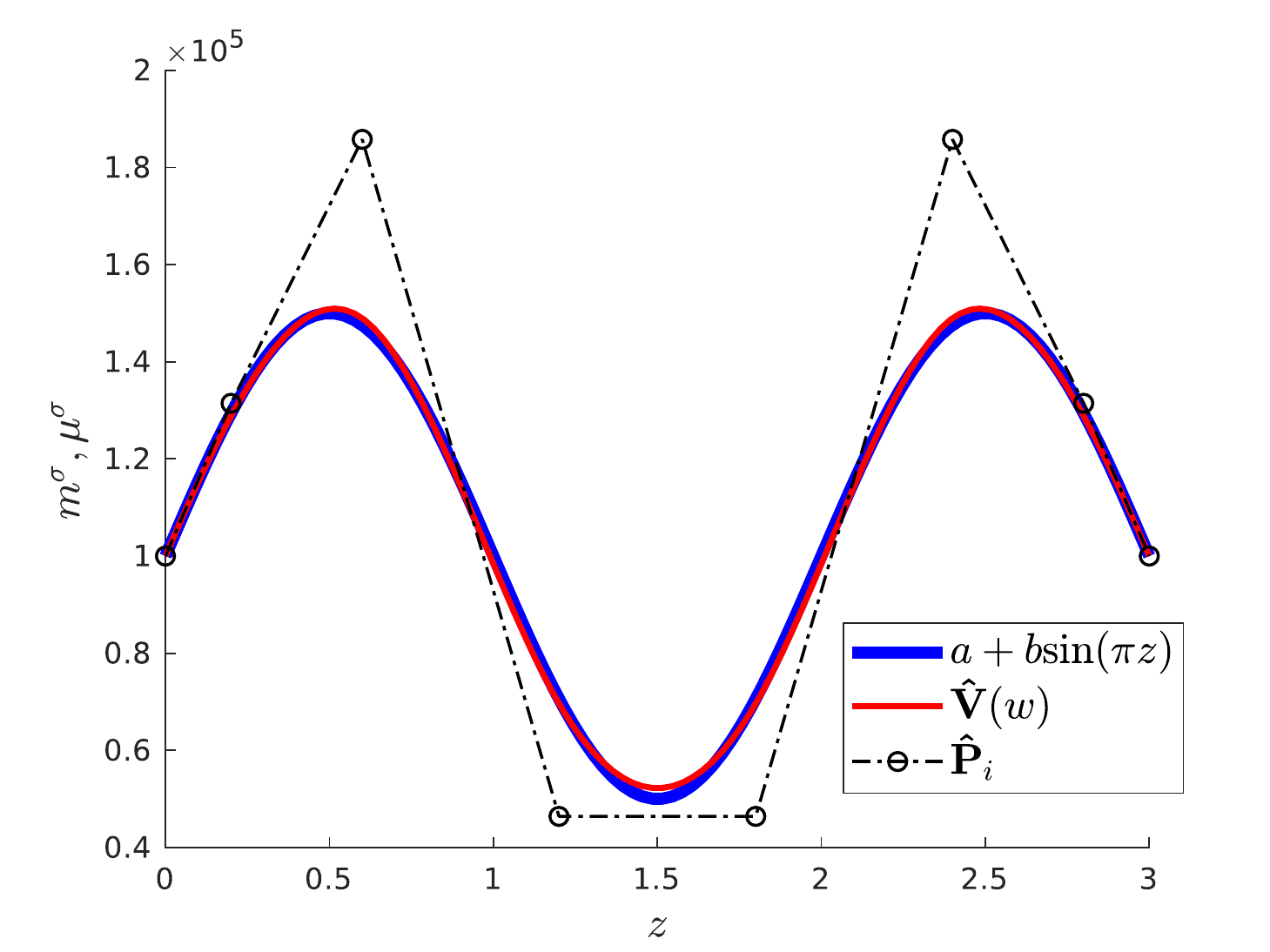}
             \caption{}
             \label{fig:cuboidLsMatlab}
        \end{subfigure}	   
	\end{tabular}
	\caption{(a) Dimensions of the cuboid. (b) Least squares fitting of $E(z)$ yielding the material coordinates $\mu^E_i$. }	
	\label{fig:cuboidSetup}
\end{figure}

\begin{figure}[H]
	\centering
	\begin{subfigure}{0.33\textwidth}
        \centering
         \includegraphics[width=\textwidth]{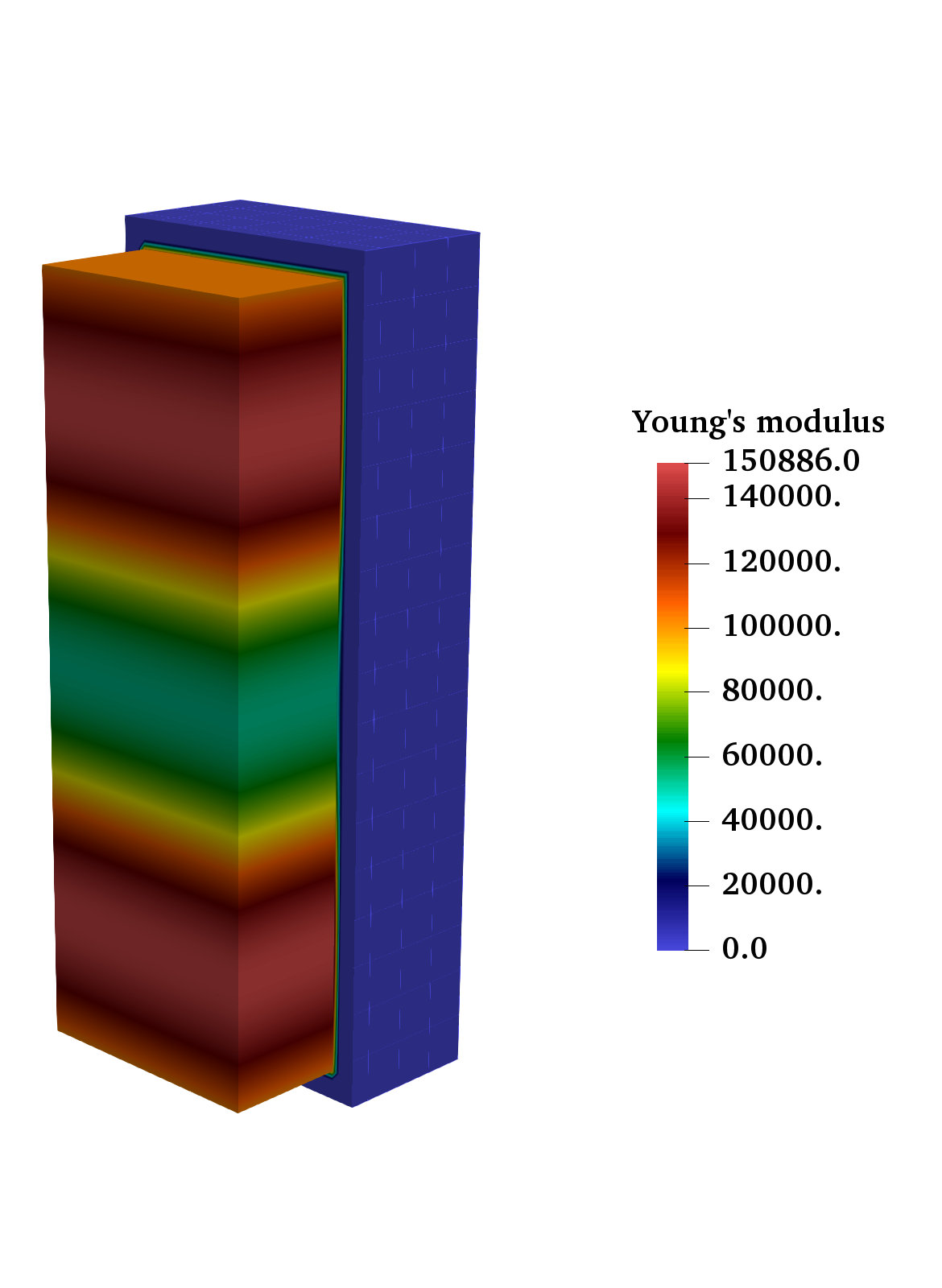}
         \caption{}
         \label{fig:cuboidMaterial}
    \end{subfigure}
    \begin{subfigure}{0.58\textwidth}
		\centering 
		\includegraphics[width=\textwidth]{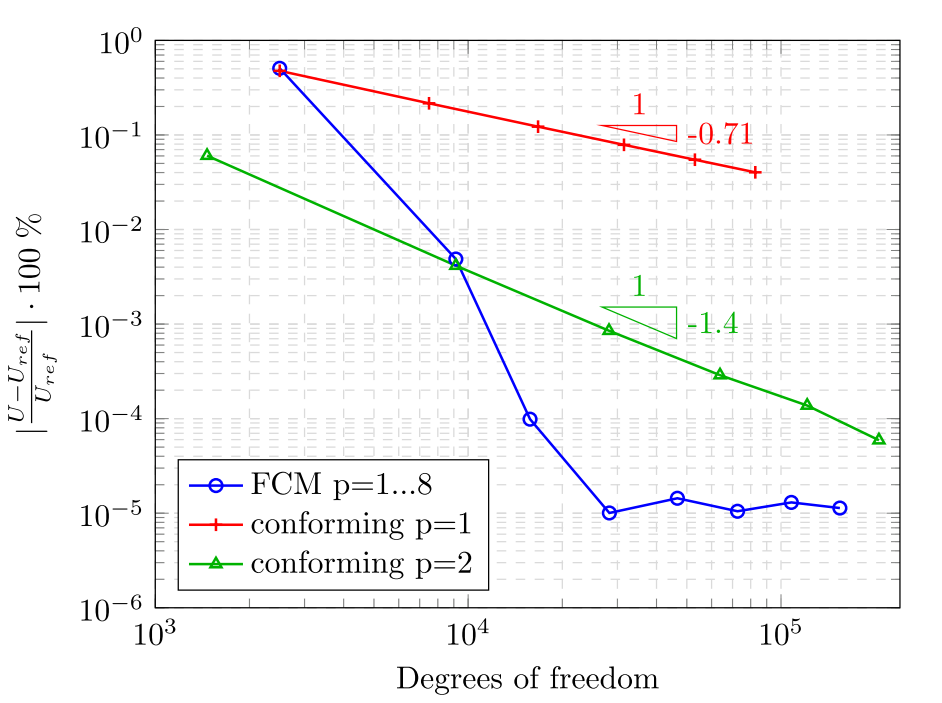}
		\caption{}	
		\label{fig:convergenceStudy}
	\end{subfigure}
	\caption{(a) Young's modulus evaluated on integration points inside the cuboid and in the fictitious domain. (b) Relative error in the strain energy for polynomial degrees p = 1...8.}
\end{figure}

As a benchmark problem, the cuboid with varying material distribution in $z-$direction is chosen\footnote{\blue{Remark: This benchmark example is chosen to be of simple shape to able to obtain a highly precise reference solution. Yet, as the structure is embedded in a larger domain, the solution is non-trivial for immersed boundary methods.}}. The cuboid is a trivariate B-spline and is created with GuIrit~\cite{Elber2019}. As the spline basis functions are initially linear in each direction, they are not able to represent the material function $E(z)$. For this reason, a degree elevation to $r=3$ and subsequent multiple knot insertions in $z-$direction were carried out, yielding a knot-vector of $W=[0,0,0,0,0.2,0.4,0.6,0.8,1,1,1,1]$. The control points in $z-$direction are depicted in Figure~\ref{fig:cuboidDimension}. The cuboid has assigned a constant Poisson ratio of $\nu = 0.3$. The functionally graded Young’s modulus is given as an analytical function  
\begin{equation}
	E(z) = 10^6 + 5\cdot10^4 \cdot sin(z\pi) \,.
\end{equation}
The material 'coordinates' $\mu_i^E$ of the control points are computed using least squares with $n_{LS} = 100$ sample points, according to Equation~(\ref{eq:leastSquares}) (see Figures~\ref{fig:cuboidLsMatlab}, and~\ref{fig:cuboidMaterial})
\begin{equation}
	\mu^E = [100000,\,131438,\,185772,\,46415,\,46415,\,185772,\,131438,\,100000] \, .
\end{equation}

\noindent For the simulation, the cuboid is embedded into a slightly larger fictitious domain \blue{($\pm0.1$ in each direction, ergo $1.2 \times 1.2 \times 3.2$)} and discretized by $6 \times 6\times 16$ finite cells \blue{deploying hierarchic Lengendre shape functions}. Homogeneous Dirichlet boundary conditions are applied in $x-$direction on the left, in $y-$direction on the front, and in $z-$direction on the bottom surface using the penalty method. The cuboid is loaded on the top surface with a traction of $f=-1000$ in $z-$direction.

\blue{To prove the validity of the FCM for multi-material FGM, a convergence study is carried out. The polynomial degrees of the Legendre ansatz function are increased from $p=1...8$, and the corresponding strain energies are compared to a reference solution $U_{ref}$, which was computed by a boundary-conforming $p-$FEM analysis. To minimize integration errors, a composed integration is used, which can accurately recover the cuboid's exact shape -- similar to the smart octree~\cite{Kudela2016}. In order to compare the convergence behavior of the FCM with the standard FEM, two additional convergence studies using $h-$refinement are carried out on boundary conforming FEM discretizations, with polynomial degrees of $p=1$ and $p=2$, respectively.}

\blue{As depicted in Figure~\ref{fig:convergenceStudy}, the FCM shows a pre-asymptotic exponential convergence until it reaches the numerical precision of the underlying Irit library at $p=4$, whereas the $h-$studies show algebraic convergence -- as expected\footnote{\blue{Remark: Since the relative error is stated in percent, the actual precision is the order of $10^{-7}$, which corresponds to the accuracy of the geometric modeler Irit which uses single precision.}}. Obviously, in terms of degrees of freedom, the FCM outperforms classical $h-$versions.}

\begin{figure}[H]
	\centering
    \includegraphics[width=\textwidth]{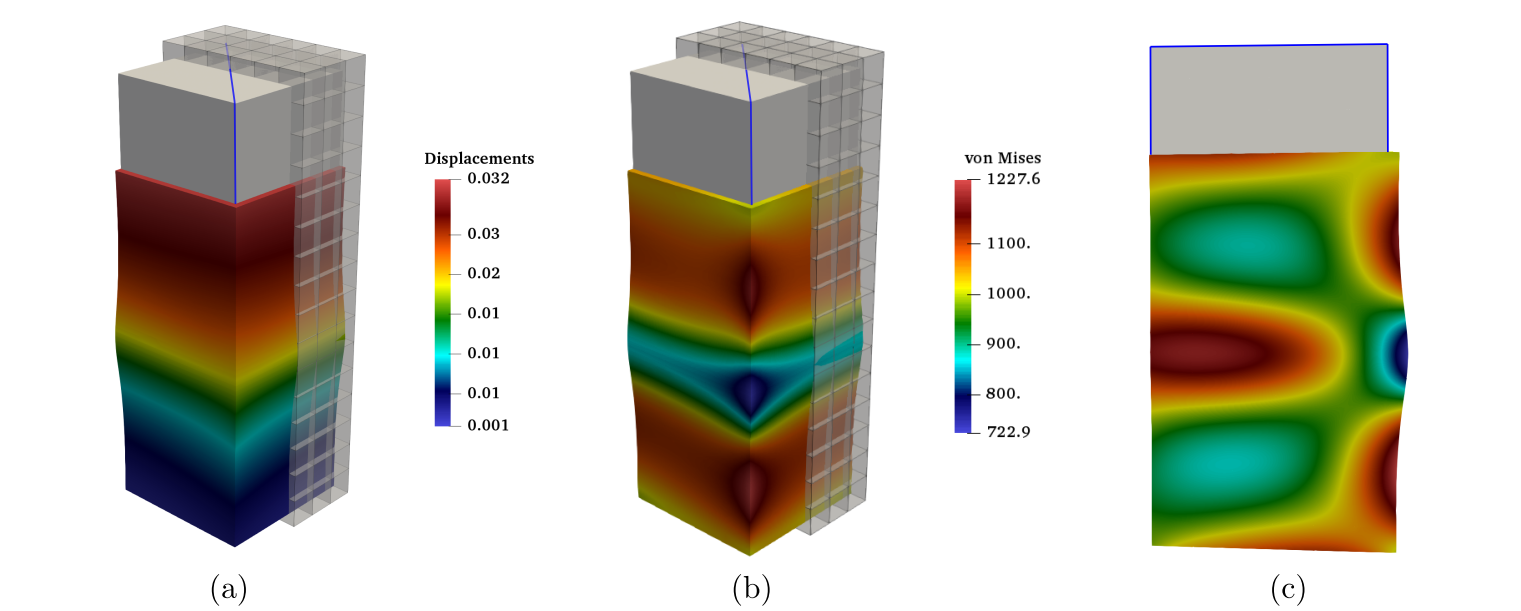}
	\caption{\red{(a) Displacements and (b) von Mises stresses warped around the undeformed cuboid (grey block) embedded into the finite cell mesh. (c) Cross section (aligned to the blue frame) showing the von Mises stresses inside the volume. The deformation is scaled by a factor of 20.}}
	\label{fig:cuboidResults}
\end{figure}

\noindent Figure \ref{fig:cuboidResults} shows the displacements and the von Mises stresses of the deformed cuboid. As expected, the regions of lower stiffness are undergoing a \red{larger} deformation.

\subsubsection{Example 2: Curved thermal shielding tile}\label{sec:Example2}
The second example \blue{demonstrates the applicability for practical applications. To this end,} three curved thermal shielding tiles are simulated. Such tiles are needed for high-temperature applications, such as re-entrance shielding for spacecraft or the inner coating of fusion power plants. The tiles consist of a load-carrying zone made of titanium $\mathrm{Ti}$ and an insulating zone made of porous silica $\mathrm{SiO_2}$ with a porosity of 70\%. Both materials have similar melting points of $\Theta_{Ti} = 1.668^{\circ}C$ for titanium and $\Theta_{SiO_2} = 1.710^{\circ}C$ for silica, which allows a fabrication with additive manufacturing using e.g. powder bed laser melting.

Particular focus is laid on the continuity of the transition zone between these materials. The first discontinuous tile consists of two distinct domains where both domains are assumed to be homogeneous titanium and silica, respectively, i.e., there is no transition zone. \blue{Hence, the first tile is not an FGM, but a composite material.} The material is changed $C^0-$continuously in the second tile, and $C^1-$continuously in the third tile. To evaluate the stresses under a heat load, a coupled simulation is carried out. An initial thermal simulation provides the temperature distribution, which is then used to apply thermal strains for the subsequent thermo-elastic simulation. Consequently, the model will deform due to the different thermal expansion ratios. However, this deformation is hindered by the different Young's moduli in the transition zone, then leading to internal stresses.

The underlying V-model consists of one V-cell and was generated by extruding a curved two-dimensional B-spline surface $5\,cm$ in $z-$direction. \blue{The control point mesh of the curved surface is defined as follows\footnote{\blue{Remark: Blank columns indicate a new row of control points in $x-$direction}}
\begin{equation}
\bm{P}^{surface}_i = \begin{bmatrix}
0&  0&  0&  0&& 4&  4&  7&  7&& 8&  8& 14& 14&& 12& 12& 21& 21  \\
0& 10& 20& 30&& 0& 10& 20& 30&& 0& 10& 20& 30&&  0& 10& 20& 30  \\
0&  0&  7&  7&& 0&  0&  7&  7&& 0&  0&  0&  0&&  0&  0&  0&  0
\end{bmatrix}\,.
\end{equation}
The knot vectors in $x-$ and $y-$ direction read $U=V=[0,\,0,\,0,\,0.5,\,1,\,1,\,1]$. Consequently, the surface has polynomial degrees of $p_x = p_y = 2$. }

\begin{figure}[H]
	\begin{tabular}{c c}
	 	\begin{subfigure}{0.47\textwidth}
             \centering
             \includegraphics[width=\textwidth]{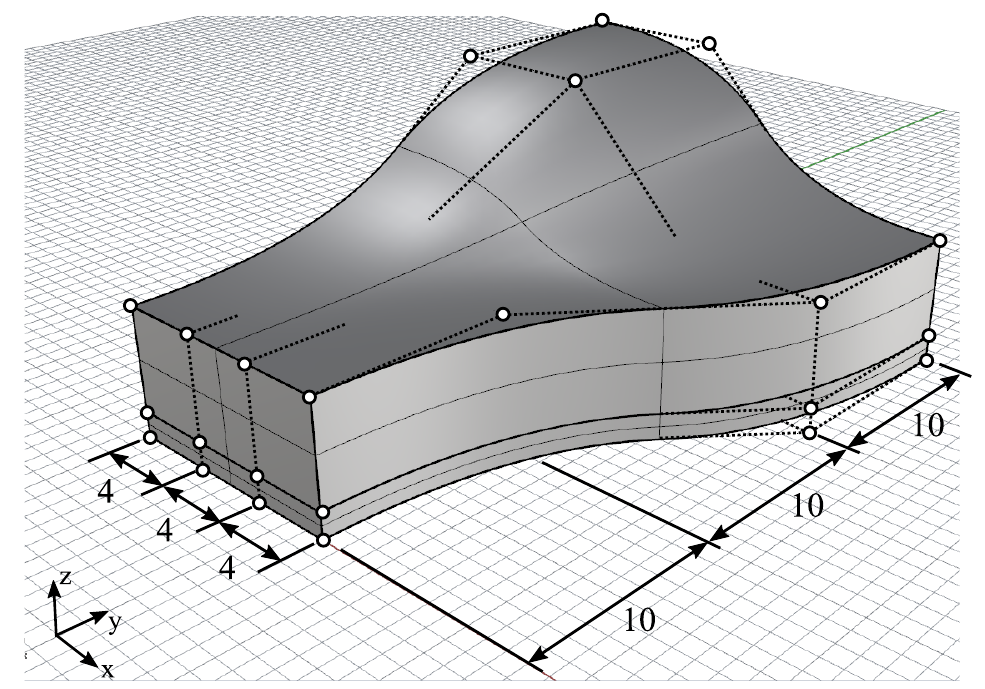}
             \caption{}
        \end{subfigure}	 
	&
		\begin{subfigure}{0.47\textwidth}
             \centering
             \includegraphics[width=\textwidth]{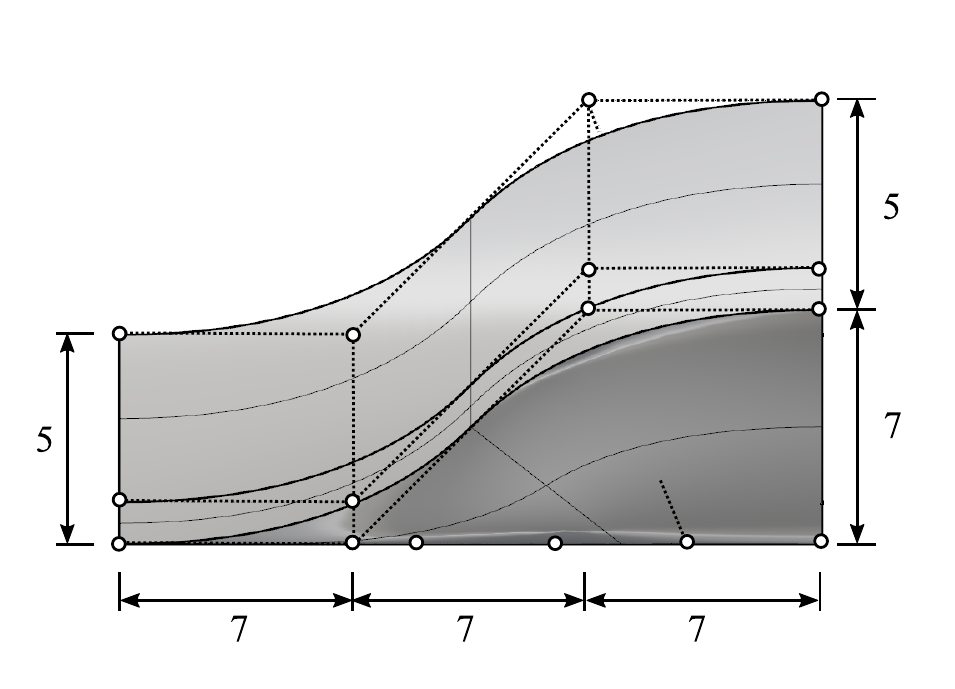}
             \caption{}
        \end{subfigure}	  
	\end{tabular}
	\caption{Model dimensions (in $cm$) and control point mesh of the discontinuous tile in (a) isometric view, and (b) from the back side. }	
	\label{fig:curvedTileSetup}
\end{figure}

\begin{figure}[H]
	\begin{tabular}{c c}
	 	\begin{subfigure}{0.47\textwidth}
             \centering
             \includegraphics[width=\textwidth, trim={9cm 3cm 1cm 0cm}, clip]{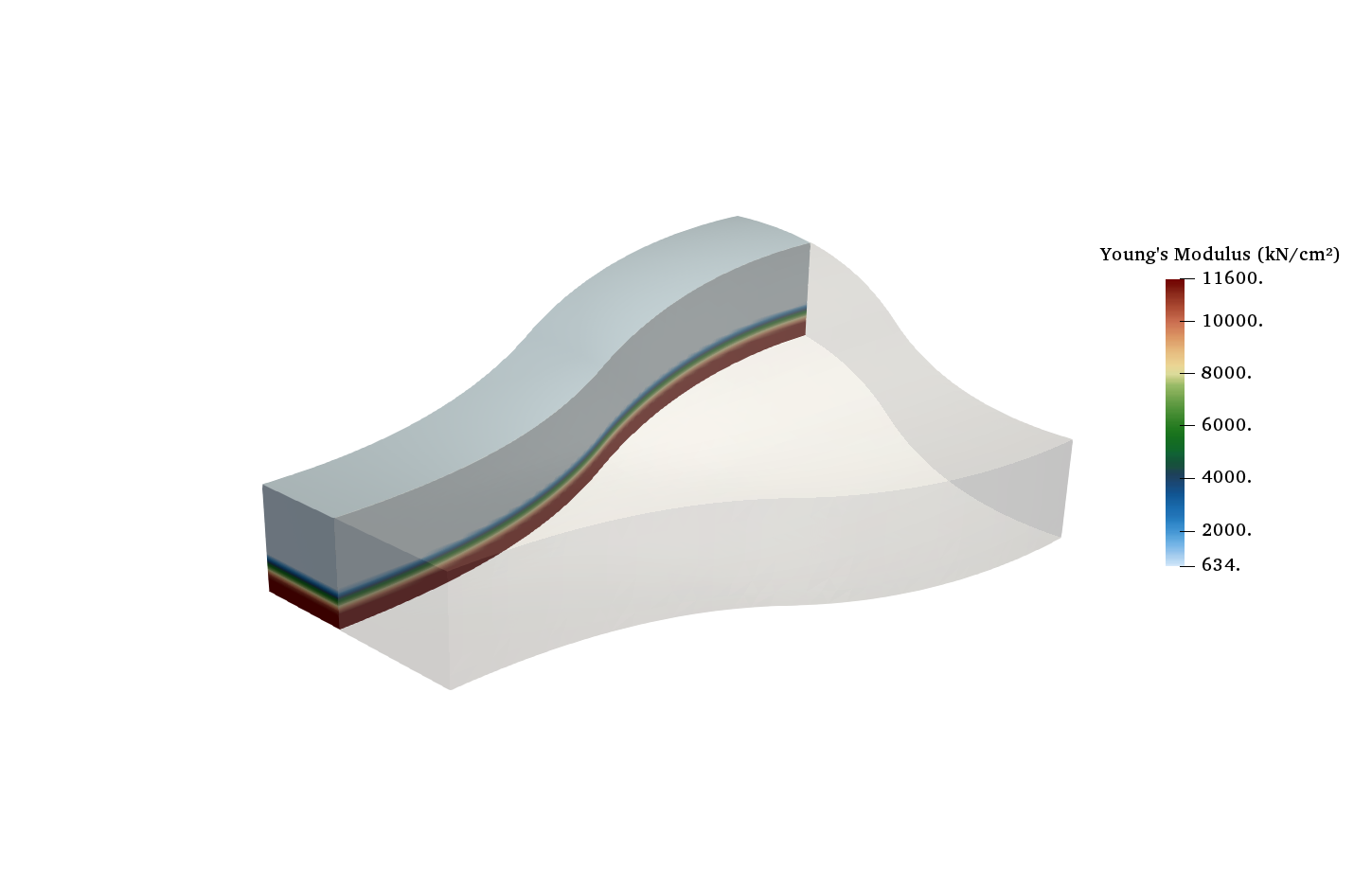}
             \caption{}
        \end{subfigure}	 
	&
		\begin{subfigure}{0.47\textwidth}
		\centering  
		\includegraphics[width=\textwidth]{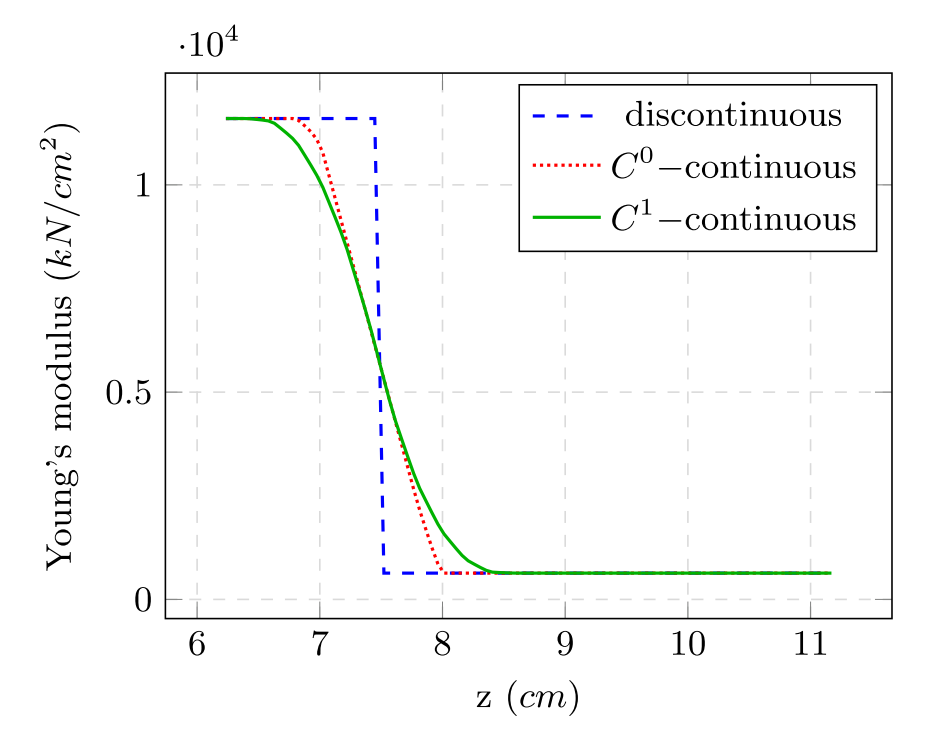}
        \caption{}		
        \end{subfigure}	  
	\end{tabular}
	\caption{Material distribution of the Young's modulus (a) inside the $C^0-$continuous tile and (b) plotted at $x=5\,cm$, $y=25\,cm$ over the thickness. }	
	\label{fig:YoungsModulus}
\end{figure}

\blue{The extrusion yields a V-cell with a polynomial degree of $p_z = 1$ and a knot vector $W=[0,\,0,\,1,\,1]$. Thus, the volumetric control point mesh consists of twice the initial mesh of the surface, where the second half of the control points have an offset of $\mathrm{d}z = 5$ cm}. For the $C^1-$ continuous tile, the polynomial degree in $z-$direction is \blue{increased to} $p_z = 2$. To construct the discontinuous tile, the V-model was split at $\Delta z_{div} = 1.25\,cm$ using knot-insertion. The knot vectors and \blue{the offsets of the control points in $z-$direction} for all tiles read as follows
\begin{flalign}
	 W_{Discont.} &= [0,\,0,\,0.25,\,0.25,\,1,\,1]\\
	 W_{C^0} &= [0,\,0,\,0.15,\,0.35,\,1,\,1]\\
	 W_{C^1} &= [0,\,0,\,0,\,0.05,\,0.29,\,0.5,\,1,\,1,\,1]\\
	 \mathrm{d}z_{Discont.} &= [0,\,1.25,\,1.25,\,5]\\
	 \mathrm{d}z_{C^0} &= [0,\,0.75,\,1.75,\,5]\\
     \mathrm{d}z_{C^1} &= [0,\,0.2,\,0.8,\,1.7,\,3.6,\,5] \, .
\end{flalign}
The resulting material distributions are depicted in Figure~\ref{fig:YoungsModulus} exemplary for the Young's modulus. The other material properties are distributed similarly. The parameters for the B-splines were chosen such that the integral of the material over the thickness is equal for all three tiles. Figure~\ref{fig:curvedTileSetup} shows the outer dimensions of the tiles in $cm$. 

\noindent To perform the coupled simulation, four different material parameters are required for both materials (see Table~\ref{tab:materialProperties}). The properties were taken from AZO Materials and averaged if necessary~\cite{AZOMaterials2020}. Due to the porosity of the silica, the respective Young's modulus $E_{SiO_2}$ and the thermal conductivity $\kappa_{SiO_2}$ must be adapted. This is implemented with the Gibson-Ashby criteria\blue{, which provide simple formulas to estimate the properties based of the porosity~\cite{Gibson1982, Pabst2018}
\begin{flalign}
	\kappa_r &= (1-\phi)^{3/2}\,,\\
	E_r &= (1-\phi)^{2}\,,
\end{flalign}
where $\phi$ is the porosity (in this example $\phi = 0.7$) and $\kappa_r$ and $E_r$ are the weighting factors for the thermal conductivity and Young's modulus, respectively.} In contrast, the Poisson's ratio $\nu_{SiO_2}$ and the thermal expansion $\alpha_{SiO_2}$ require no adjustment~\cite{Pabst2015}.

\begin{table}[H]
	\centering
	\begin{tabular}{|c|c|c|c|c|}
		\hline
		Property & Symbol & Titanium & Silica (70\% porosity) & Units \\[0.5ex] 
		\hline\hline
		Young's Modulus & $E$ & $11,600$ & $634$ & $kN/cm^2$ \\
		Poisson's Ratio & $\nu$ &$0.36$ & $0.17$ & $-$ \\
		Thermal conductivity& $\kappa$ & $0.216$ & $2.3\cdot10^{-3}$ & $W/cmK$ \\
		Thermal expansion& $\alpha$ & $8.6\cdot10^{-6}$ & $6.5\cdot10^{-7}$ & $1/K$ \\
		\hline
	\end{tabular}
	\caption{Material properties of titanium and porous silica for the coupled simulation.}
	\label{tab:materialProperties}
\end{table}

\noindent The simulation uses $16\times23\times9$ finite cells with a polynomial degree of $p=3$ and an integration subdivision depth of $n=3$. For the preceding heat simulations, Dirichlet boundary conditions are applied with a prescribed heat of $1000^{\circ}C$ on the top surface and $20^{\circ}C$ on the bottom surface. The resulting temperature inside the tiles is then transferred as a body strain to perform a thermo-elastic simulation. Additionally, the tiles are clamped at the bottom surface. \red{Since the higher-order shape functions are not able to represent jumps in the material distribution, }the simulations of the tile with the discontinuous material distribution are carried out on two separate meshes \red{-- one for each domain --}, which are 'glued' together in a weak sense along their coupling surface\blue{, using the penalty method~\cite{Elhaddad2018}}. \red{Both meshes are equally discretized with $16\times23\times9$ finite cells. Thus, on each mesh, only material jumps from the physical into the fictitious domain appear, which can be decently represented by the FCM.} 

\begin{figure}[!htb]
	\begin{tabular}{c c}
	 	\begin{subfigure}{0.47\textwidth}
             \centering
             \includegraphics[width=\textwidth, trim={10cm 5cm 3cm 5cm}, clip]{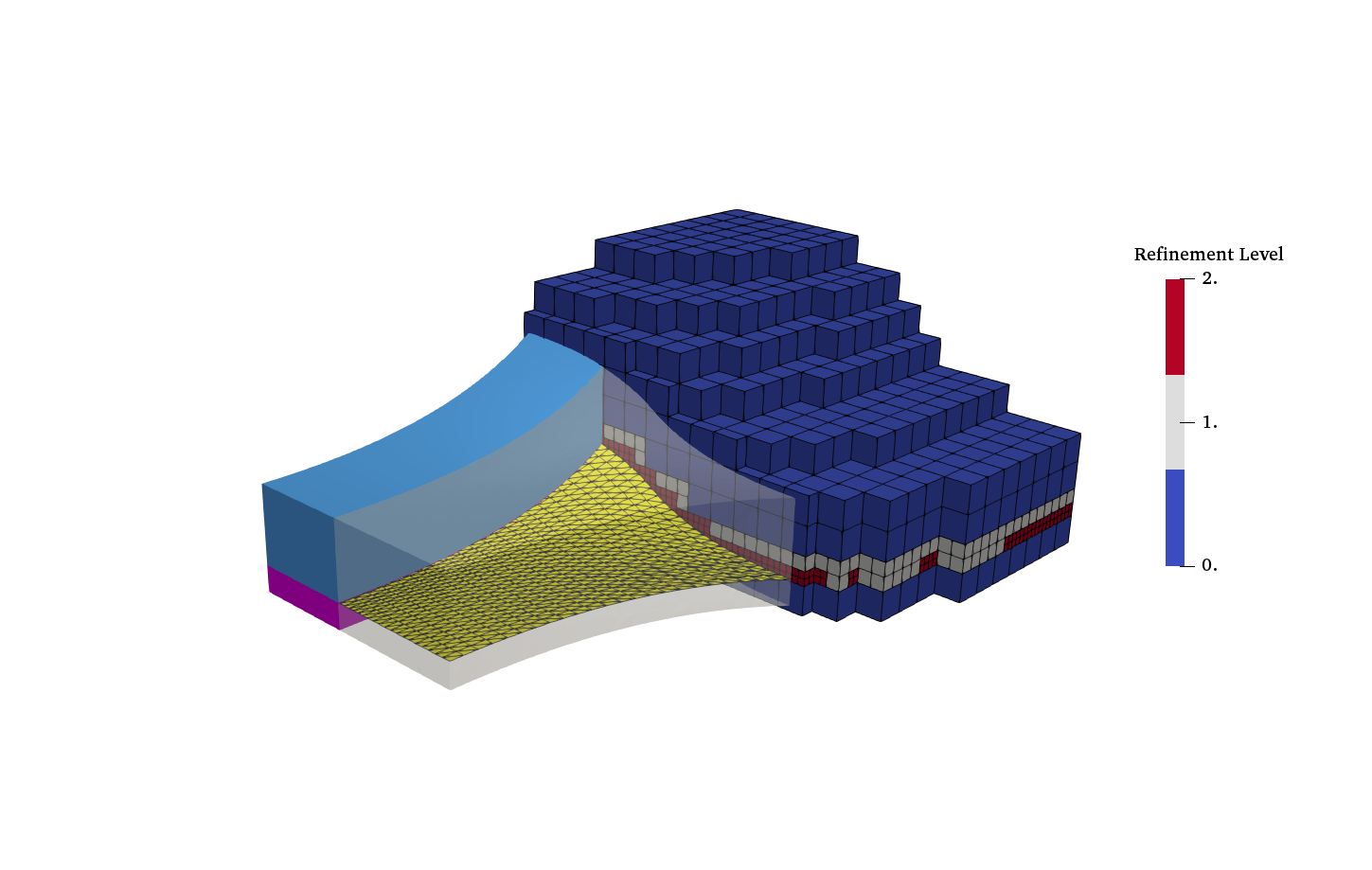}
             \caption{}
        \end{subfigure}	 
	&
		 	\begin{subfigure}{0.47\textwidth}
             \centering
             \includegraphics[width=\textwidth, trim={10cm 5cm 3cm 5cm}, clip]{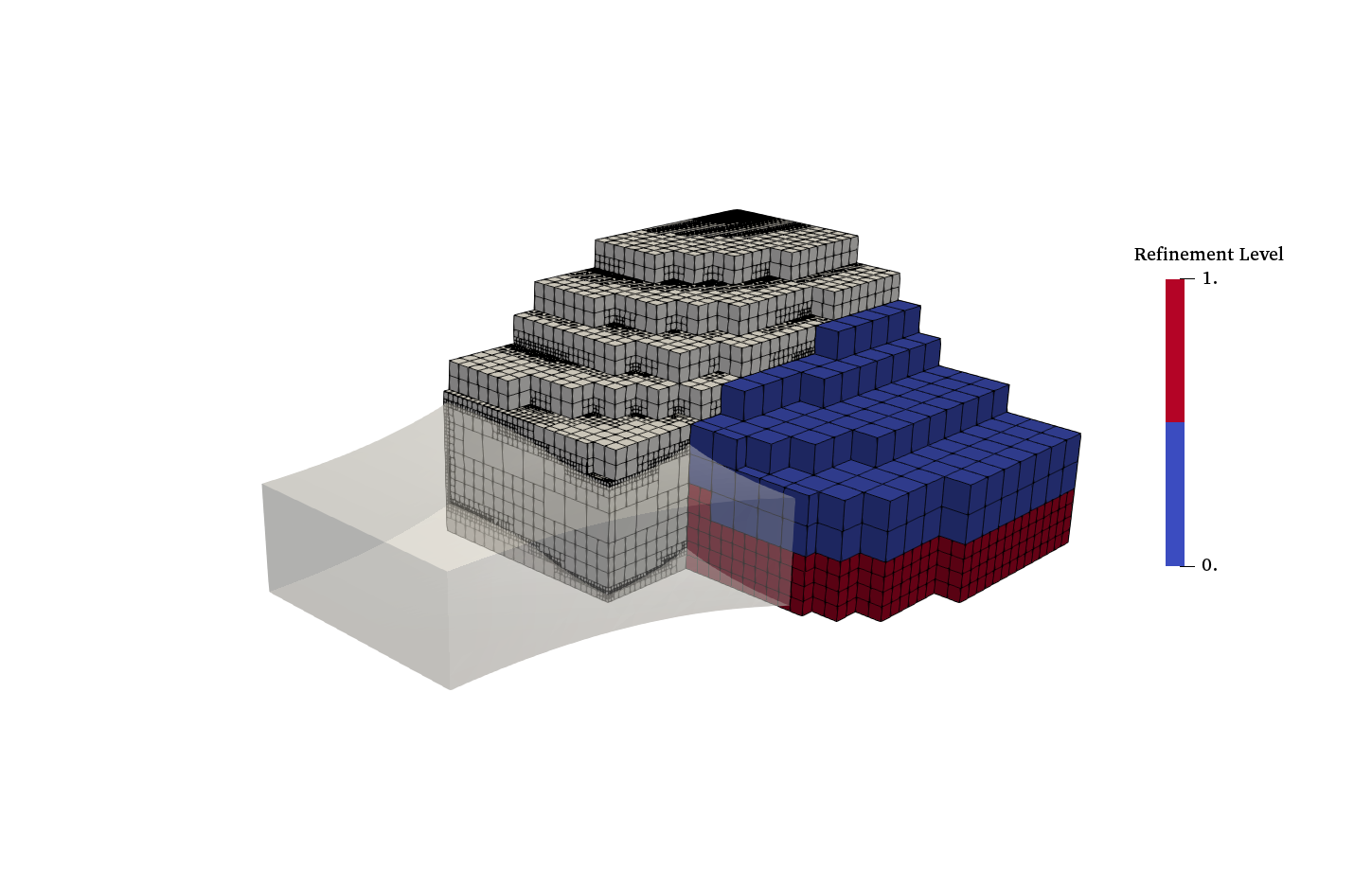}
             \caption{}
        \end{subfigure}	  
	\end{tabular}
	\caption{Discretizations of (a) the discontinuous tile: The mesh is refined twice around the coupling surface (yellow), which divides the upper (light blue) and lower domain (purple). (b) The $C^0-$continuous tile: The FCM mesh is refined once in the transition zone (cells in blue are unrefined, and cells in red are refined once). The grey mesh in the background corresponds to the octree for the integration. The $C^1-$continuous tile is meshed and refined analogously.}	
	\label{fig:curvedTileMesh}
\end{figure}

In order to resolve the critical regions, \blue{the finite cell mesh is refined using $h-$refinement. One $h-$refinement step yields eight subcells for each (refined) finite cell, which can then be further refined in a subsequent refinement step. This kind of refinement introduces hanging nodes between refined and unrefined cells. The multi-level \textit{hp}-method~\cite{ Zander2016} resolves the resulting incompatibilities between the shape functions. For the discontinuous tile, both meshes are refined twice towards the coupling surface -- meaning the individual finite cells are refined with a minimum of 15 and a maximum of 64 subcells. For the continuous tiles, all finite cells that are intersecting the respective transition zones are refined once (see Figure~\ref{fig:curvedTileMesh})}.

\begin{figure}[H]
	\begin{tabular}{c c}
	 	\begin{subfigure}{0.47\textwidth}
             \centering
             \includegraphics[width=\textwidth, trim={6cm 5cm 3cm 5cm}, clip]{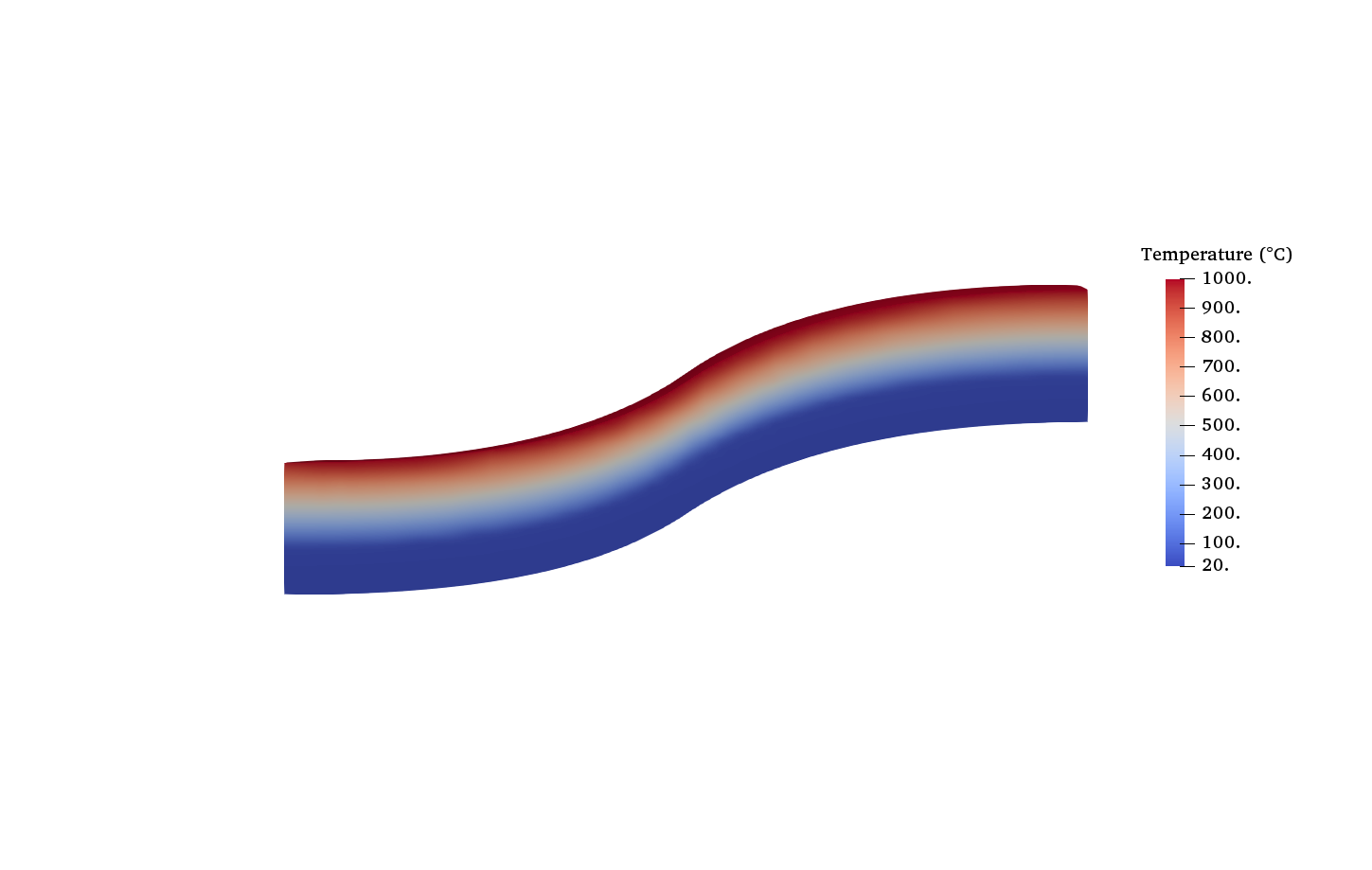}
             \caption{}
        \end{subfigure}	 
	&
		 	\begin{subfigure}{0.47\textwidth}
             \centering
             \includegraphics[width=\textwidth, trim={6cm 5cm 1cm 5cm}, clip]{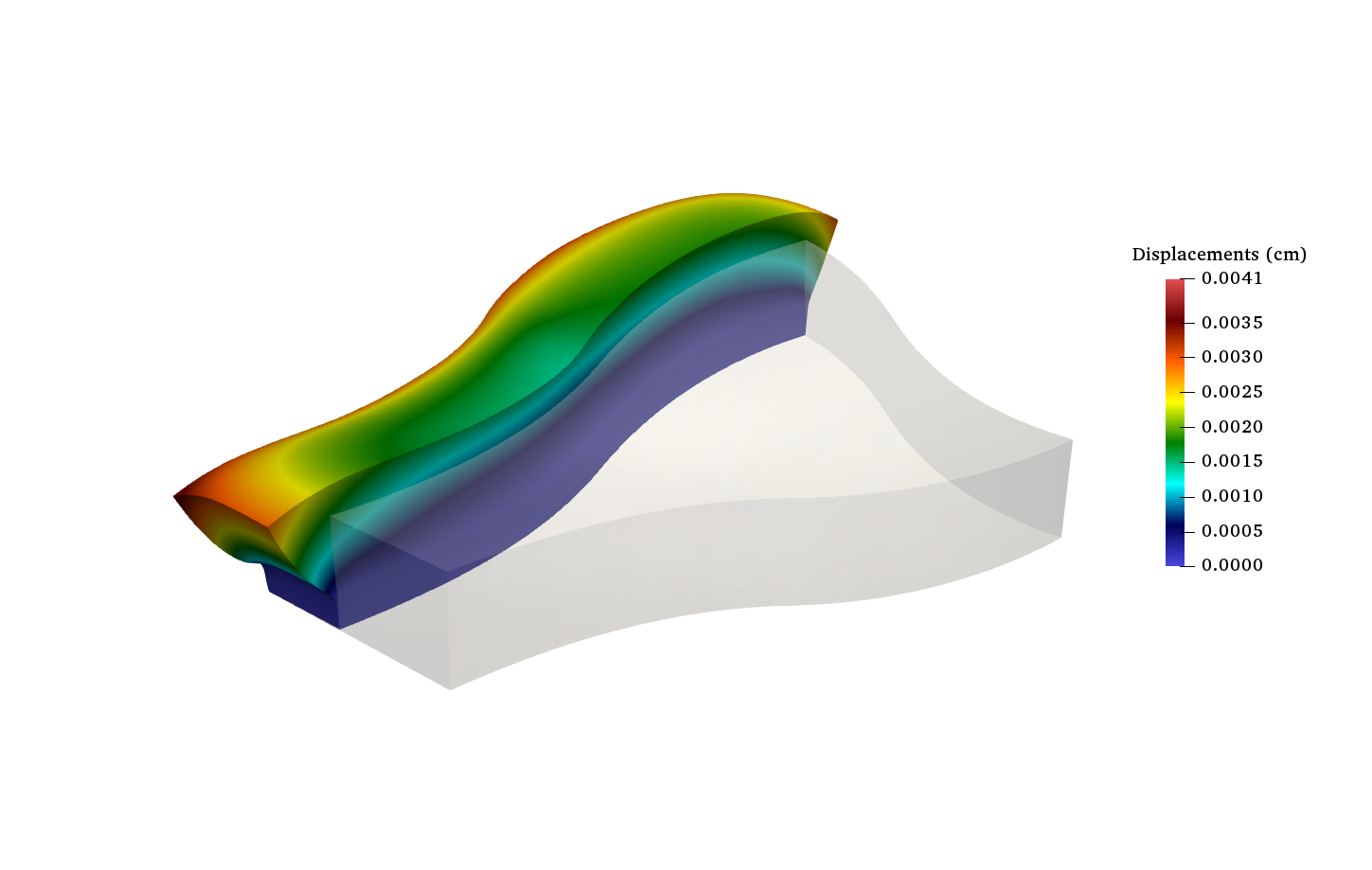}
             \caption{}
        \end{subfigure}	  
	\end{tabular}
	\caption{$C^0-$continuous tile: (a) Temperature distribution and (b) displacements warped by a scaling factor of 1000.}	
	\label{fig:curvedTileHeatDisplacement}
\end{figure}

\begin{figure}[H]
	\begin{tabular}{c c}
	 	\begin{subfigure}{0.47\textwidth}
             \centering
             \includegraphics[width=\textwidth, trim={10cm 8cm 2cm 8cm}, clip]{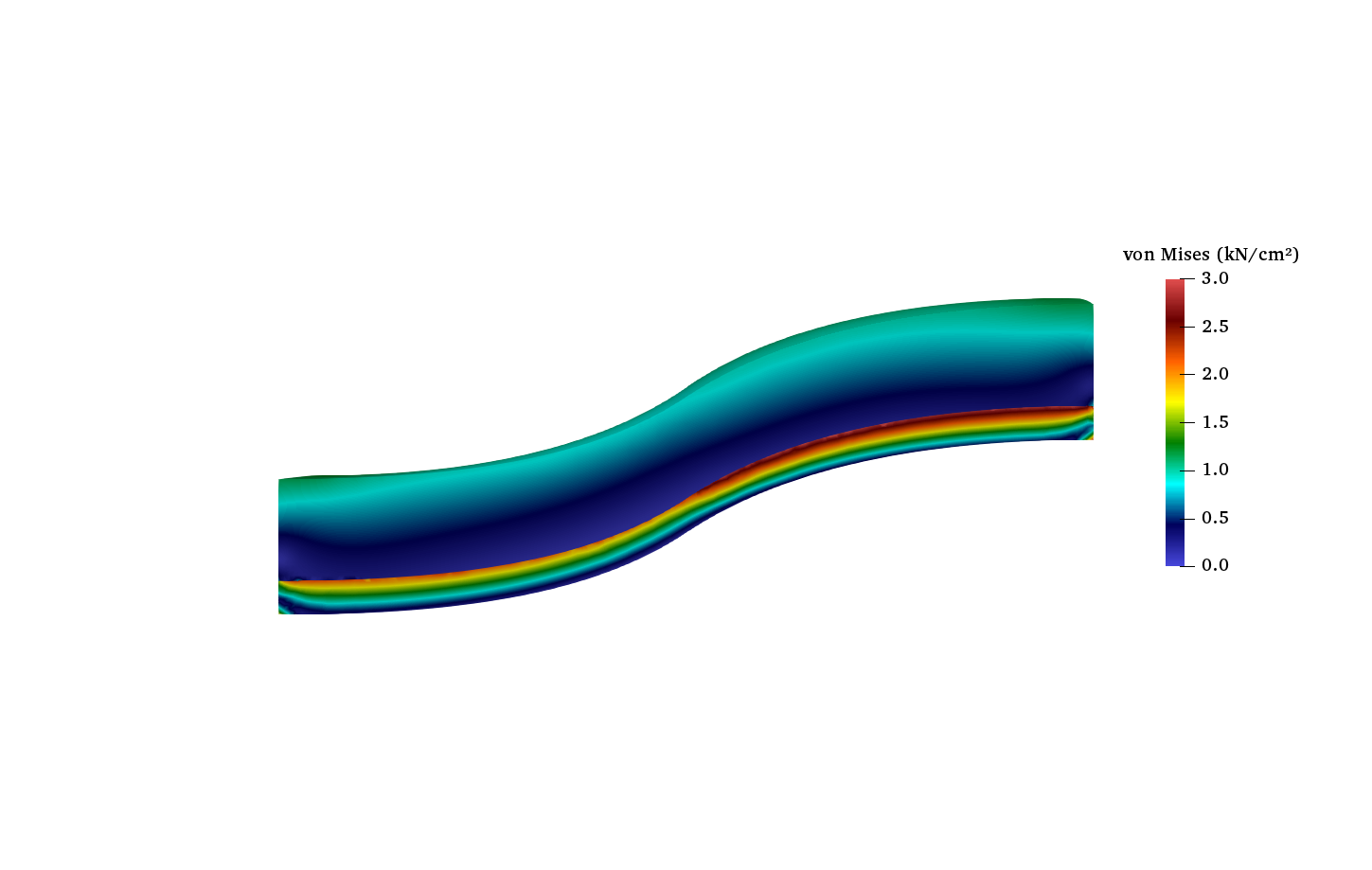}
             \caption{}
        \end{subfigure}	 
	&
		 	\begin{subfigure}{0.47\textwidth}
             \centering
             \includegraphics[width=\textwidth, trim={10cm 8cm 2cm 8cm}, clip]{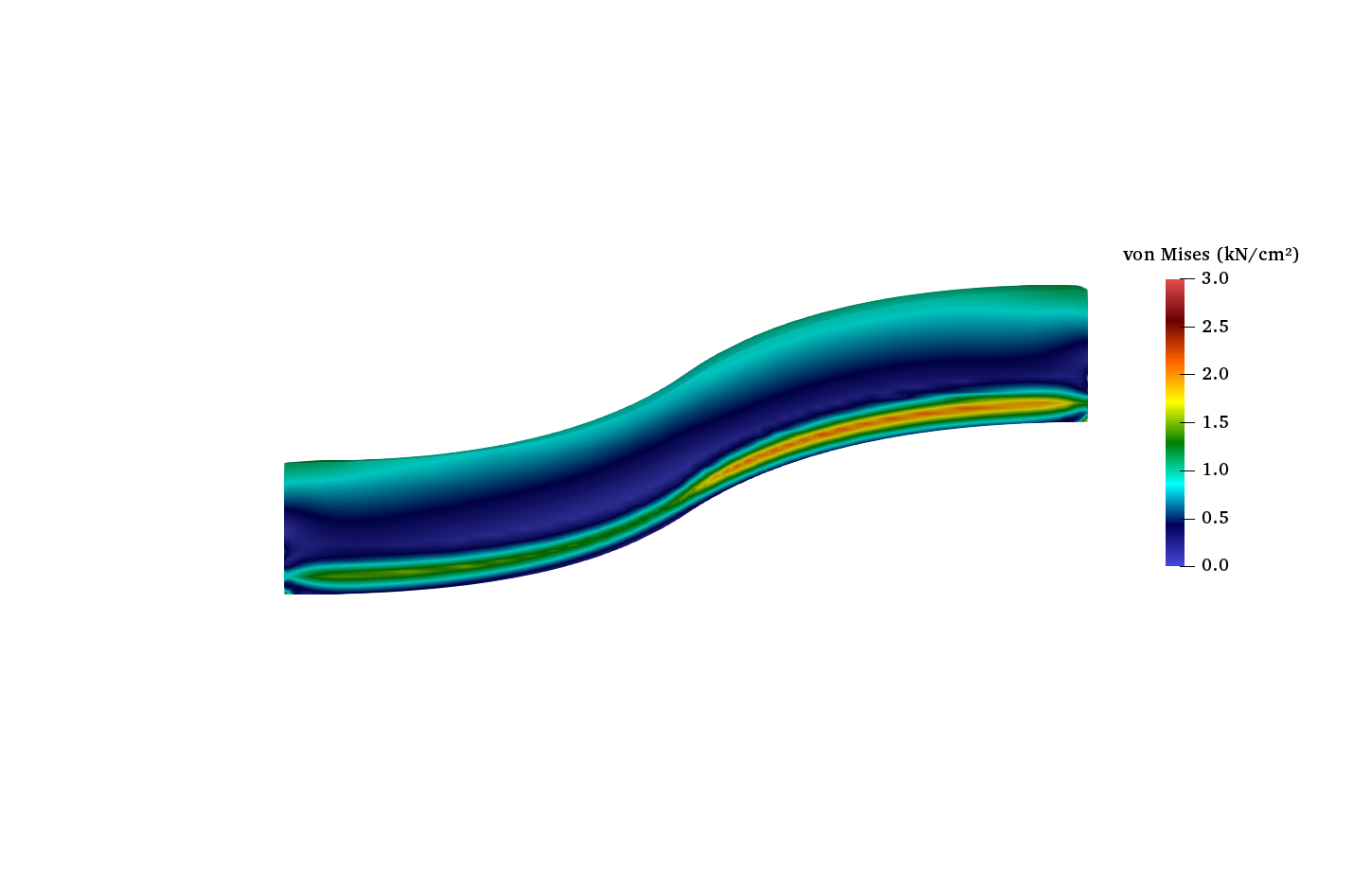}
             \caption{}
        \end{subfigure}	  
	\end{tabular}
	\caption{Von Mises stresses of the (a) discontinuous and (b) $C^0-$continuous thermal shielding tile. The stress distribution of the $C^1-$continuous tile looks very similar to the $C^0-$continuous tile.}	
	\label{fig:curvedTileStresses}
\end{figure}

\begin{figure}[H]
\centering
	\includegraphics[width=\textwidth]{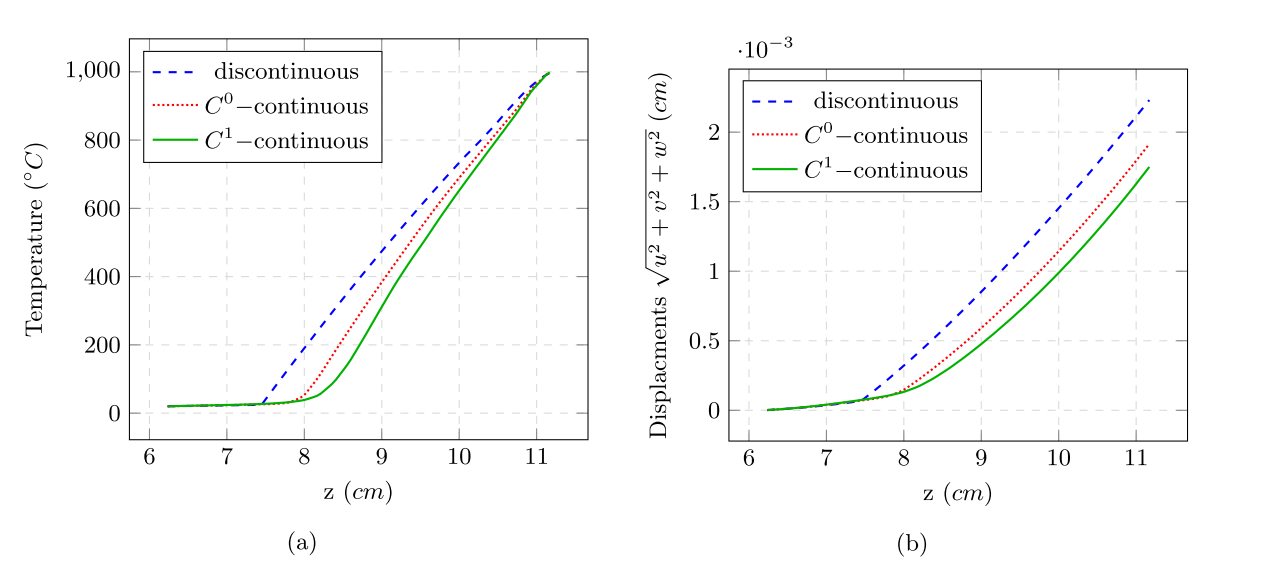}
	\caption{Comparison of (a) the temperature and (b) the displacements of the discontinuous and continuous tiles at $x=5\,cm$, $y=25\,cm$ over the thickness.}	
	\label{fig:curvedTileHeatTempDiagrams}
\end{figure}

\begin{figure}[H]
    \centering
    \includegraphics[width=\textwidth]{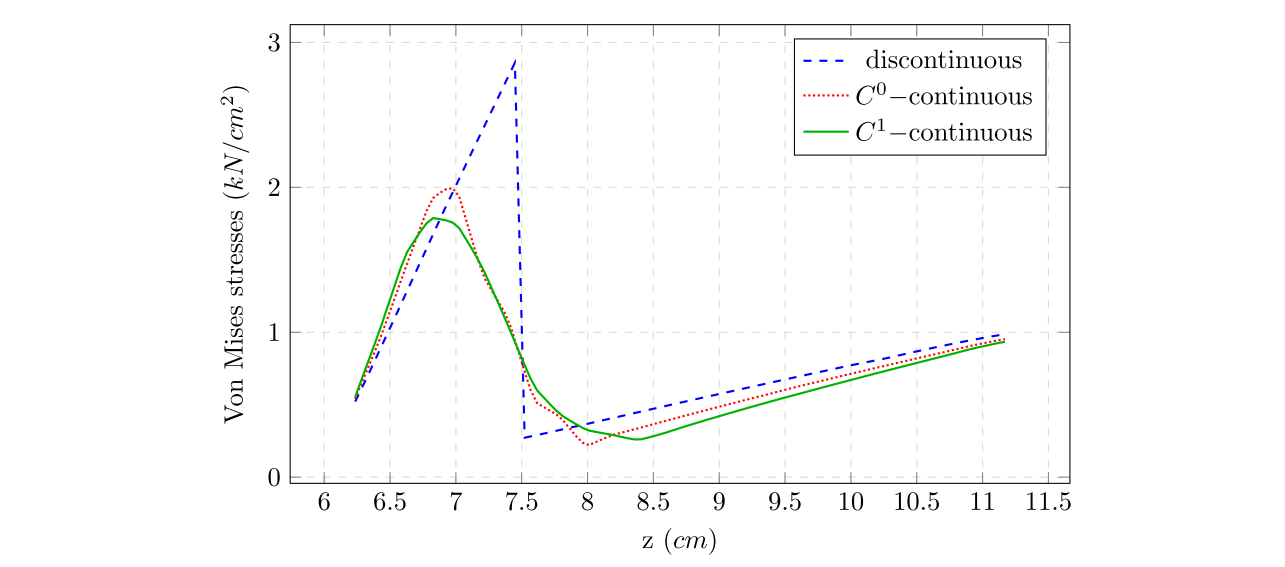}
	\caption{Comparison of the von Mises stresses of the discontinuous and continuous tiles at $x=5\,cm$, $y=25\,cm$ over the thickness.}	
	\label{fig:curvedTileStressDiagram}
\end{figure}

To visualize the results inside the tiles, a cut through the model is investigated at $x = 5\,cm$. Figure~\ref{fig:curvedTileHeatDisplacement} shows the temperature distribution and displacements of the $C^0-$continuous tile. The temperature and the displacement distributions are almost identical for all tiles. More relevant are the stress distributions. As can be seen in Figure~\ref{fig:curvedTileStresses}, a stress concentration occurs at the coupling surface of the discontinuous tile. Figures~\ref{fig:curvedTileHeatTempDiagrams} and~\ref{fig:curvedTileStressDiagram} plot the temperature distribution, displacements, and stresses over the height at $x=5\,cm$ and $y=25\,cm$. 

\noindent The discontinuous material distribution yields a $C^0-$continuous heat and displacement distribution, which then entails a discontinuous stress distribution with a maximum peak at the interface region. This stress concentration is critical as it will potentially cause delamination. The $C^0-$continuous material distribution, on the other hand, ensures a continuous and much smaller stress distribution throughout the entire domain. This effect can be augmented further by using a $C^1-$continuous material distribution. Continuous materials, on the other hand, involve a larger heat flux. For the 1D case, the thermal resistance is reduced to approximately $86\%$ for the $C^0-$continuous and approximately $75\%$ for the $C^1-$continuous material with respect to the discontinuous material distribution.

\subsubsection{Example 3: Anisotropic microstructure}\label{sec:Example3}
\begin{figure}[H]
	\begin{tabular}{c c}
	 	\begin{subfigure}{0.5\textwidth}
             \centering
             \includegraphics[width=\textwidth]{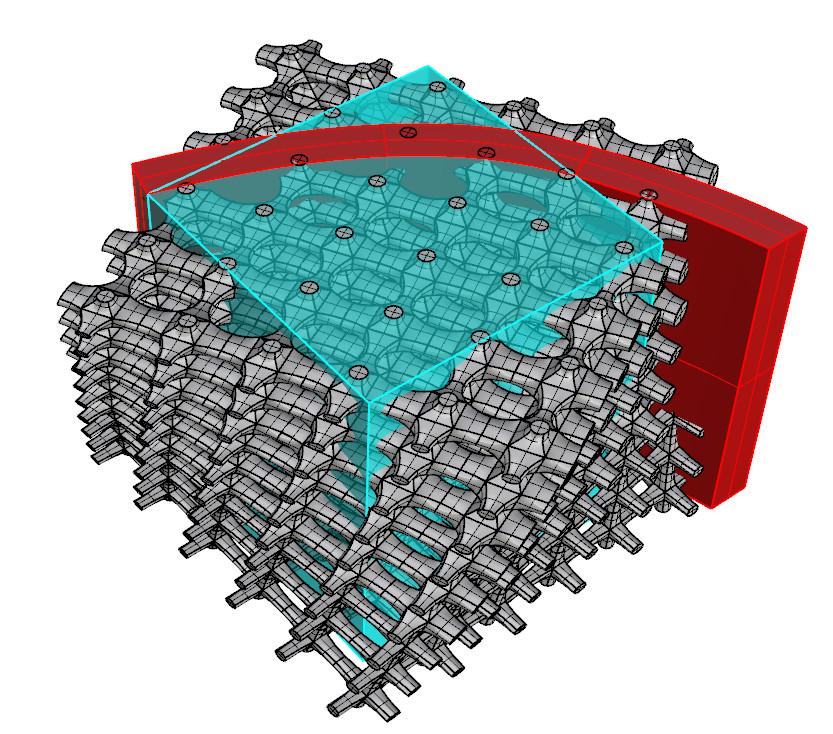}
             \caption{}
        \end{subfigure}	 
	& \quad
		 	\begin{subfigure}{0.36\textwidth}
             \centering
             \includegraphics[width=\textwidth]{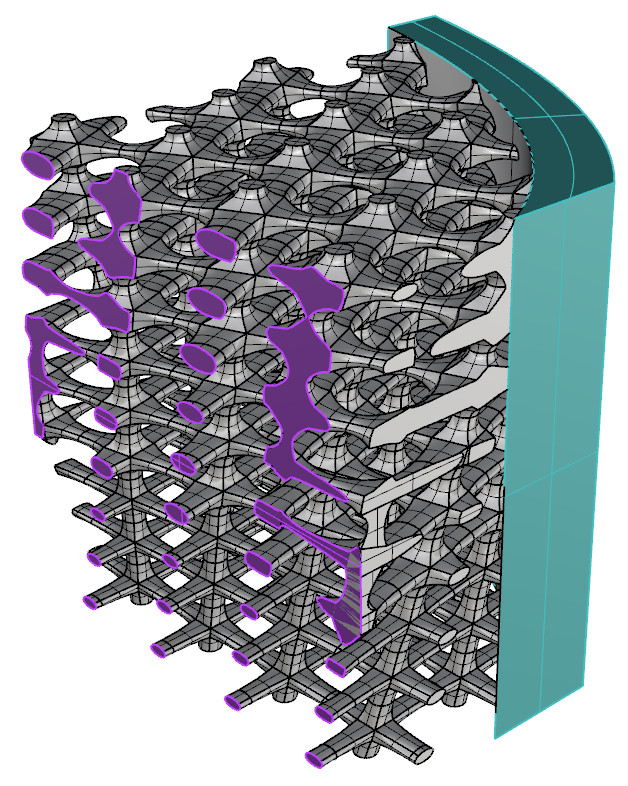}
             \caption{}
             \label{fig:MS_BC}
        \end{subfigure}	  
	\end{tabular}
	\caption{(a) Selection of the computational domain (turquoise). An outer shell (red) is embedded into a microstructure. (b) The intersection of the microstructure with $\Omega_{\cup}$ leading to the physical domain $\Omega_{phy}$. Boundary conditions are applied to the highlighted intersection surfaces.}	
	\label{fig:MS_Selection}
\end{figure}

\blue{The third example addresses the second kind of functionally graded materials -- namely single-material FGM. For this, a linear-elastic simulation of the continuously changing} microstructure, depicted in Figure~\ref{fig:microStructure}, is carried out. It resembles a porous, foam-like microstructure stiffened by an outer shell. To generate this model, a continuously changing microstructure is created with GuIrit. \blue{Different unit tiles -- each composed of seven trivariate B-splines -- are used to tile a parametrically described ruled body (see Figure~\ref{fig:complexPrimitives}). The unit tiles have a growing stiffness from bottom to top, realized by an increasing diameter of the rod in $x-$direction\footnote{\blue{Remark: Due to the rotation of the ruled body, the stiffer direction is changing from bottom to top}}. The resulting microstructure consists of $6\times 6 \times 9$ unit tiles and an overall number of 2268 trivariate B-splines, or V-cells. A direct simulation on this V-model leads to unreasonably high runtimes due to the complexity of the inverse mapping. However, since in this example, the FGM is not modeled within the individual V-cells, but as a single-material continuously changing microstructure, it is possible to carry out a simulation significantly faster on an auxiliary B-rep model. To this end, the V-cells' B-spline surfaces are extracted, and inner surfaces, between consecutive V-cells, are deleted. The resulting B-rep model consists of 8064 B-spline surfaces. With a B-rep CAD tool (Rhinoceros\textsuperscript{\textregistered}), the shell is added as a B-rep volume and combined with the microstructure using the Boolean union operation. Subsequently, the microstructure on the outer side of the shell is trimmed away using the trimming operation with the shell volume's outer surface.} Finally, the computational model is extracted with a Boolean intersection \blue{with the computational domain}. Figure~\ref{fig:MS_Selection} depicts the selection of the computational domain and the final model with the respective surfaces for the boundary conditions. 

\noindent For the simulation, homogeneous Dirichlet boundary conditions are applied on the cutting planes of the shell (see Figure~\ref{fig:MS_BC} -- highlighted in turquoise). The top and bottom surface fix the displacements in $x-$ and $z-$direction, and the front and back surface restrict the displacements in $x-$ and $y-$direction. Dirichlet boundary conditions of $\Delta u = 0.1$  are applied on the outer surfaces on the left side (see Figure~\ref{fig:MS_BC} -- highlighted in purple). All boundary conditions are enforced with the penalty method. A Young's modulus of $E=100\, GPa$ and a Poisson's ratio of $\nu=0.3$ are chosen for $\forall \, \bm{x} \in \Omega_{phy}$. The simulation uses $20\times20\times20$ finite cells, employing Legendre polynomials of degree $p=4$. The subdivision depth of the octree for the integration is set to $n=4$. 

\begin{figure}[H]
	\centering
    \includegraphics[width=\textwidth]{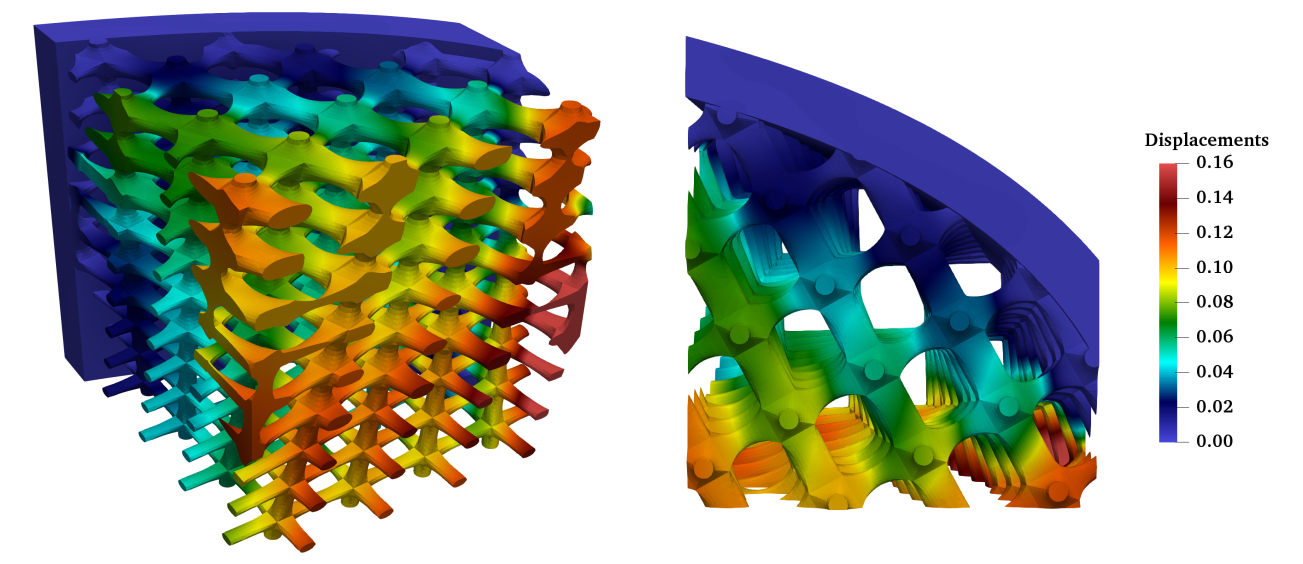}
	\caption{Displacements. }	
	\label{fig:MS_Displacements}
\end{figure}

\begin{figure}[H]
	\centering
    \includegraphics[width=\textwidth]{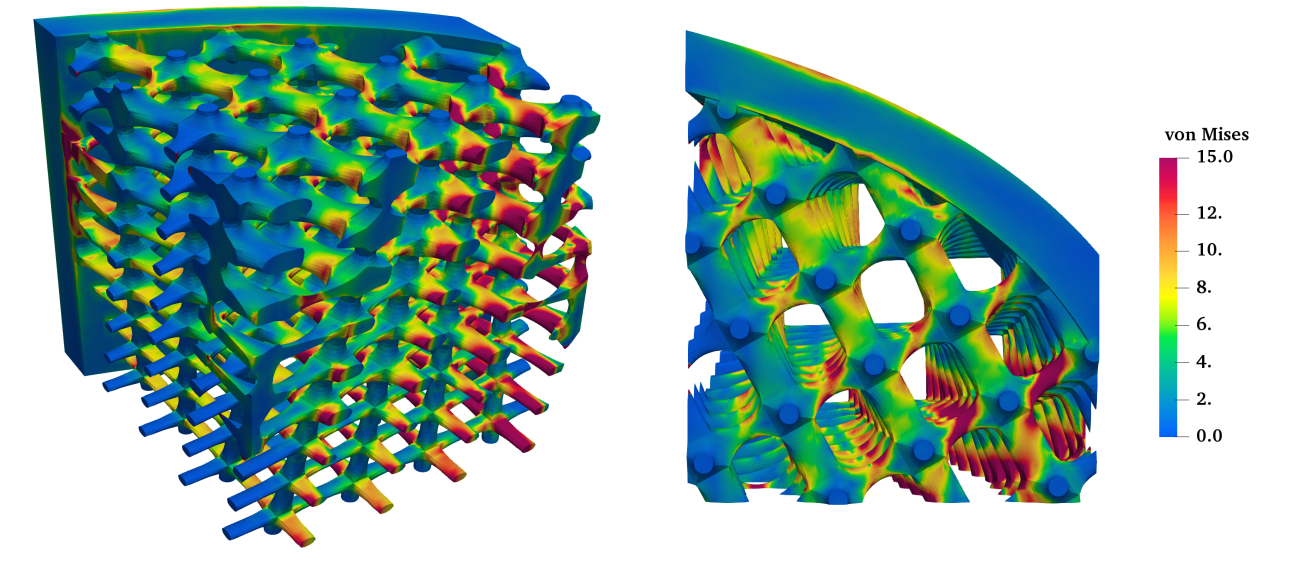}
	\caption{Von Mises stresses. }	
	\label{fig:MS_Stresses}
\end{figure}

Figures~\ref{fig:MS_Displacements} and~\ref{fig:MS_Stresses} show the displacements and the von Mises stresses. Certainly, such a fully resolved simulation is slower than the numerical homogenization presented in Section~\ref{sec:constructiveFGM} – especially because homogenization in the linear case allows the creation of a look-up table. However, the discussed fully resolved model can be used to verify the homogenization. Homogenization is addressed in the following Examples~\ref{sec:Example4}, and~\ref{sec:Example5}. Note, since the shape functions are badly suited to represent holes inside one finite cell, meaning 'material--void--material' \cite{Coradello2020}, the microstructure needs to be resolved with many finite cells. A remedy can be local enrichment, as presented in~\cite{Legrain2012}.

\subsubsection{Example 4: Material characterization database for unit tiles}\label{sec:Example4}
\blue{A fully resolved numerical simulation of a microstructure -- as presented in Example~\ref{sec:Example3} -- is computationally very demanding in both memory consumption and simulation time. For large-scale microstructures (as in Example~\ref{sec:Example5}), fully-resolved computations need to be carried out on a high-performance computer, or might even be not applicable. A remedy is offered by homogenization. As explained in Section~\ref{sec:homogenization}, for a functionally graded microstructure it is sufficient to compute the effective material tensors $\bm{C}_{Ti}^*$ only for several representative unit tiles, and interpolate the material properties in-between, according to the parametrization of microstructure.}

\blue{Two parameters are used to characterize the unit tiles in the Examples~\ref{sec:Example3}, \ref{sec:Example4} and \ref{sec:Example5}, the diameter of the rod in $x-$direction and rotation angle around the $z-$axis. In order to compute the respective microscopic material behaviors, homogenization simulations are carried out for unit tiles with three different configurations of the diameter of the rod in $x-$direction (\O~0.2 mm, \O~0.3 mm, and \O~0.4 mm), yielding the unrotated, effective material tensors $\bm{C}_{Ti}^*$.}

\blue{For the homogenization simulations, the} material of the microstructure is considered to be steel with a Young's modulus of $E=210\,GPa$, and a Poisson's ratio of $\nu = 0.3$. Each tile is discretized with $11\times11\times11$ finite cells of polynomial degree $p=5$. For the domain integration, the moment-fitting approach~\cite{Hubrich2017} with the depth of an underlying octree of $d=6$ is chosen. As the structures under consideration are\blue{, in good approximation,} geometrically periodic, periodic boundary conditions are the natural choice for transferring the macroscopic quantities to the microscopic unit cells.

\begin{figure}[H]
	\centering
	\begin{tabular}{c c c}
		\begin{subfigure}{0.3\textwidth}
			\centering
			\includegraphics[width=\textwidth]{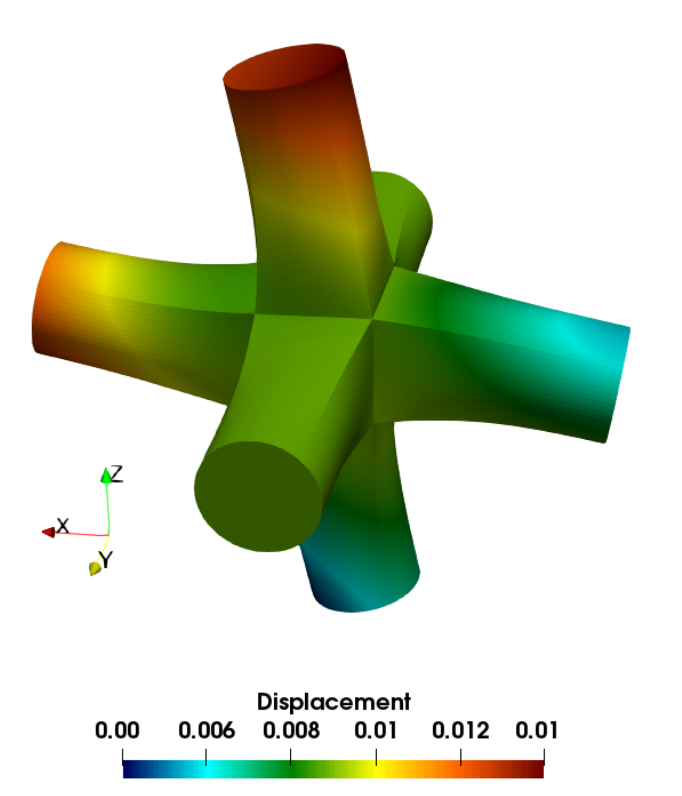}
			\caption{ Tile 1}
		\end{subfigure}	 
		&
		\begin{subfigure}{0.3\textwidth}
			\centering
			\includegraphics[width=\textwidth]{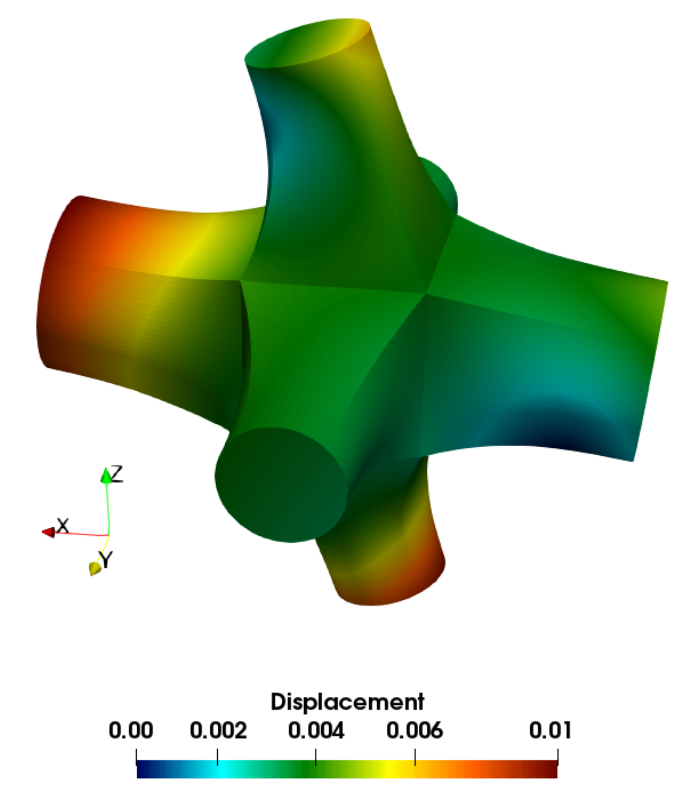}
			\caption{ Tile 2}
		\end{subfigure}	 
		& 
		\begin{subfigure}{0.3\textwidth}
			\centering
			\includegraphics[width=\textwidth]{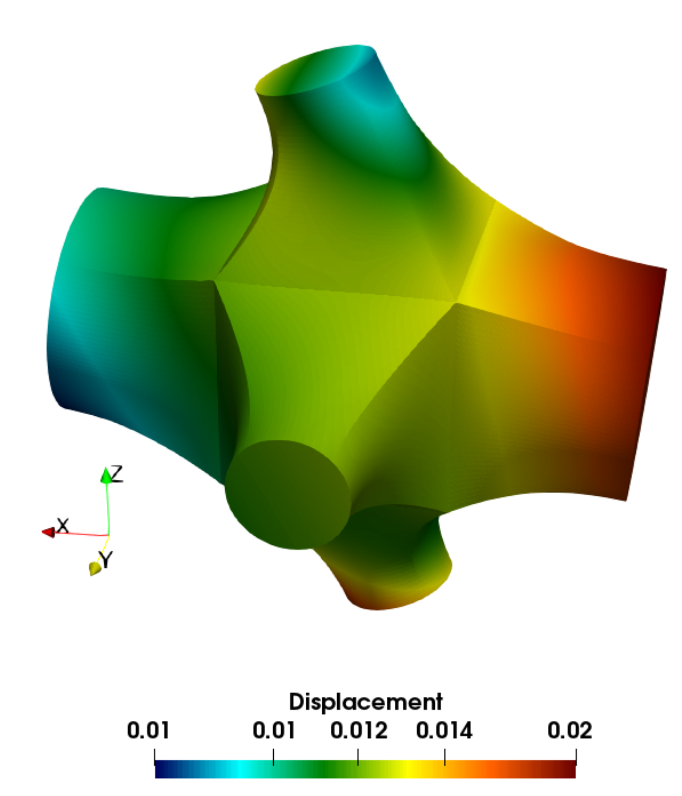}
			\caption{Tile 3}
		\end{subfigure}	  
	\end{tabular}
	\caption{\blue{Homogenization simulation with periodic boundary conditions:} Displacement field of the warped tiles with a scale factor $s=10$.}
	\label{fig:PBCTiles}
\end{figure}

\blue{Figure~\ref{fig:PBCTiles} shows the displacement fields under shear load for the unit tiles in the unrotated configuration. The resulting homogenized material tensors for the tiles 1, 2 and 3 are summarized in the Equations~(\ref{eq:Ctile1}),~(\ref{eq:Ctile2}) and~(\ref{eq:Ctile3}), respectively. One can identify different material behaviors, which is expected due to the respective unit tiles' geometrical features. The orientation and the thickness of the rods have an important effect on the final material behavior.} Tile 1 shows a cubic macroscopic material symmetry with three independent elasticity coefficients~\cite{Cowin2007}, namely $C_{11}, C_{12}$ and $C_{44}$

\begin{equation}
\bm{C}^*_{T1} = 
\begin{bmatrix}
7895.81 & 432.89  & 432.89       & 0.00  & 0.00  & 0.00  \\
432.89  & 7895.81 & 432.89       & 0.00  & 0.00  & 0.00  \\
432.89  & 432.89  & 7895.81      & 0.00  & 0.00  & 0.00  \\
0.00    & 0.00    & 0.00         & 200.71& 0.00  & 0.00  \\
0.00    & 0.00    & 0.00         & 0.00  & 200.71& 0.00  \\
0.00    & 0.00    & 0.00         & 0.00  & 0.00  & 200.71	\\
\end{bmatrix}
\label{eq:Ctile1}
\end{equation}

\noindent Due to the stiffer direction in $x-$direction, tile 2 and 3 show a tetragonal effective material symmetry with $C_{11}, C_{22}, C_{44}, C_{55}, C_{12}$ and $C_{23}$ as independent entries:

\begin{equation}
\bm{C}^*_{T2} = 
\begin{bmatrix}
18246.81 & 1026.56  & 1026.56      & 0.00  & 0.00  & 0.00  \\
1026.56  & 11066.80 & 659.81       & 0.00  & 0.00  & 0.00  \\
1026.56  & 659.81   & 11066.80     & 0.00  & 0.00  & 0.00  \\
0.00     & 0.00     & 0.00         & 769.49& 0.00  & 0.00  \\
0.00     & 0.00     & 0.00         & 0.00  & 590.69& 0.00  \\
0.00     & 0.00     & 0.00         & 0.00  & 0.00  & 769.49	\\
\end{bmatrix}
\label{eq:Ctile2}
\end{equation}

\begin{equation}
\bm{C}^*_{T3} = 
\begin{bmatrix}
33809.00 & 2037.73  & 2037.73      & 0.00  & 0.00  & 0.00  \\
2037.73  & 14770.28 & 997.14       & 0.00  & 0.00  & 0.00  \\
2037.73  & 997.14   & 14771.08      & 0.00  & 0.00  & 0.00  \\
0.00    & 0.00    & 0.00         & 2022.10& 0.00  & 0.00  \\
0.00    & 0.00    & 0.00         & 0.00  & 1375.86& 0.00  \\
0.00    & 0.00    & 0.00         & 0.00  & 0.00  & 2022.17	\\
\end{bmatrix}
\label{eq:Ctile3}
\end{equation}

\noindent \blue{The material tensors $\bm{C}_{Ti}'$ for the second changing parameter -- the rotation around the $z-$axis -- can be computed by a coordinate transformation, and thus require no homogenization simulations.} The Bond-Transformation matrices~\cite{Bond1943} can be used to rotate the effective elasticity tensor by a matrix-matrix multiplication. Assume the following ordering of the macroscopic stresses $\sigma_{ij}^M$ and strains $\varepsilon_{ij}^M$ in the Voigt notation
\begin{equation}
\begin{bmatrix}
\sigma^M_{11}\\\sigma^M_{22}\\\sigma^M_{33}\\\sigma^M_{12}\\\sigma^M_{23}\\\sigma^M_{13}\\
\end{bmatrix} = \begin{bmatrix}
C^*_{11} & C^*_{12} & C^*_{13} &C^*_{14} & C^*_{15} & C^*_{16}\\
C^*_{12} & C^*_{22} & C^*_{23} &C^*_{24} & C^*_{25} & C^*_{26}\\
C^*_{13} & C^*_{23} & C^*_{33} &C^*_{34} & C^*_{35} & C^*_{36}\\
C^*_{14} & C^*_{24} & C^*_{34} &C^*_{44} & C^*_{45} & C^*_{46}\\
C^*_{15} & C^*_{25} & C^*_{35} &C^*_{45} & C^*_{55} & C^*_{56}\\
C^*_{16} & C^*_{26} & C^*_{36} &C^*_{46} & C^*_{56} & C^*_{66}
\end{bmatrix} 
\begin{bmatrix}
\varepsilon^M_{11}\\\varepsilon^M_{22}\\\varepsilon^M_{33}\\\varepsilon^M_{12}\\\varepsilon^M_{23}\\\varepsilon^M_{13}\\
\end{bmatrix} \, .
\label{eq:effectiveMaterial}
\end{equation}

\noindent Then, the transformation of the effective elastic tensor reads as follows
\begin{equation}
\bm{C}' = \bm{M} \bm{C}^* \bm{N}^{-1} \, ,
\label{eq:transformationRotation}
\end{equation}

\noindent where $\bm{C}^*$ is the effective elasticity tensor, $\bm{C}'$ is the effective elasticity tensor in rotated coordinates, and $\bm{M}$ and $\bm{N}$ are the Bond-stress and the Bond-strain transformation matrices, respectively. \blue{For the rotation around the $z-$axis, the} Bond strain and stress matrices are defined as follows
\begin{equation}
M = 
\begin{bmatrix}
cos^2(\alpha) & sin^2(\alpha) & 0 & sin(2\alpha)  & 0 & 0\\
sin^2(\alpha) & cos^2(\alpha) & 0 & -sin(2\alpha) & 0 & 0\\
0 & 0 & 1.0 & 0 & 0 & 0\\
-\frac{sin(2\alpha)}{2} & \frac{sin(2\alpha)}{2} & 0 & cos(2\alpha) & 0 & 0 \\
0 & 0 & 0 & 0 & cos(\alpha) & -sin(\alpha)\\
0 & 0 & 0 & 0 & sin(\alpha) & cos(\alpha)
\end{bmatrix} 
\end{equation}

\begin{equation}
N = 
\begin{bmatrix}
cos^2(\alpha) & sin^2(\alpha) & 0 & \frac{sin(2\alpha)}{2}  & 0 & 0\\
sin^2(\alpha) & cos^2(\alpha) & 0 & -\frac{sin(2\alpha)}{2}  & 0 & 0\\
0             & 0             & 1.0 & 0 & 0 & 0\\
-sin(2\alpha)  & sin(2\alpha)  & 0 &cos(2\alpha) & 0 & 0 \\
0             & 0             & 0 & 0 & cos(\alpha) & -sin(\alpha)\\
0             & 0             & 0 & 0 & sin(\alpha) & cos(\alpha)
\end{bmatrix} 
\end{equation}

\noindent \blue{In the Appendix, Section~\ref{sec:appendixEffectiveMaterialTensors} presents the respective independent material tensor entries $C_{ii}$ of the three unit tiles for arbitrary rotations around the $z-$axis, following Equation~(\ref{eq:transformationRotation}).}

Given a set of different (an-)isotropic unit tiles that can be used to construct such microstructures, it is possible to create a look-up table of homogenized materials, which can then be used to simulate different macroscopic load cases. Table~\ref{tab:lookUpTable} is a snippet of such a look-up table, and it shows the effective elasticity tensors for the two varying material properties. The material properties in-between can be interpolated. \blue{This Table~\ref{tab:lookUpTable}, will be used in the following Example~\ref{sec:Example5} to compute a large-scale microstructure with interpolated homogenized material properties.}

\begin{table}[!htb]
\renewcommand{\arraystretch}{1.2} 
\centering
\begin{tabular}{cc|c|c|c|c} \cline{3-5}
 & & \multicolumn{3}{ c| } {Rotation around the $z-$axis} \\ \cline{3-5} 
 & & $0\si{\degree}$ & $22.5\si{\degree}$ & $45\si{\degree}$          \\ \cline{1-5}
\multicolumn{1}{|c }{\multirow{3}{*}[-25mm]{\rotatebox[origin=c]{90}{Diameter of the rod in $x-$direction}} } & 
\multicolumn{1}{|c|}{\raisebox{-12mm}{$0.2\,mm$}} &
	\raisebox{-27mm}{\includegraphics[height=30mm]{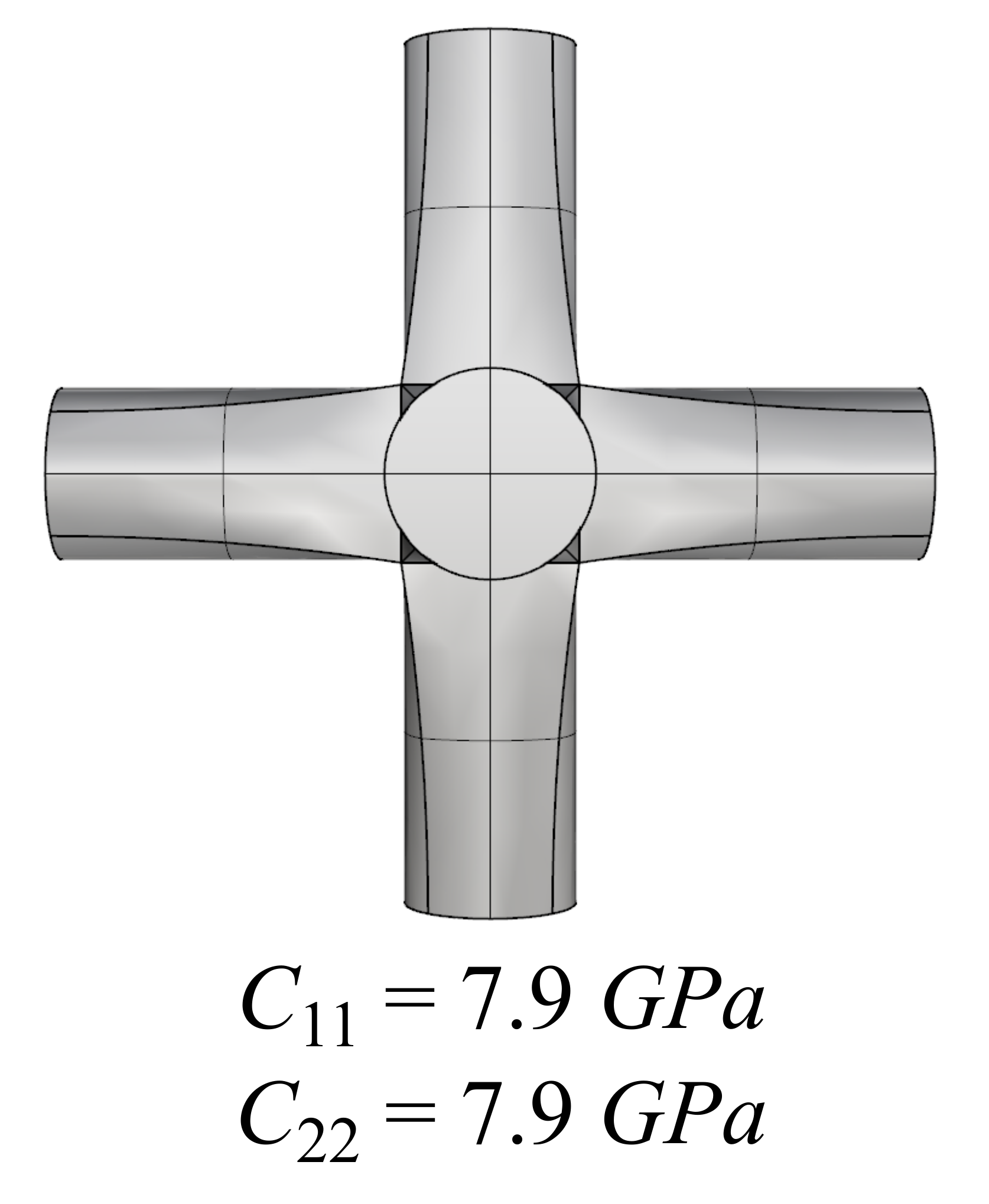}} & 
	\raisebox{-27mm}{\includegraphics[height=30mm]{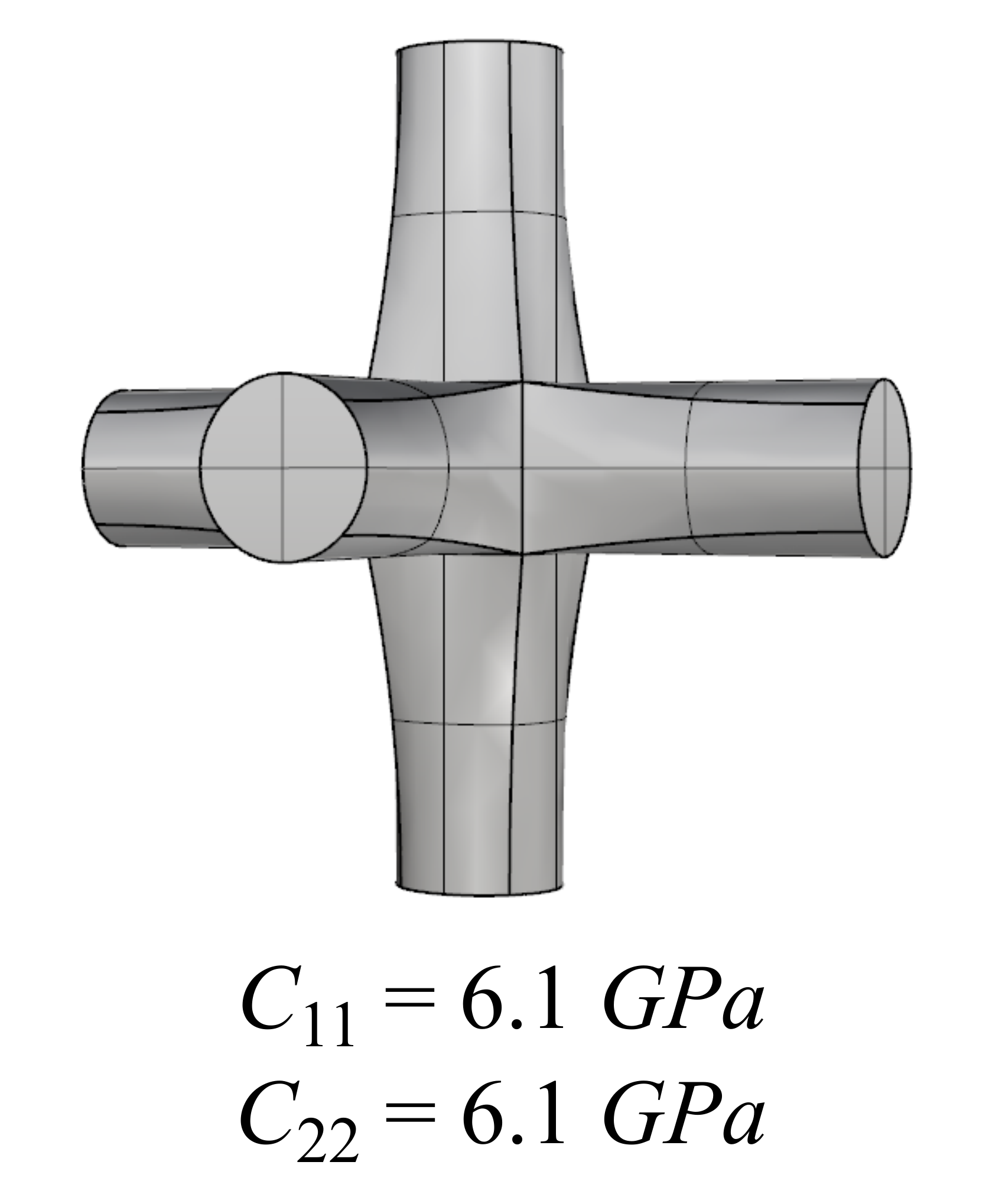}} & 
	\raisebox{-27mm}{\includegraphics[height=30mm]{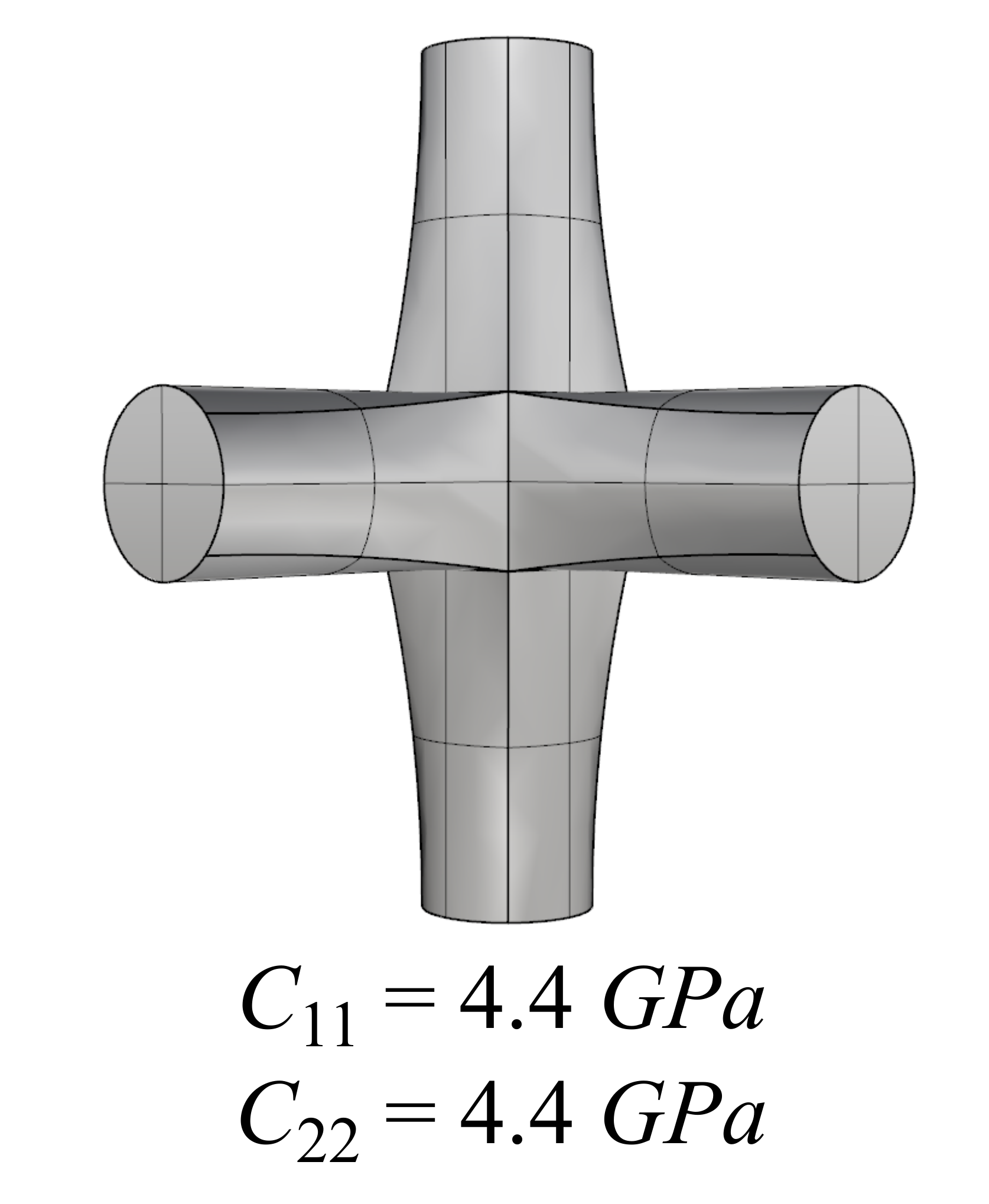}} \\ \cline{2-5}
\multicolumn{1}{ |c }{} & \multicolumn{1}{ |c| }{\raisebox{-12mm}{$0.3\,mm$}} & 
	\raisebox{-27mm}{\includegraphics[height=30mm]{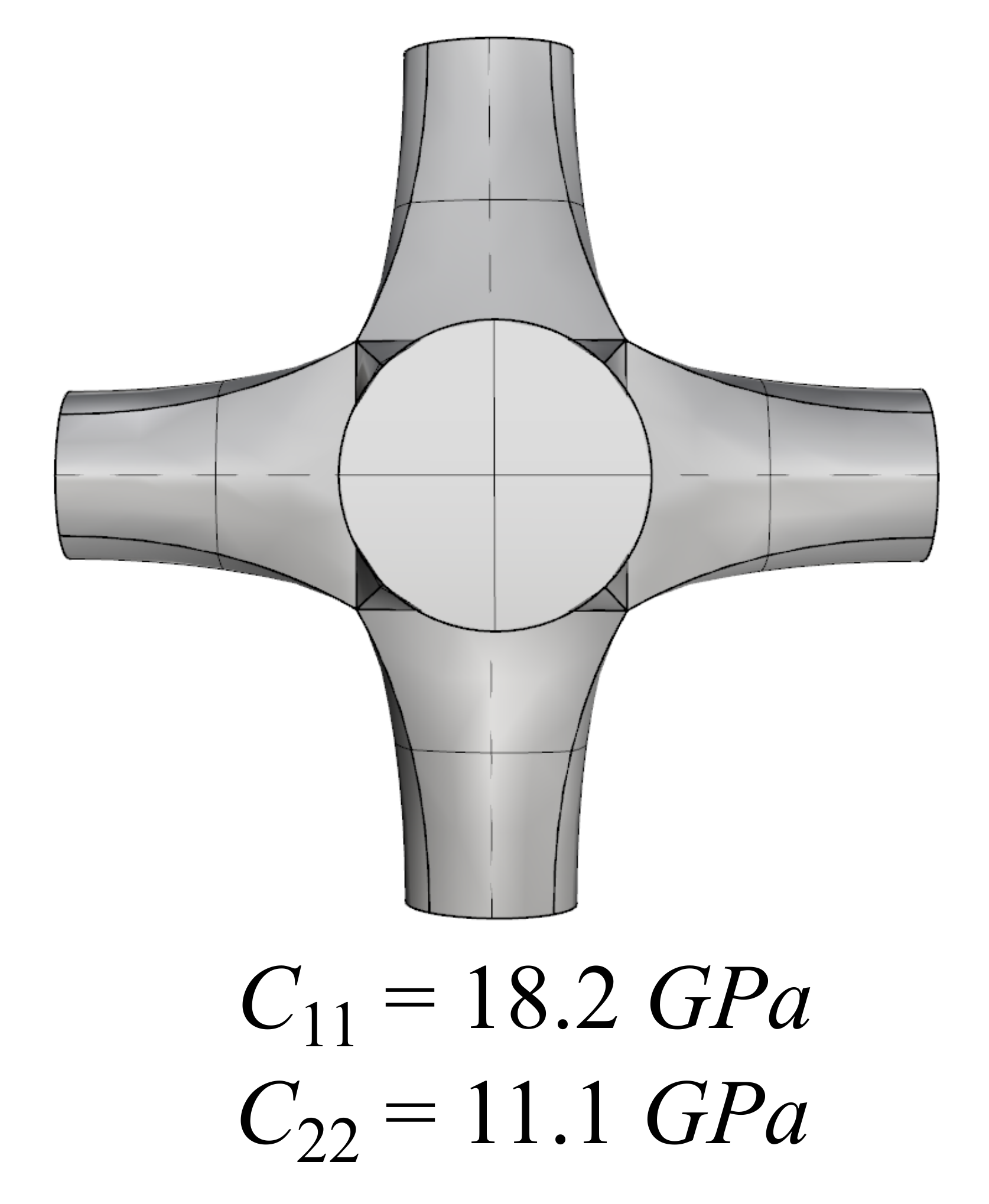}} & 
	\raisebox{-27mm}{\includegraphics[height=30mm]{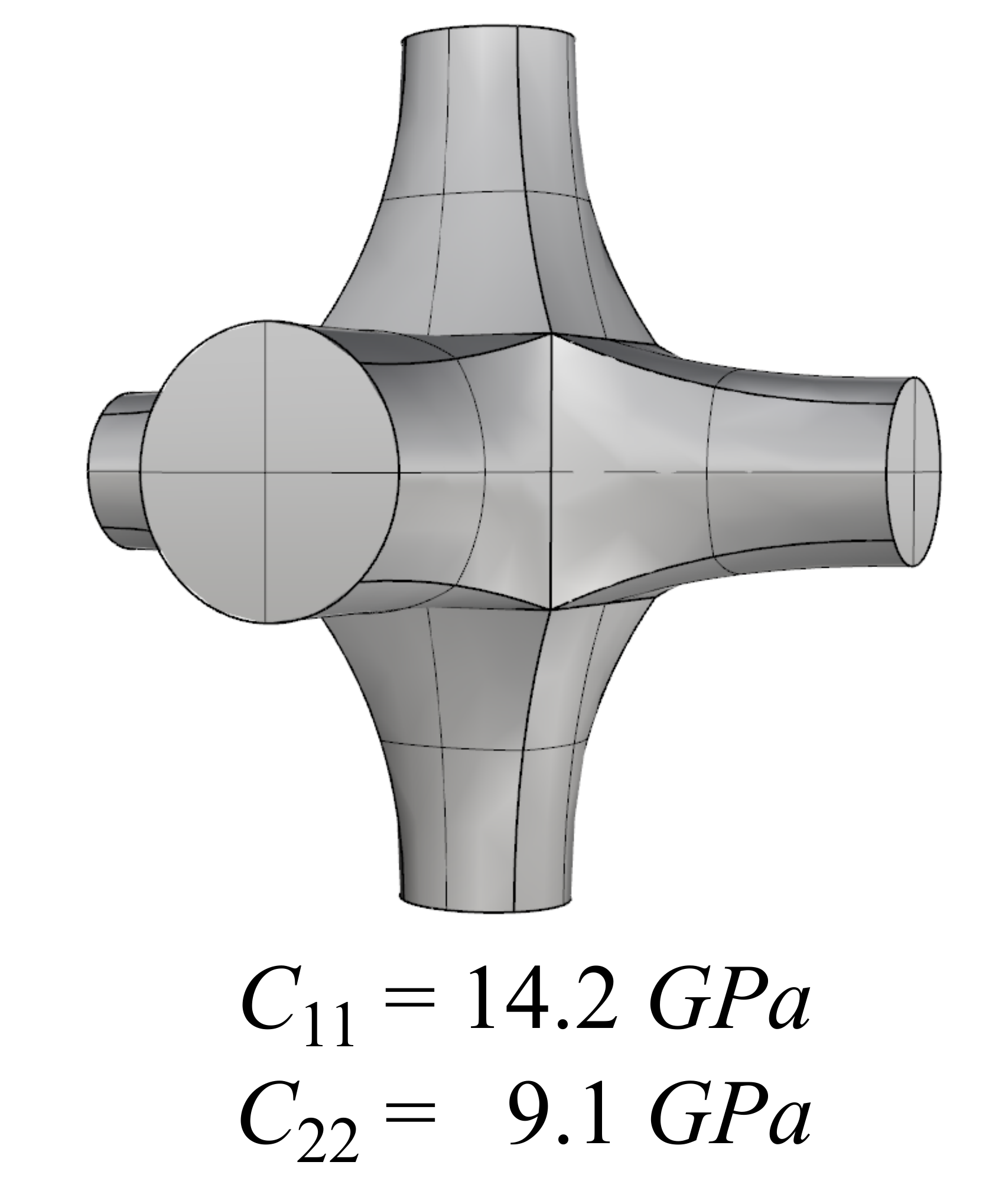}} & 
	\raisebox{-27mm}{\includegraphics[height=30mm]{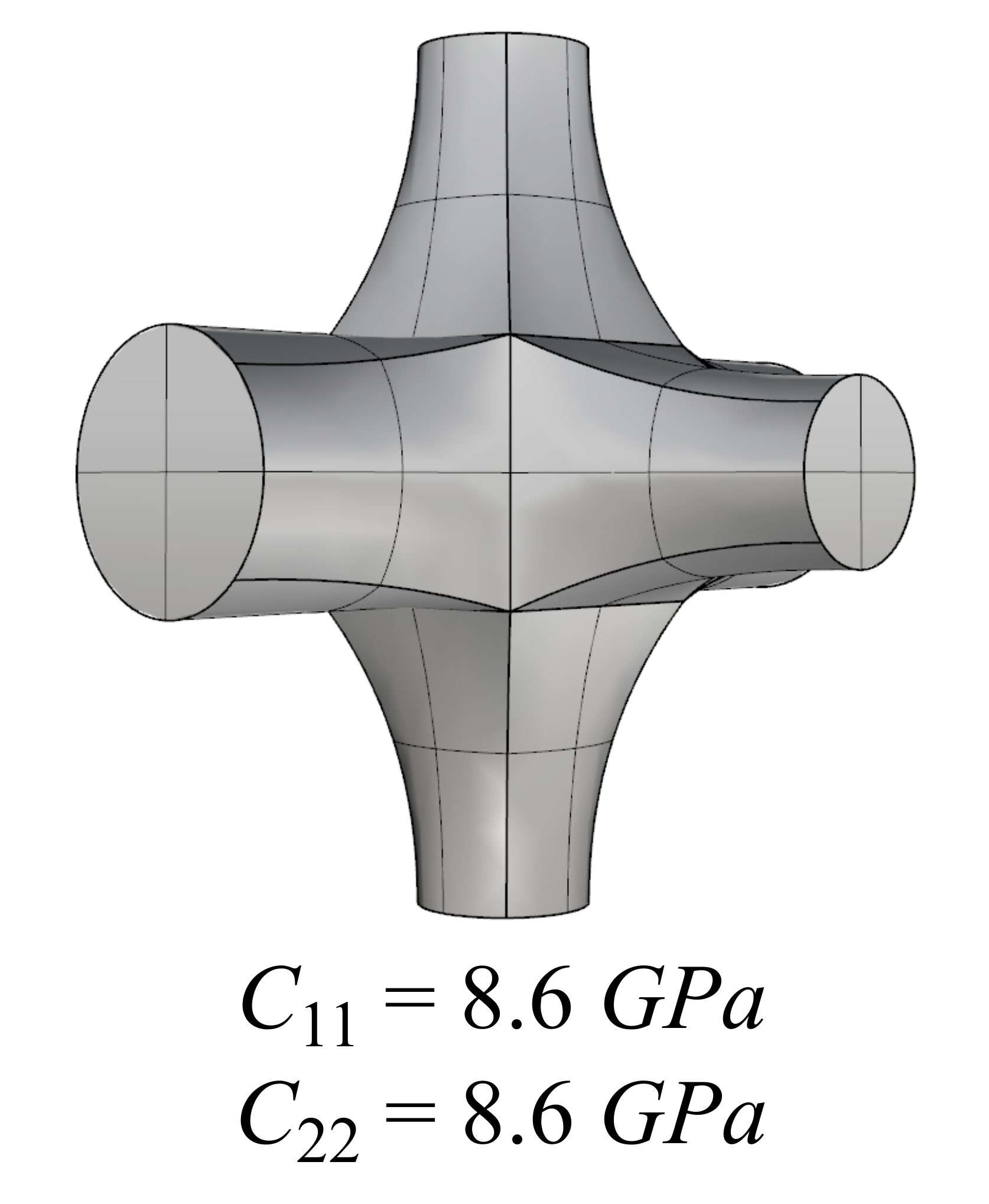}} \\ \cline{2-5}
\multicolumn{1}{ |c }{} & \multicolumn{1}{ |c| }{\raisebox{-12mm}{$0.4\,mm$}} & 
	\raisebox{-27mm}{\includegraphics[height=30mm]{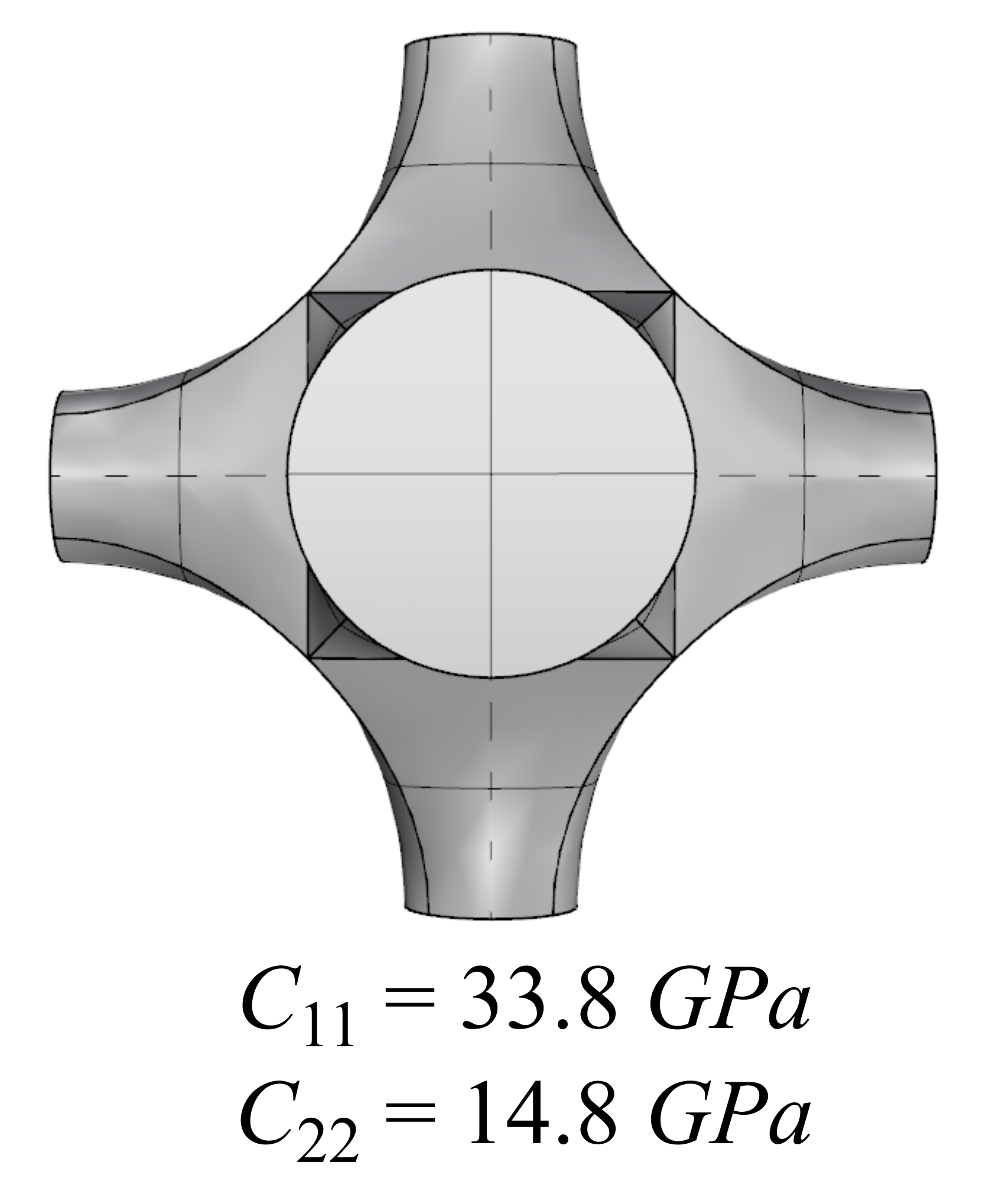}} & 
	\raisebox{-27mm}{\includegraphics[height=30mm]{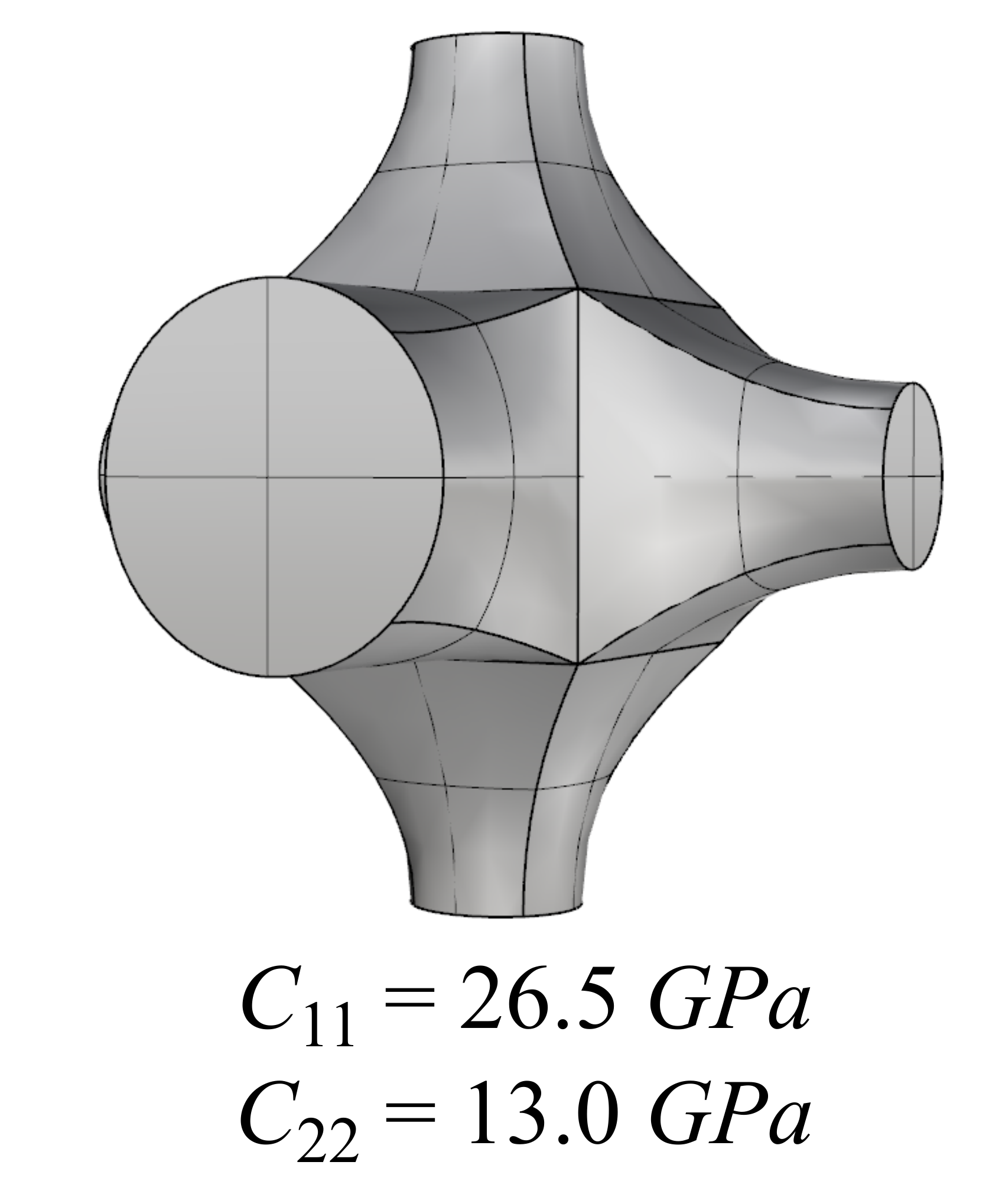}} & 
	\raisebox{-27mm}{\includegraphics[height=30mm]{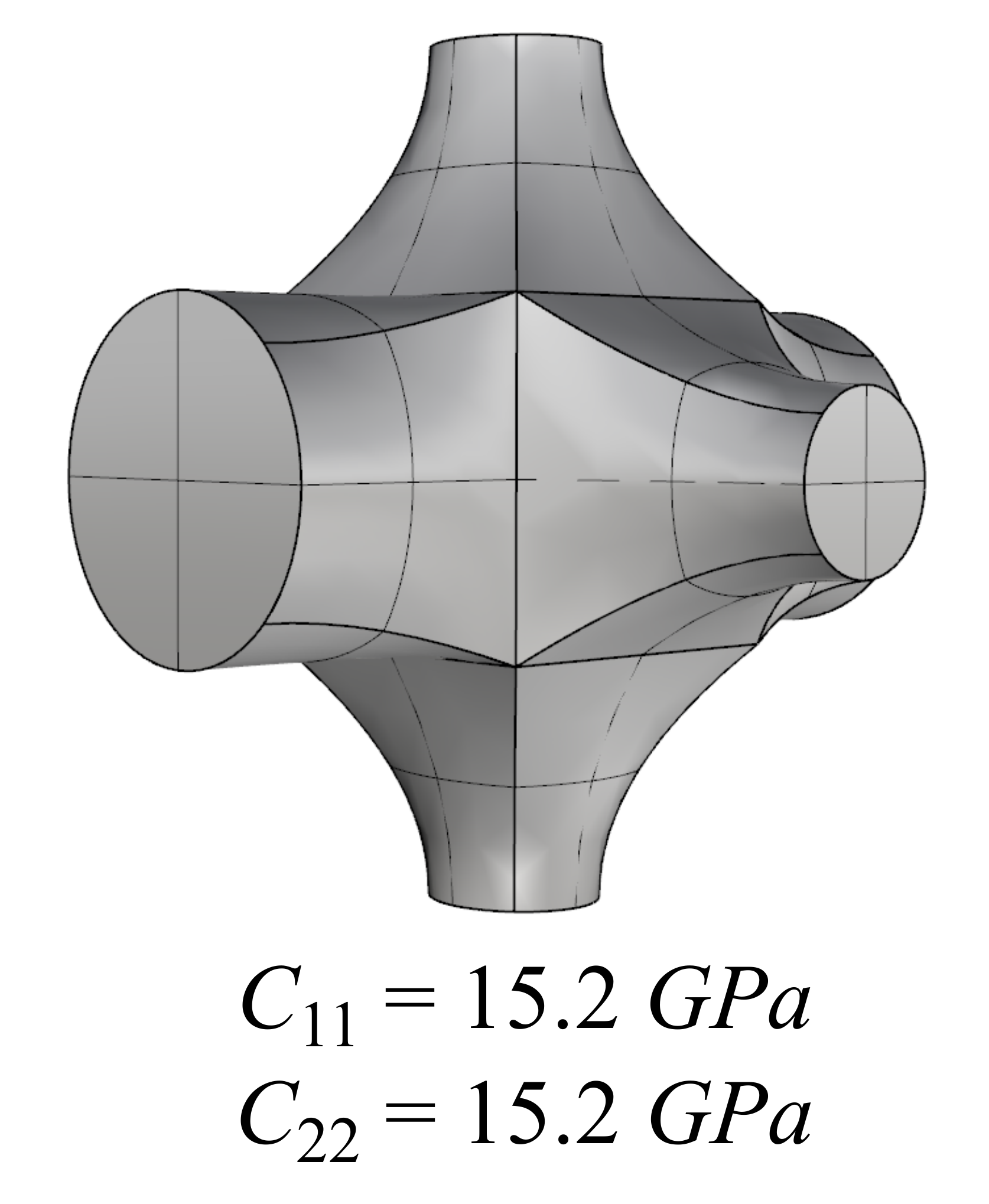}} \\ \cline{1-5}
\end{tabular}
\caption{Exemplary look-up table for the effective elasticity tensors (here represented by $C_{11}$ and $C_{22}$) for changing diameters of the rod in $x-$direction and rotations around the $z-$axis.}
\label{tab:lookUpTable}
\end{table}
    
\subsubsection{Example 5: Homogenized microstructure} \label{sec:Example5}
\begin{figure}[H]
	\centering
    \includegraphics[width=0.75\textwidth]{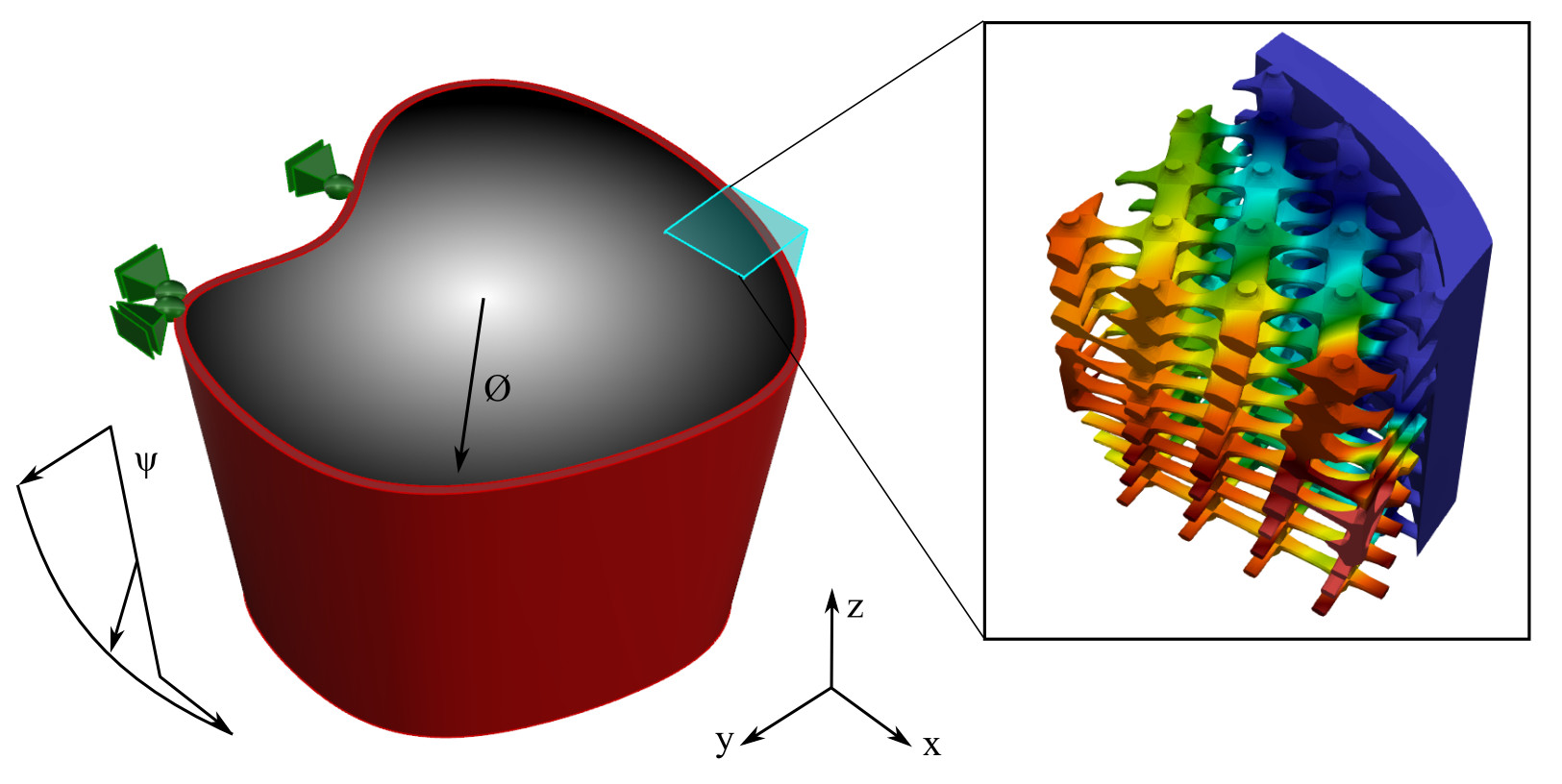}
    \caption{Structure consisting of a solid shell (red) and a homogenized microstructure (gray scale).}
    \label{fig:HomogenFGMModel}
\end{figure}

Consider the model of Example~\ref{sec:Example3} to be a part of a larger structure (see Figure~\ref{fig:HomogenFGMModel}). Based on the material database for the homogenized unit tiles (see Table~\ref{tab:lookUpTable}), it is possible to simulate such a structure \blue{with a homogenized material. Similar to Example~\ref{sec:Example3}, the corresponding geometric parts are modeled as B-rep models.} For the simulation, the model is subdivided into an outer shell and an infill. The shell is considered to be of solid isotropic material. In contrast, the infill is a homogenized microstructure which continuously changes the two \blue{known} properties: the rotation angle $\psi$ around the $z-$axis varies from $0^{\circ}$ at the bottom to $90^{\circ}$ at the top and the thickness of the rod \O~increases from the center $z-$axis of the infill (\O~= 0.2 mm) towards the interface of the shell (\O~= 0.4 mm). A uni-axial compression state is achieved by applying a uniform displacement of $\Delta z = -1.0$ on the top surface and restricting the displacements in the $z-$direction on the bottom surface. Three additional point-bearings block the rigid body motions.

\noindent The simulation uses $15\times 15\times 15$ high-order Legendre finite cells with a polynomial degree of $p=4$. For the integration, moment-fitting with the depth of an underlying octree of $d=4$ is chosen. At the interface between shell and infill, \blue{one $h-$refinement step is carried out} to capture the material discontinuity. As the unit tiles' homogenization was carried out with periodic boundary conditions, the behavior at the interface between shell and infill is not captured precisely. However, the affected domain is small compared to the overall structure. Thus, the introduced error is negligible. If, however, the microscopic stress state at the transition from the micro-tiles to the shell is of interest, then a geometrically resolving simulation as in Example~\ref{sec:Example3} can be performed.
 
\begin{figure}[H]
	\begin{tabular}{c c}
		\begin{subfigure}{0.45\textwidth}
			\centering
			\includegraphics[width=\textwidth]{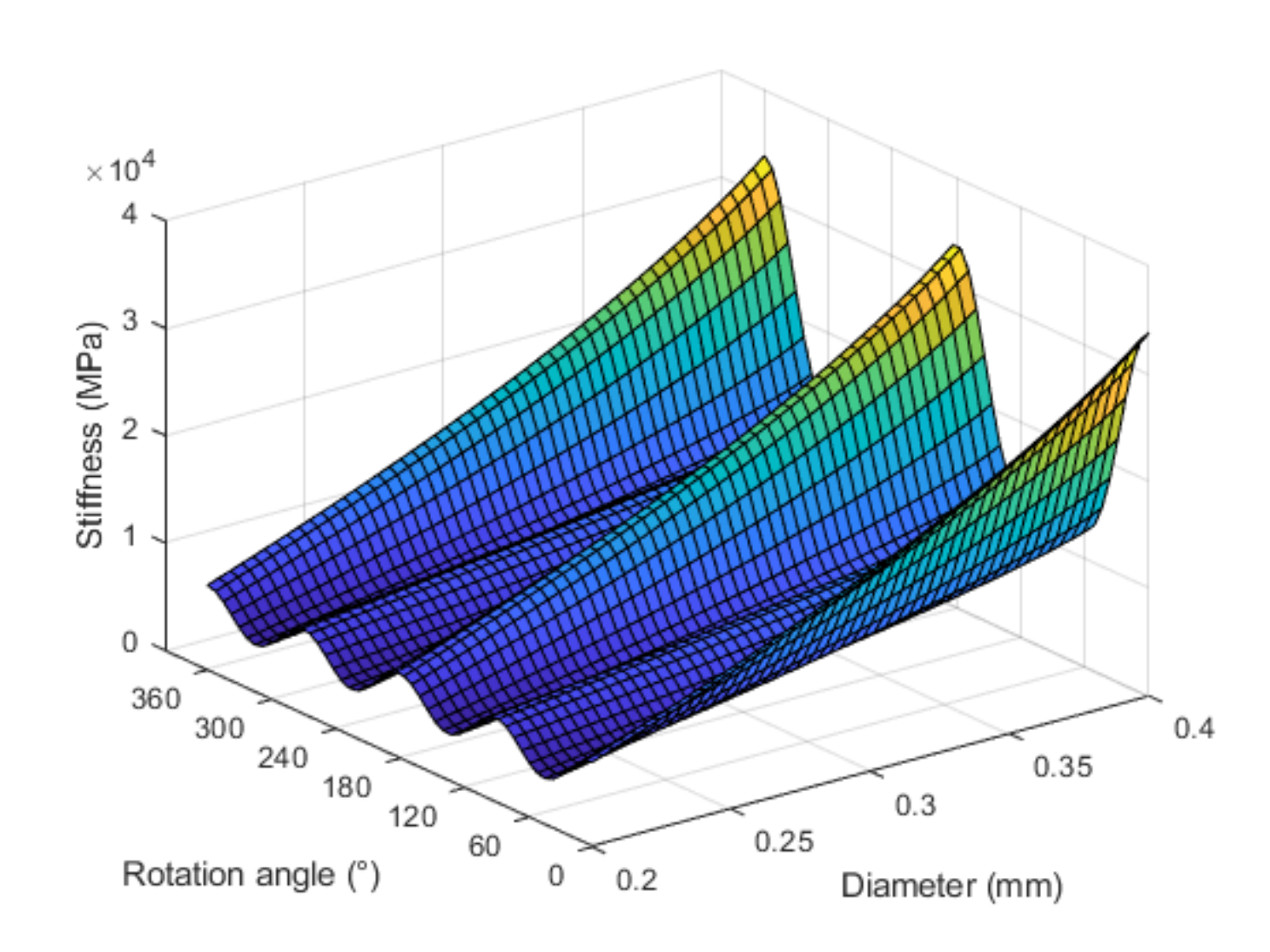}
			\caption{$C_{11}$}
		\end{subfigure}	 &
		\begin{subfigure}{0.45\textwidth}
			\centering
			\includegraphics[width=\textwidth]{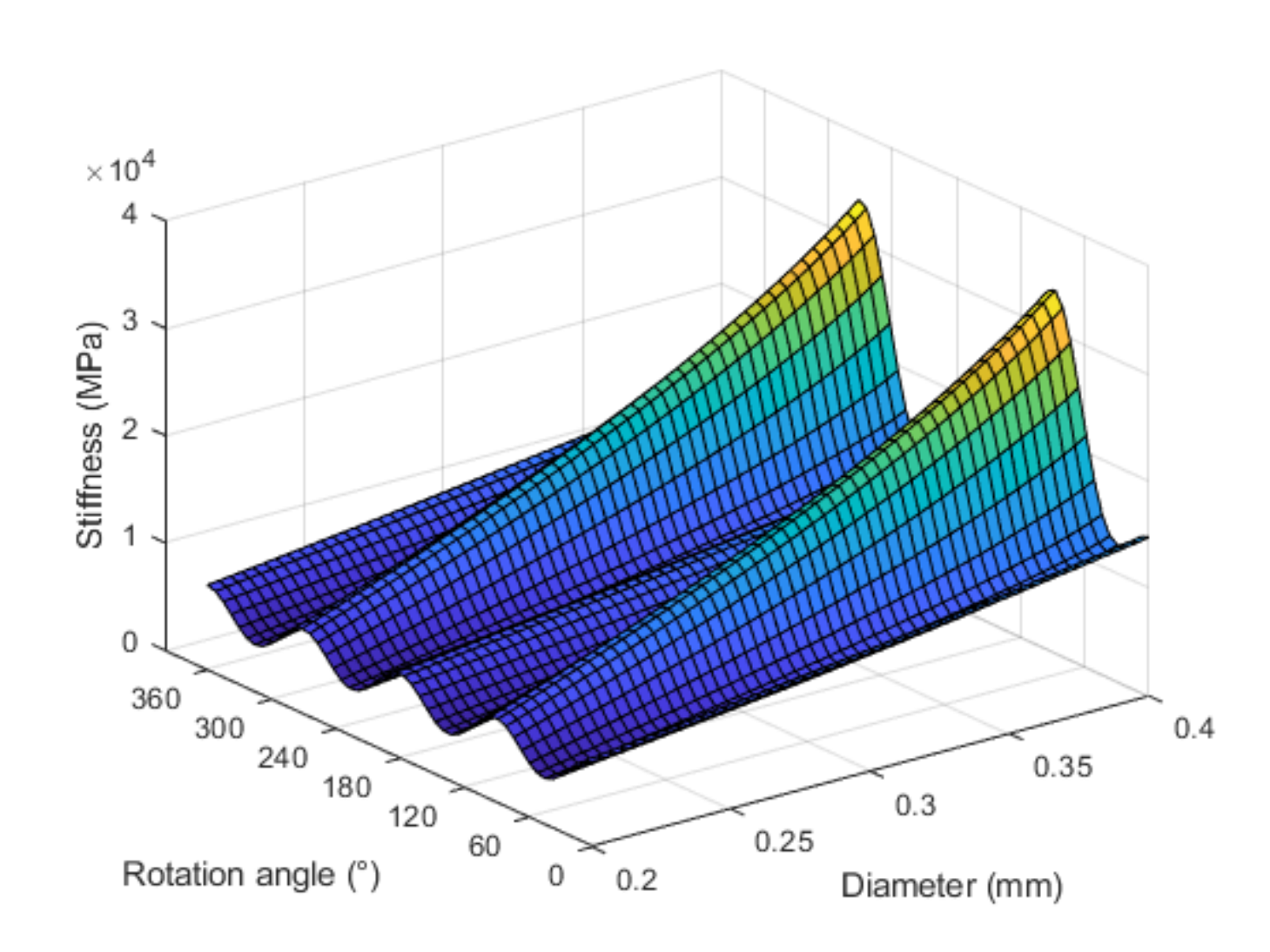}
			\caption{$C_{22}$}
		\end{subfigure}
	\end{tabular}
	\caption{ Spline based interpolation of the material coefficients $C_{11}$ and $C_{22}$.}
	\label{fig:CoeffInterp}
\end{figure}

A total of 13 independent material coefficients are required to evaluate the material tensor of the continuously changing microstructure. To this end, the material coefficients \blue{that were computed in Example~\ref{sec:Example4} and that are stored in} a look-up table (see Table~\ref{tab:lookUpTable}) are interpolated using spline fitting. Figure~\ref{fig:CoeffInterp} exemplary shows the interpolation for the material coefficients $C_{11}$ and $C_{22}$ of the homogenized material tensor shown in Equation (\ref{eq:independentCii}).
\begin{equation}
\bm{C}(\varnothing(\bm{x}), \psi(\bm{x})) = 
\begin{bmatrix}
C_{11} & C_{12} & C_{13} & C_{14} & 0  & 0   \\
  	   & C_{22} & C_{23} & C_{24} & 0  & 0   \\
       &        & C_{33} & C_{34} & 0  & 0   \\
       &        &        & C_{44} & 0  & 0   \\
       &        &        &        & C_{55}& C_{56} \\
symm.  &        &        &        &       & C_{66}
\end{bmatrix}
\label{eq:independentCii}
\end{equation}

\begin{figure}[H]
	\begin{tabular}{c c}
		\begin{subfigure}{0.47\textwidth}
			\centering
			\includegraphics[width=\textwidth, trim={13cm 2cm 8cm 2cm}, clip]{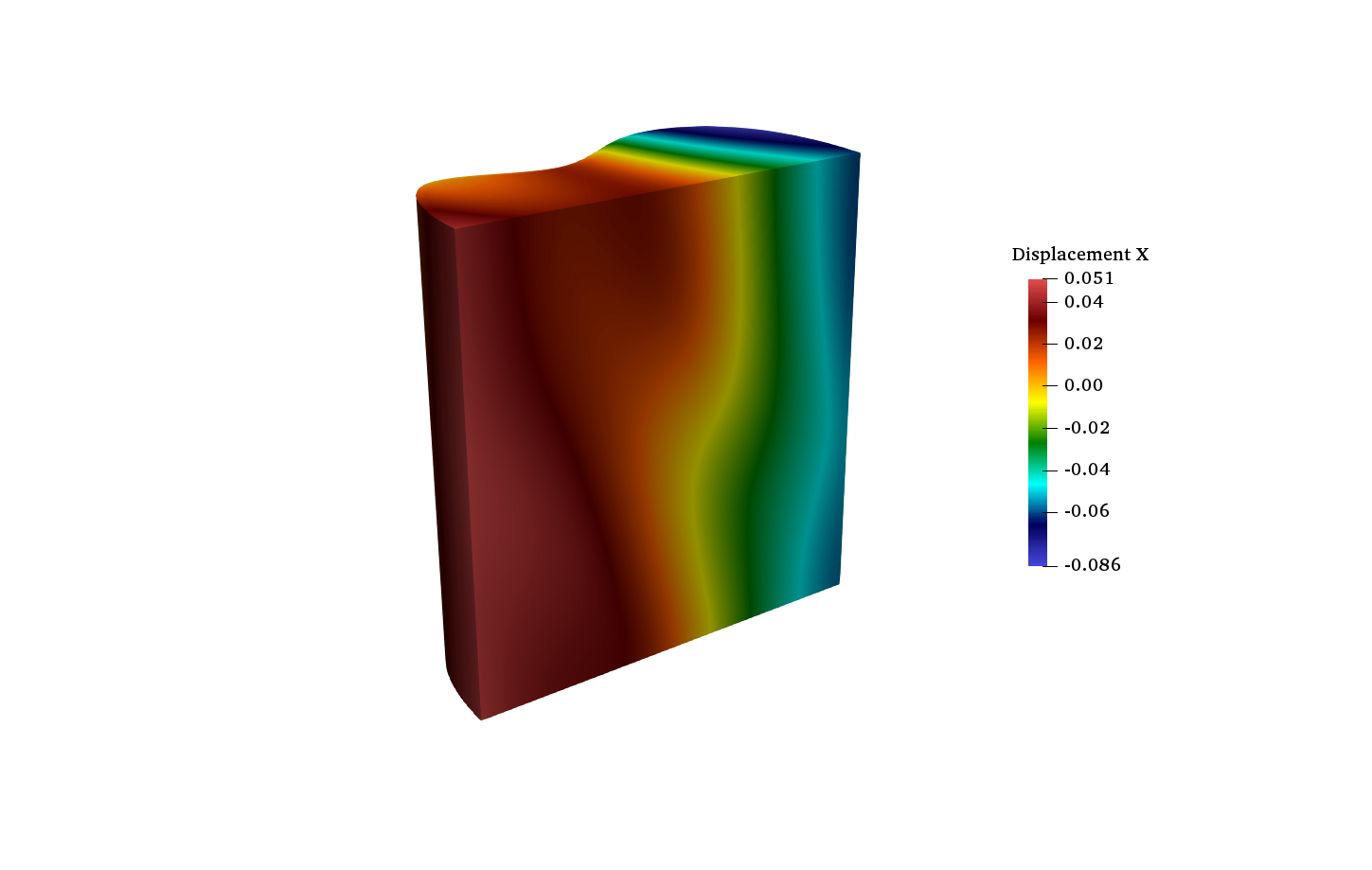}
			\caption{}
		\end{subfigure}	 
		&
		\begin{subfigure}{0.47\textwidth}
			\centering
			\includegraphics[width=\textwidth, trim={13cm 2cm 8cm 2cm}, clip]{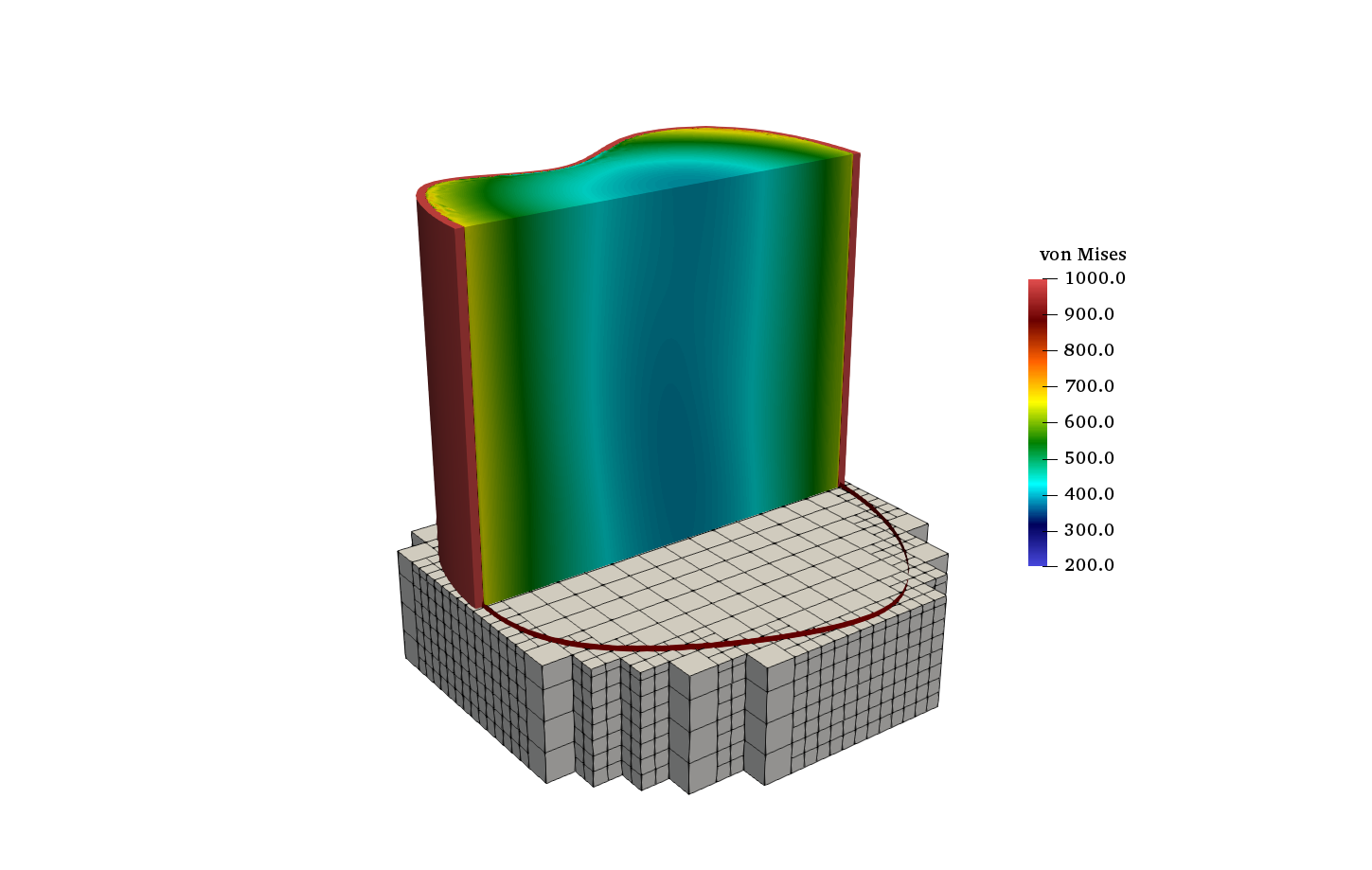}
			\caption{}
		\end{subfigure}
	\end{tabular}
	\caption{(a) Displacements in $x-$direction and (b) von Mises stresses with the finite cell mesh.}
	\label{fig:HomogenFGMResults}
\end{figure}

\noindent Figure~\ref{fig:HomogenFGMResults} shows the displacements in $x-$direction and von Mises stresses of the structure under uni-axial compression $z-$direction. The load is mainly transferred through the stiffer shell, yet the infill's contribution cannot be neglected. Due to the uni-axial compression, the rotation angle $\psi$ of the microstructure has only little influence. The thickness of the rod \O, on the other hand, can be deducted directly from the stress field of the infill.

\begin{figure}[H]
	\begin{tabular}{c c c}
		\begin{subfigure}{0.3\textwidth}
			\centering
			\includegraphics[width=\textwidth]{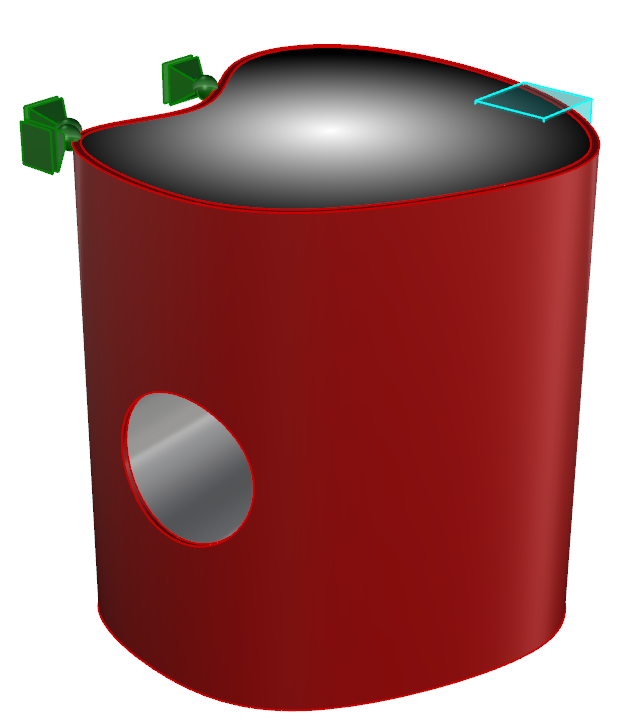}
			\caption{}
		\end{subfigure}	 
		&
		\begin{subfigure}{0.3\textwidth}
			\centering
			\includegraphics[width=\textwidth]{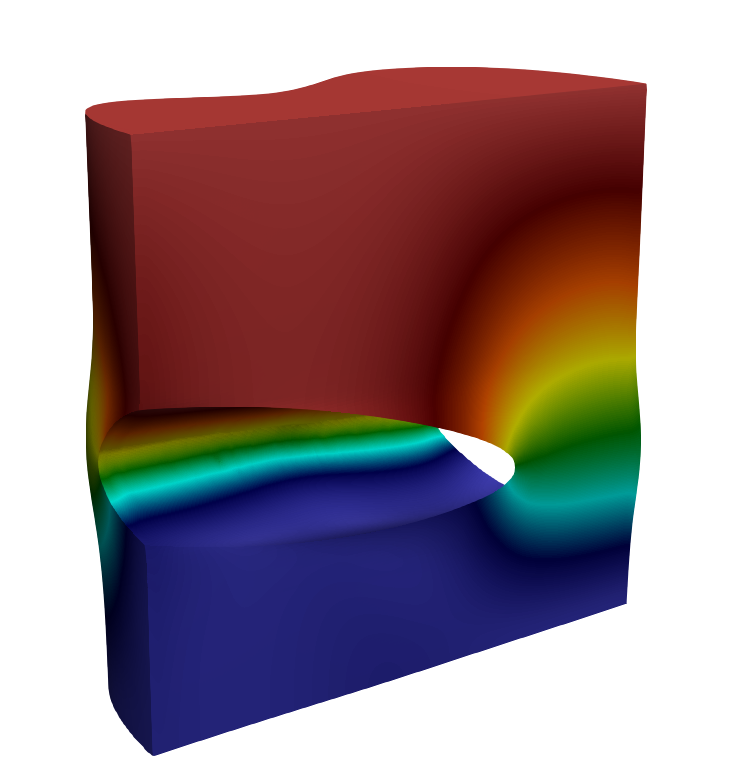}
			\caption{}
		\end{subfigure}
		&
		\begin{subfigure}{0.3\textwidth}
			\centering
			\includegraphics[width=\textwidth]{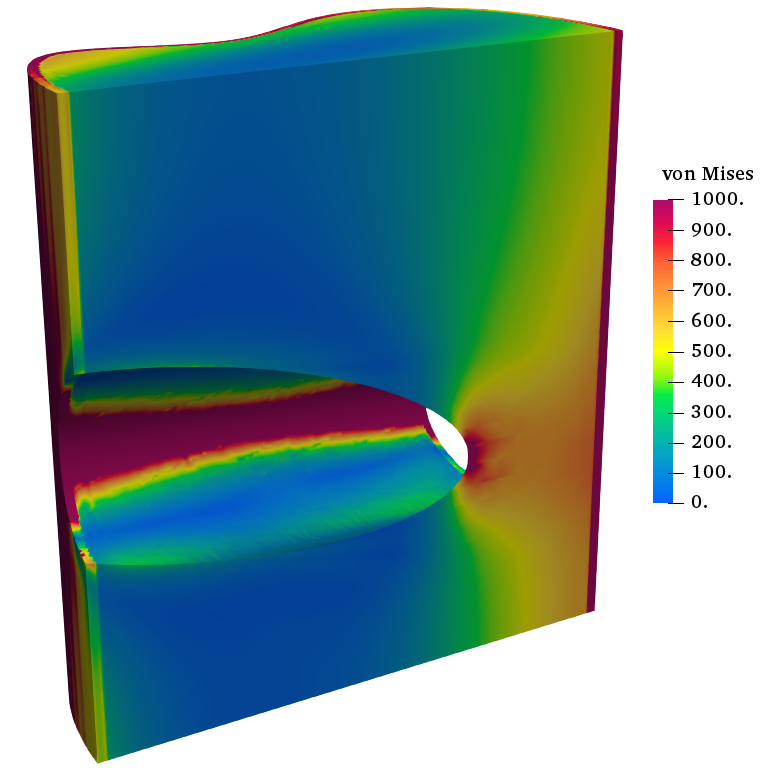}
			\caption{}
		\end{subfigure}
	\end{tabular}
	\caption{Structure with hole: (a) model, (b) displacements in $z-$direction (warped by a factor of $s=2$), and (c) von Mises stresses.}
	\label{fig:HomogenFGMHoleResults}
\end{figure}

\noindent It should be noted that a geometrical change does not influence the overall workflow. Even a topological change does not lead to a re-meshing, as it would be required for a simulation with classical FEM or IGA. In order to illustrate such a topological change, a hole is drilled through the structure (see Figure~\ref{fig:HomogenFGMHoleResults}). In the context of the FCM, a cylinder is subtracted with a Boolean difference. As can be seen, the infill contributes less to the load transfer, and high stress concentrations appear at the hole walls.

\section{Conclusions} \label{sec:conclusion}
\blue{In this paper, three novel methodologies were presented: (a) At first, the FCM was extended to V-models, as novel CAD representation form. As V-rep is based on a tri-variate spline-formulation, the inversion -- that is necessary for the point inclusion test -- turns out to be costly, in particular in cases where due to the geometric complexity of the model a large number of integration points has to be used. In these cases the definition of an auxiliary B-rep model using ray tracing for the point membership test turns out to improve the computational performance significantly.}

\blue{(b) Secondly, the FCM was extended to multi-material FGM. For this, the dimension of the V-cells' control points was increased to carry material information, as well. During the integration -- apart from the point-inclusion test -- also the material properties are retrieved. The spline-based description of the V-cells renders the V-rep framework perfectly suitable to model smooth material distributions. Yet, also rapidly changing materials can be represented using knot-insertion.}

\blue{(c) Finally, an efficient method for the simulation of large-scale single-material FGM -- in this case continuously changing microstructures -- was presented. Using the microstructures' parametric description, representative unit tiles can be selected on which homogenization simulations provide effective material properties. Material properties for adjacent parameter sets are then interpolated using these values. Although this approach allows the efficient simulation of large-scale microstructures, two problems arise. Firstly, depending on the microstructure's complexity and the amount of varying geometrical features, the number of representative unit tiles might become large. As for each of these unit tiles, an individual homogenization simulation needs to be carried out. Thus, these structures can become demanding in memory consumption as well as in computational time. And secondly, the homogenization simulations with periodic boundary conditions provide only precise microstructure results, yet not at the interface to another material or a free surface. However, provided this interface or surface area is small compared to the overall domain. Considering that such kind of boundary layer effects usually vanish rapidly away from the interface, the error is not dominant.}

\section{Declarations}
\subsection{Availability of data and materials}
The geometric models simulated and analysed during the current study are either reproducible with the provided information, or available from the corresponding author on reasonable request.

\subsection{Competing interests}
The authors declare that they have no competing interests.

\subsection{Funding}
We gratefully acknowledge the support of the German Research Foundation (DFG) under Grant No. Ra 624/22-2. We would also like to thank the German Research Foundation (DFG) for its support through the TUM International Graduate School of Science and Engineering (IGSSE), GSC 81. Furthermore, we gratefully acknowledge the support of the Transregional Collaborative Research Centre SFB/TRR 277 ”Additive Manufacturing in Construction. The Challenge of Large Scale”, funded by the German Research Foundation (DFG).

\subsection{Authors' contributions}
BW was the corresponding author who wrote the central part of the paper, integrated the Irit geometry kernel into the Adhoc++ FCM framework, and carried out most of the simulations. NK wrote the section on homogenization and carried out the simulation and classification of the unit-tiles. SK was responsible for the content regarding the finite cell method and implementation issues. ER guided the paper's general structure and contents, cross-checked the results, and proposed most examples. GE provided the geometry kernel and assistance for its access. Additionally, he was responsible for the content of the V-reps. All authors read and approved the final manuscript.

\subsection{Acknowledgements}
We acknowledge the contributions of the research groups at the chair of Computation in Engineering regarding the development of the finite cell method framework \textit{Adhoc++} and at the Center for Graphics and Geometric Computing concerning the development of the geometry kernel \textit{Irit} and CAD software \textit{GuIrit}.

\section{Appendix}
\begin{figure}[H]
	\begin{tabular}{c c c}
		\begin{subfigure}{0.3\textwidth}
			\centering
			\includegraphics[width=\textwidth]{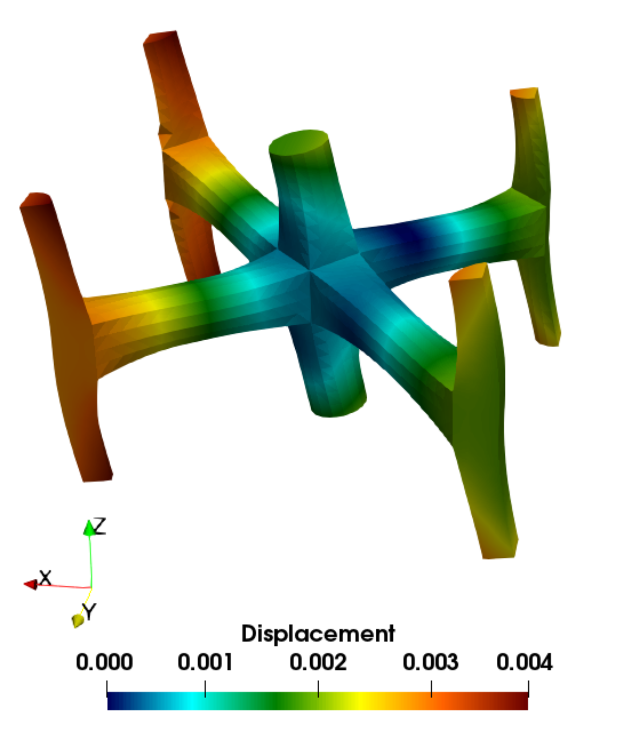}
			\caption{ Tile 1}
		\end{subfigure}	 &
		\begin{subfigure}{0.3\textwidth}
			\centering
			\includegraphics[width=\textwidth]{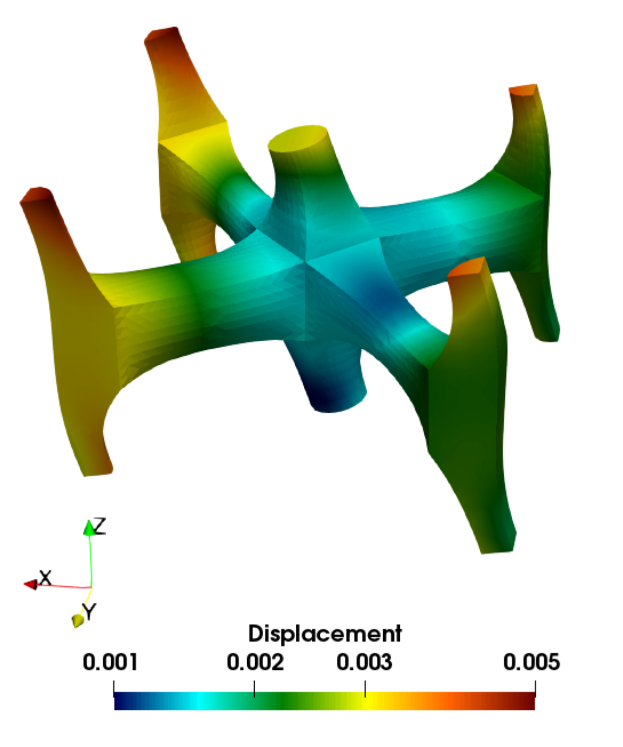}
			\caption{ Tile 2}
		\end{subfigure}	  &
		\begin{subfigure}{0.3\textwidth}
			\centering
			\includegraphics[width=\textwidth]{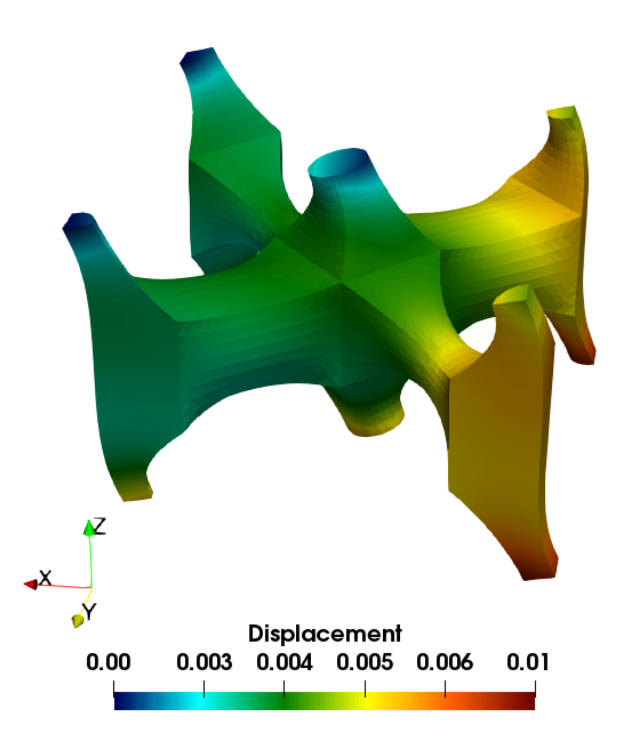}
			\caption{Tile 3}
		\end{subfigure}	  
	\end{tabular}
	\caption{ Displacement field of the warped rotated tiles with a scale factor $s=10$.}
	\label{fig:PBCRotatedTiles}
\end{figure}

\paragraph{Effective material tensors of unit tiles}\label{sec:appendixEffectiveMaterialTensors}
\blue{This section is an extension of Example~\ref{sec:Example4} and provides material data, which is used in Example~\ref{sec:Example5}.}
\blue{The following polar diagrams depict the independent entries $C_{ii}$ of the three material tensors (of the unit tiles) for an arbitrary rotation around the $z-$axis. The values are computed with the Bond transformation matrices~\cite{Bond1943}, according to Equation~(\ref{eq:transformationRotation}). Thus, at an angle of $0\si{\degree}$ the value equals the corresponding entry of the respective unrotated material tensor $\bm{C}^*_{Ti}$ of Example~\ref{sec:Example4}. Additionally, for a rotational degree of $45\si{\degree}$ the results are numerically verified (see Figure~\ref{fig:PBCRotatedTiles}).}

\begin{figure}[H]
	\centering
	\includegraphics[width=\textwidth]{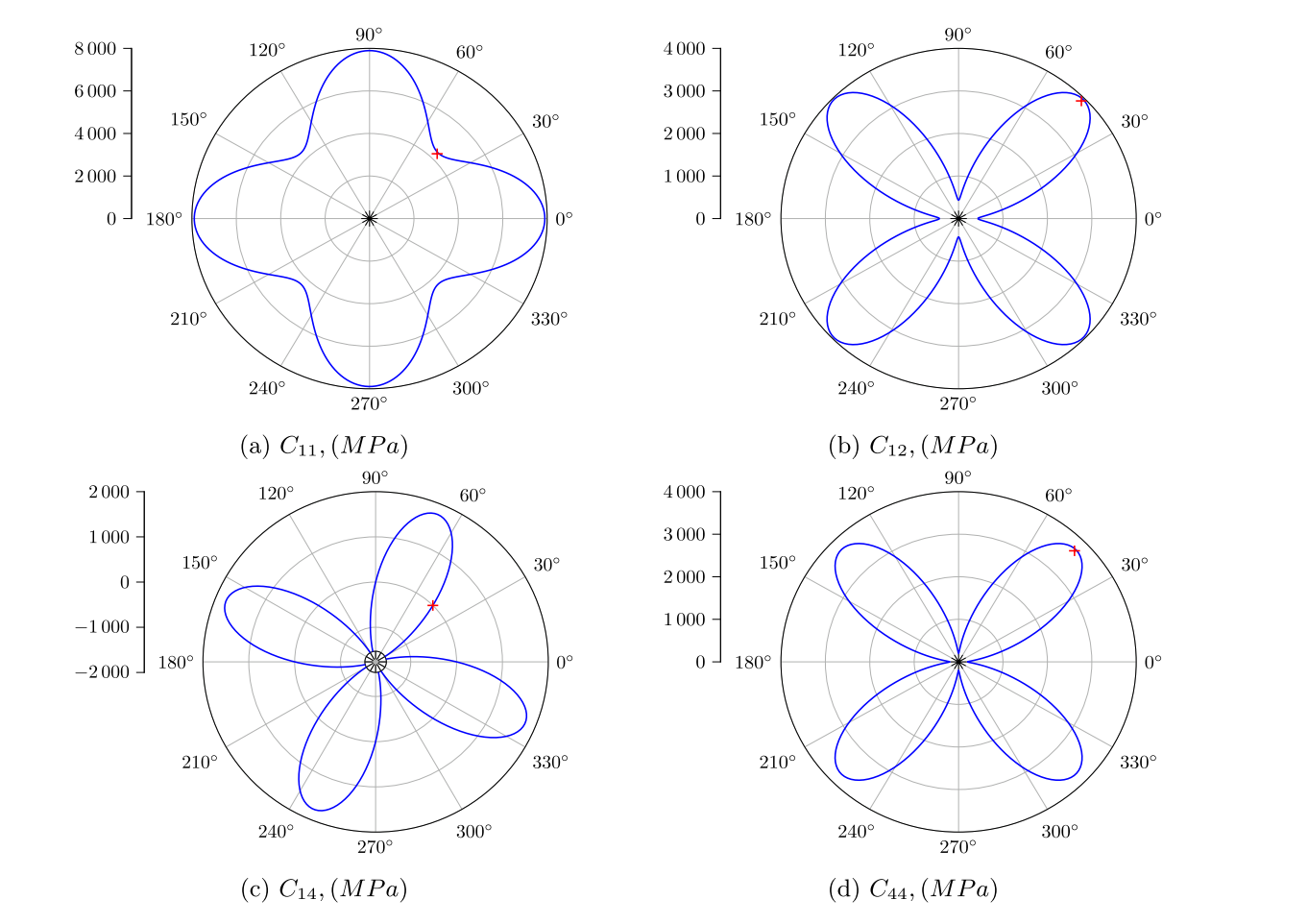}
	\caption{ Independent elastic constants for tile 1 under rotation around the $z-$axis.}
	\label{fig:RotationTile1}
\end{figure}

\noindent A rotation of tile 1 around the $z-$axis does not influence the third, fifth, and sixth columns, neither on the respective rows of the effective tensor. The coefficient $C_{11}$ equals $C_{22}$ due to the geometrical symmetry in $x-$ and $y-$direction. $C_{14}$ and $C_{24}$ are of equal magnitude but have opposite signs. Figure~\ref{fig:RotationTile1} shows the remaining independent material constants with respect to the rotational angle. The results of the numerical simulation at $45\si{\degree}$ are indicated with red crosses.

\begin{figure}[H]
	\centering
	\includegraphics[width=0.98\textwidth]{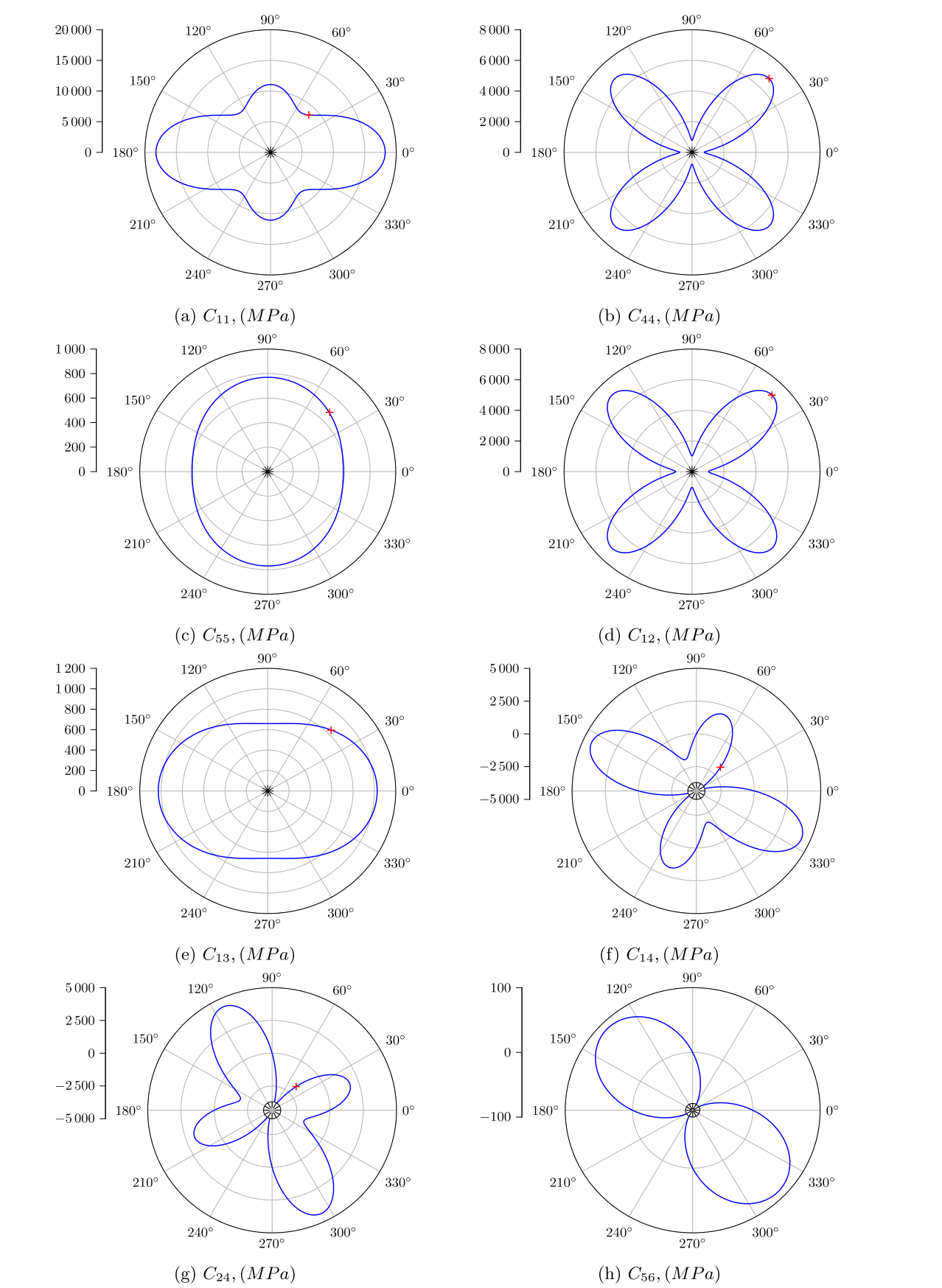}
	\caption{ Independent elastic constants for tile 2 under rotation around the $z-$axis.}
	\label{fig:RotationTile2}
\end{figure}

\noindent For tile 2, only the coefficient $C_{33}$, which corresponds to the stiffness in $z-$direction, remains unchanged under rotation around the $z-$axis. All other entries are affected by the altered symmetry. Considering a rotation angle of $90^\circ$, it is noteworthy that the coefficients $C_{11}$ and $C_{22}$ are switched concerning the initial position. The same holds for the coefficient pairs $C_{55}$-- $C_{66}$, and $C_{13}$--$C_{23}$. The rest of the independent material parameters are depicted in Figure~\ref{fig:RotationTile2}. Again, the results of the numerical simulation at $45^{\circ}$ are marked with red crosses.

\begin{figure}[H]
	\centering
	\includegraphics[width=0.98\textwidth]{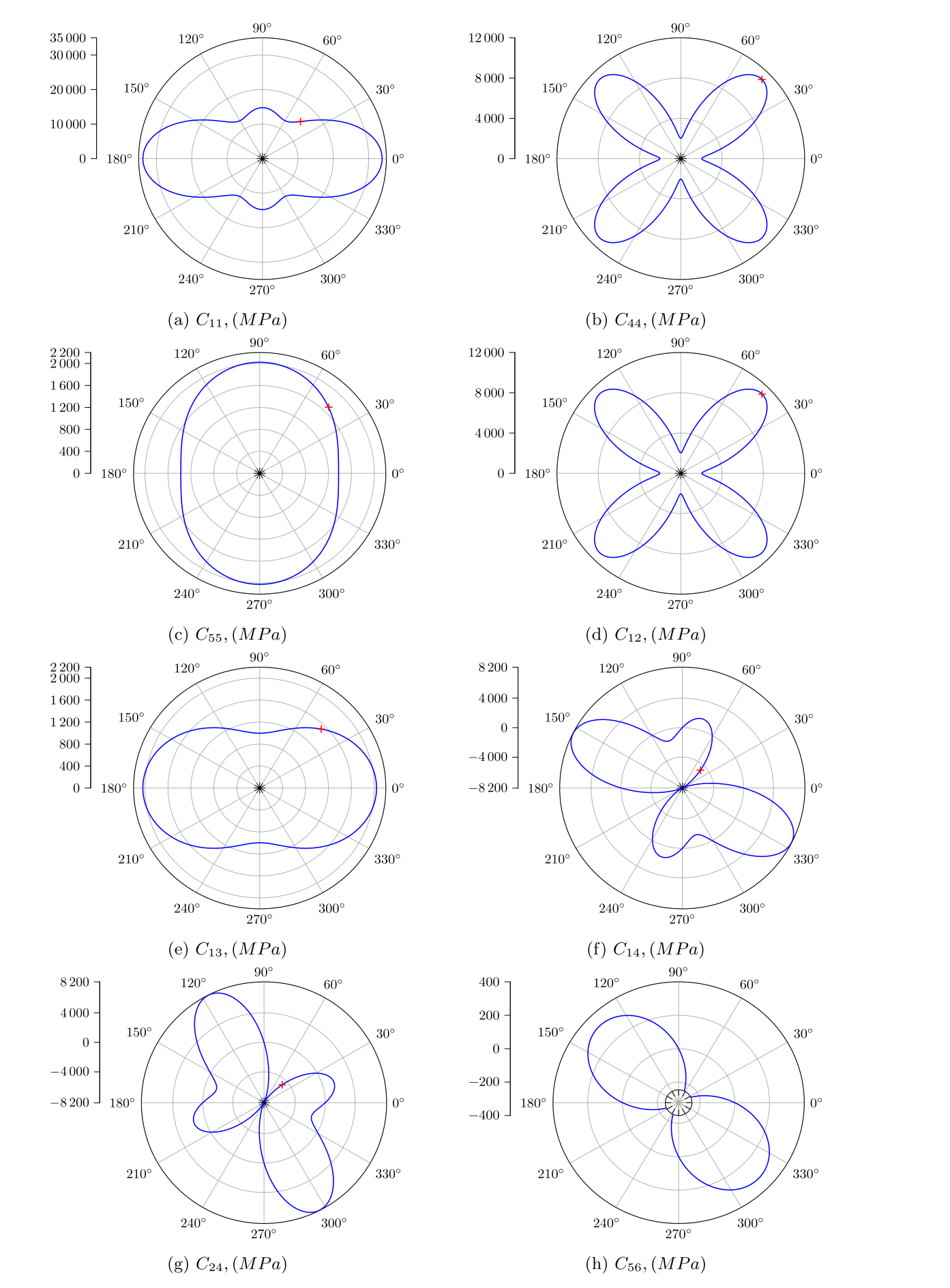}
	\caption{ Independent elastic constants for tile 3 under rotation around the $z-$axis.}
	\label{fig:RotationTile3}
\end{figure}

\noindent Tile 3 exhibits similar material symmetries as the second tile. Figure~\ref{fig:RotationTile3} shows the material coefficients. Again, the results of the numerical simulation at $45^{\circ}$ are marked with red crosses.

\bibliographystyle{ieeetrDoi}
\bibliography{library}

\end{document}